\documentclass{article}
\usepackage{graphicx} 
\usepackage{tabularx}
\usepackage{amsmath,amsthm}
\usepackage{amssymb}
\newtheorem{theorem}{Theorem}
\usepackage{algorithm}
\usepackage[T1]{fontenc}
\usepackage{hyperref}
\usepackage[utf8]{inputenc}
\usepackage[numbers]{natbib}
\usepackage{breqn}
\usepackage{float}
\hypersetup{hidelinks}
\usepackage{enumitem} 
\usepackage[font=small,labelfont=bf]{caption}
\usepackage{algpseudocode}

\usepackage{amsmath,amssymb,amsthm}
\usepackage{graphicx}
\usepackage{float}
\usepackage{booktabs}
\theoremstyle{plain}
\newtheorem{definition}{Definition}[section]
\newtheorem{proposition}{Proposition}[section]

\newtheorem{lemma}[theorem]{Lemma}
\usepackage{cleveref}
\newtheorem{corollary}[theorem]{Corollary}
\theoremstyle{definition}
\newtheorem{axiom}[theorem]{Axiom}

\usepackage[a4paper,
            left=4cm,   
            right=3.5cm,  
            top=2.5cm,    
            bottom=2.5cm  
           ]{geometry}

\title{Markov Blanket Density and

Free Energy Minimization}
\author{Luca M. Possati}
\date{July 2025}

\begin{document}

\maketitle

\begin{abstract}
\noindent

This paper introduces a novel geometrical framework for understanding the conditions under which the Free Energy Principle (FEP) can be meaningfully applied. Departing from formulations that treat the FEP as a universal variational imperative governing all self-organizing systems, we define a spatially continuous field called the \emph{Markov Blanket Density} $\rho(x) \in [0,1]$. This quantity measures, at each point in a spatial domain $\Omega \subset \mathbb{R}^n$, the degree of conditional independence between internal and external states given a local blanket. Rather than assuming the existence of a Markov blanket as a binary structural feature of the system, we treat conditional independence as a graded and spatially variable property. We show that the standard components of the FEP---including generative modeling, variational inference, and action as free energy minimization---are only definable in regions where $\rho(x) < 1$, i.e., where conditional independence fails and mutual information across the internal-external divide remains non-zero. In regions where $\rho(x) = 1$, no inferential coupling is possible, and the variational free energy becomes ill-defined or vacuous. This implies that the FEP is not a general law but a locally valid phenomenological regularity, contingent on the informational geometry of the environment. The model introduces a differential-geometric structure on the informational field $\rho(x)$, defining an information-theoretic metric, a Laplacian curvature operator, and a log-transformed free energy landscape. In this setting, epistemic dynamics are reinterpreted as gradient flows on $\rho(x)$, and agent behavior becomes a response to informational topography. This approach provides a new foundation for the FEP as a situated, spatially constrained phenomenon, and offers a principled framework for modeling inference in distributed, multi-agent, or weakly bounded systems where classical blanket structures are unstable or undefined. The result is a shift in the explanatory architecture of the FEP: from a foundational principle imposed on all systems, to an emergent regime of behavior permitted by the structure of the environment. We argue that this reframing offers a more falsifiable, flexible, and geometrically grounded interpretation of inference and self-organization, with implications for formal epistemology, situated cognition, and the design of epistemic systems in both natural and artificial contexts.
\end{abstract}

\section{Introduction}

The Free Energy Principle (FEP) provides a powerful framework to understand how agents (i.e., self-organizing systems such as living systems) maintain their structure by minimizing variational and expected free energy \cite{1, 2, 3, 4, 9, 20, 31, 37}. Central to FEP is the Markov blanket, traditionally viewed as a discrete boundary separating internal states from external environmental states. However, this binary view limits our ability to model nuanced interactions and spatial dynamics.

In this paper, we introduce "Markov blanket density" as a continuous scalar field quantifying the degree of conditional independence between internal and external states at every spatial point relative to an observer and their scale of observation. Blanket strength is thus measured by how effectively bkanket states mediate interactions, structuring space into continuous gradients of coupling. Preferred states become regions of optimal coupling rather than purely internal homeostatic targets. Although the fundamental idea—that agents naturally move towards regions of lower Markov blanket density (greater coupling)—is intuitive, the paper offers originality through: (a) Shifting from discrete partitions to a continuous scalar field, allowing nuanced spatial modeling; (b) Rigorous mathematical formalization capturing precise, verifiable dynamics through spatial gradients; (c) Practical applicability, providing a robust framework for empirical predictions and novel simulations.

We think that classic active inference fails to properly account for the spatial dimension, collapsing it into a notion of space as an empty, passive, and predictable “environment.” In doing so, active inference cannot fully grasp the concept of affordance, reducing it to a set of predictions about the environment. Essentially, active inference remains captive to a lab‐based perspective, where space adapts to hypotheses rather than hypotheses adapting to space. The point is that space is complex, as are affordances. The very unity of perception and action depends on that complexity.

Through detailed mathematical analysis, this paper demonstrates how free energy minimization dynamics depend on variations in Markov blanket (MB) density, including scenarios that invert typical inference dynamics. By bridging ecological and embodied perspectives with formal variational inference, this work advances our understanding of the embodied mind as actively embedded within dynamically structured informational environments. The purpose of this paper is to introduce an informational geometry of inferential space: not only “what an agent does,” but “where it does it,” how much it can infer, at what speed, and with what access to information. The central thesis is that the FEP should not be treated as a universal law, but rather as an emergent regime of behavior permitted by the geometry of $\rho(x)$. We formalize this framework by constructing an informational metric, a free energy landscape defined via log-transform of $\rho(x)$, and a dynamic formulation in which agent trajectories follow informational gradients. This reinterpretation positions the FEP not as a principle to be imposed \emph{a priori}, but as a local affordance of the informational structure of the environment—one that may fail to hold, shift over time, or vary across systems. In so doing, we aim at extendind the explanatory power of the FEP to a broader class of systems, including weakly bounded agents, distributed cognitive architectures, and informationally heterogeneous environments. Our aim is not to reject the FEP, but to recover it as a special case within a more general geometric and epistemic framework—one that better accommodates the spatial, graded, and context-sensitive nature of real-world inference. In other words, the FEP emerges as a local property of regions with good informational permeability.

As regards the structure of this paper, Section 2 introduces some key concepts from active inference and the notion of the Markov blanket. Section 3 introduces the central thesis of the paper, connecting it to existing literature and offering several conceptual clarifications. Beginning with Section 4, the mathematical framework is laid out. A clear definition of Markov blanket density is provided, followed by an operational definition in Section 4.4—namely, an algorithm for identifying and computing the Markov blanket density at each point in a hypothetical space. Three theorems (Sections 5–9) follow, illustrating the relationship between free energy minimization and Markov blanket density. Sections 10–11 present a fourth theorem, which demonstrates how the temporal evolution of free energy minimization depends on and is continually modulated by the Markov blanket density. Section 12 addresses the problem of how an active inference agent learns the space and how Markov blanket density is incorporated into the agent’s inferential dynamics. Section 13 offers a complete axiomatic derivation of the Markov blanket density concept, showing how the FEP can be seen as a local emergent effect of this density.

Throughout the paper, a series of figures illustrates the results of various simulations. The appendices cover technical details.

\section{The Free Energy Principle}

\subsection{A basic outline}

The FEP is a mathematical framework rooted in statistical physics, information theory, and variational inference techniques from machine learning \cite{9, 22}. It provides a unifying account of self-organizing systems by interpreting their dynamics in terms of the minimization of variational free energy. In particular, consider a random dynamical system that satisfies the following conditions:
\begin{itemize}
    \item it exhibits a degree of ergodicity, allowing time-averaged behavior to approximate ensemble statistics;
    \item it possesses a pullback attractor, that is, a set of states toward which the system tends over time — its "preferred" or most frequently occupied states;
    \item it admits an ergodic density that probabilistically characterizes long-term state occupancy; and
    \item it maintains a degree of separation from its environment, such that internal and external states can be distinguished (e.g., via a Markov blanket structure).
\end{itemize}
\textit{Under these assumptions}, the system's behavior can be interpreted as performing approximate Bayesian inference by minimizing a quantity known as variational free energy. In this context, the flow of states (e.g., internal states, active states) follows a gradient descent on variational free energy, which serves as an upper bound on the system’s surprisal (or self-information, see Table 1) about its sensory states. That is, even in the absence of an explicit model, the system behaves as if it were inferring the causes of its sensory inputs and acting to maintain itself within a bounded set of preferred states — thereby resisting disorder and preserving its structural and functional integrity.
\begin{table}[h]
\small
\renewcommand{\arraystretch}{1.6}
\centering
\begin{tabular}{|p{0.95\textwidth}|}
\hline
\textbf{Self-information} $I(x)$ \\ 
Surprise or informativeness of a specific outcome $x$. High for rare events, zero for certain ones. \\ 
\hspace{1em} Formal definition: $I(x) = -\log_b p(x)$, where $p(x) \in (0,1]$ and $b$ is typically 2 (bits), $e$ (nats), or 10 (Hartleys). \\
\hline
\textbf{Entropy} $H(X)$ \\
Expected uncertainty or average surprise over all outcomes of a random variable $X$. \\
\hspace{1em} Formal definition: $H(X) = -\sum_{x \in \mathcal{X}} p(x) \log_b p(x)$ \\
\hline
\textbf{Kullback–Leibler divergence} $D_{\mathrm{KL}}(P \parallel Q)$ \\
Information lost when using distribution $Q$ to approximate the true distribution $P$. \\
\hspace{1em} Formal definition: $D_{\mathrm{KL}}(P \parallel Q) = \sum_{x \in \mathcal{X}} p(x) \log_b \left( \frac{p(x)}{q(x)} \right)$, defined only if $p(x) > 0$ implies $q(x) > 0$. \\
\hline
\end{tabular}
\caption{Essential definitions of self-information, entropy, and KL divergence used in the FEP framework.}
\end{table}
As mentioned, the FEP is a mathematical modeling framework. It is not a theory seeking empirical validation, but rather a mathematical-physical formalism that can be used to generate new hypotheses or analyze data. In itself, however, it remains a purely theoretical construct, without predictive aims.

When applied to living systems (e.g., the brain), the FEP gives rise to what is known as active inference \cite{30, 9}. In this framework, a living system maintains its structural and functional integrity by resisting the natural tendency toward disorder—that is, by remaining within a bounded set of preferred states despite environmental volatility (i.e., any living system tends, on average, to move along the gradient that leads toward its attracting set, i.e., the pullback attractor). To do so, the system must possess a hierarchical generative model of the hidden causes of its sensory inputs—a probabilistic model that is continuously tested and updated through Bayesian inference, i.e., a generative model. Since exact inference is generally intractable in realistic conditions, the system performs approximate variational inference: it selects an approximate posterior distribution and updates its parameters iteratively to minimize the divergence from the true posterior. This optimization process is formally equivalent to maximizing the Evidence Lower Bound (ELBO) in machine learning. The objective of this process is to minimize a quantity known as variational free energy. Variational free energy serves as an upper bound on surprisal, or the negative log model evidence, which quantifies how unexpected sensory inputs are under the model. In formal terms, free energy is decomposed into the sum of a Kullback–Leibler divergence (between the approximate and true posterior) and a term representing log evidence (see Table 2). Minimizing free energy thus corresponds to maximizing model evidence. This process of continuously updating beliefs and actions to reduce free energy enables the system to maintain coherence and adaptivity in a changing environment. In this sense, self-organization is reframed as self-evidencing—the system acts in ways that confirm its own model of the world. Finally, the FEP asserts that "all biological systems maintain their integrity by actively reducing the disorder or dispersion (i.e., entropy) of their sensory and physiological states by minimising their variational free energy" \cite{4}.
\begin{table}[h]
  \small
  \renewcommand{\arraystretch}{1.4}
  \centering
  \begin{tabular}{|p{0.95\textwidth}|}
    \hline
    \textbf{1. Free energy as a bound on surprise}\newline
    $\mathcal{F}(o) \ge -\log p(o)$\newline
    Free energy upper‐bounds the surprisal (negative log model evidence) of sensory input. Minimizing it helps explain perception as evidence maximization.%
    \\[6pt] \hline
    \textbf{2. Free energy as a variational bound}\newline
    $\mathcal{F}(q) = \mathrm{KL}\bigl(q(s)\,\Vert\,p(s\mid o)\bigr) \;-\;\log p(o)$\newline
    Free energy is minimized when the approximate posterior $q(s)$ matches the true posterior $p(s\mid o)$. This is the essence of variational Bayesian inference.%
    \\[6pt] \hline
    \textbf{3. Free energy as energy minus entropy}\newline
    $\mathcal{F}(q) = \mathbb{E}_{q}\bigl[-\log p(o,s)\bigr] \;+\; \mathbb{E}_{q}\bigl[\log q(s)\bigr]$\newline
    Free energy is the sum of expected prediction error and the complexity of the approximate posterior. It balances accuracy and complexity.%
    \\[6pt] \hline
  \end{tabular}
  \caption{Three equivalent formulations of variational free energy.}
\end{table}

It should be emphasized that the recent literature on the FEP and active inference—from essentially \cite{1, 6, 7, 8, 39}—has introduced and developed a new formulation of the FEP, shifting from a state-based formulation to a path-based one that no longer requires the concept of a NESS. In some of the more advanced FEP formulations, it is shown how the FEP can be derived simply from Jaynes’s principle of maximum entropy or maximum caliber under the constraint that there are boundaries of certain kinds. These formulations show that the FEP is just maximum caliber with explicit boundaries. Whereas the classical formulations (Friston 2019) start from the equations of statistical physics (the Langevin or Fokker–Planck equation) and arrive at an equivalent description of the system in information-theoretic terms (a gradient descent over a free-energy functional), the more advanced approaches instead begin with maximum caliber to show how, once boundaries are imposed, it becomes the FEP. This addresses criticisms that the FEP depends on physical presuppositions (e.g., the NESS). 

\subsection{Markov blankets}

The concept of Markov blanket is crucial in the formulation of the FEP. In fact, "we assume that for something to exist it must possess (internal or intrinsic) states that can be separated statistically from (external or extrinsic) states that do not constitute the thing" \cite{1}. The existence of things implies the existence of Markov blanket, namely, "a set of states that render the internal and external states conditionally independent" \cite{1}. But what does it mean "separation" here? If the space in which the active inference agent moves is composed of nested Markov blankets, how the agent passes through these blankets and their permeability? "States of things are constituted by their Markov blanket, while the Markov blanket comprises the states of smaller things with Markov blankets within them -- and so on ad infinitum" \cite{1}.   

A Markov blanket is “a statistical partitioning of a system into internal states and external states, where the blanket itself consists of the states that separate the two” \cite{7, 1}. The Markov blanket divides the system into three groups of statistical variables: internal states, external states, and blanket states. As Friston claims, “the dependencies induced by Markov blankets create a circular causality that is reminiscent of the action-perception cycle” \cite{1}. Circular causality here means that “external states cause changes in internal states, via sensory states, while the internal states couple back to the external states through active states, such that internal and external states influence each other in a vicarious and reciprocal fashion” \cite{1}. Consequently, the internal and external states tend to synchronize over time (i.e., coupling), much like two pendulums attached to opposite ends of a wooden beam gradually swinging in unison.

The Markov blanket thus allows a certain statistical boundary to be defined between internal states and external states, which are mediated solely by active states and sensory states \cite{38}. This means that, given the Markov blanket—that is, the sensory and active states—the internal and external states are conditionally independent. In other words, once the blanket is known, knowing additional information about the external states does not further constrain or inform the internal states. This structure ensures that internal and external states remain independent while being connected only through the active and sensory states. Active and sensory states “shield” the internal states by creating a statistical boundary \cite{2, 5, 6, 7}. Put simply, internal states cannot directly affect external states but can do so indirectly by influencing active states. Likewise, external states cannot directly impact internal states but can do so indirectly by affecting sensory variables (see Table 3).  

Free energy is a functional—that is, a function of a function—that quantifies the probability distribution encoded by the internal states of the system. Importantly, this differs from surprise, which is a function of the sensory and active states on the Markov blanket itself. Put differently: free energy is a function of probabilistic beliefs (i.e., internal states) about external states—that is, expectations about the likely causes of sensory input. When these beliefs match the true Bayesian posterior, variational free energy becomes equal to surprise. Otherwise, it serves as a tractable upper bound on surprise. This is why self-organizing systems can be characterized as minimizing variational free energy, and thereby minimizing surprise, through the continuous optimization of their beliefs about what lies beyond their Markov blanket. Finally, the FEP tells us "how the quantities that define Markov blankets change as the system moves towards its variational free energy minimum" \cite{4}.

\begin{table}[h]
\small
\renewcommand{\arraystretch}{1.4}
\centering
\begin{tabular}{|l|l|p{8.2cm}|}
\hline
\textbf{Element} & \textbf{Symbol} & \textbf{Description} \\
\hline
Internal states & $I$ & Hidden states of the system that encode beliefs about external causes; not directly influenced by external states. \\
\hline
External states & $E$ & States in the environment that influence sensory states but are not directly influenced by internal states. \\
\hline
Sensory states & $S$ & States that receive input from external states and influence internal states; part of the Markov blanket. \\
\hline
Active states & $A$ & States influenced by internal states that act upon external states; part of the Markov blanket. \\
\hline
Markov blanket & $B = S \cup A$ & The boundary of the system that mediates interactions between internal and external states through sensory and active channels. \\
\hline
Conditional independence & — & Given the blanket $B$, internal and external states are conditionally independent: $p(I, E \mid B) = p(I \mid B)\, p(E \mid B)$. \\
\hline
\end{tabular}
\caption{Formal components of a Markov blanket in active inference.}
\end{table}

\section{The Space as a Continuous Gradient of Markov Blanket Strengths}

In this section we introduce the central philosophical thesis of this paper. In the following one we develop a formal demonstration. It all stems from a rather naive and abstract question: \textit{What would a space be like if every point were composed of internal and external states, i.e. had a Markov blanket? And how would an agent with a blanket of their own move in this space?} 

Free energy minimization is generally described in temporal terms: “Strictly speaking, free energy is only ever minimized diachronically—that is, over some discrete time span—as a process” \cite{2}. What role does space play in this process? The space through which an active inference agent moves is not an empty or uniform container—it is instead a structure composed by nested Markov blankets: “[...] we should be able to describe the universe in terms of Markov blankets of Markov blankets—and Markov blankets all the way up, and all the way down” \cite{2}. The key issue is how we conceptualize Markov blankets and the statistical boundaries they define. 

Classic works on active inference fails to properly account for the spatial dimension, treating space as a passive and predictable “environment.” By doing so, it cannot fully grasp the concept of affordance, reducing it to a set of predictions about that environment. In this view, affordances are not inherently part of the environment itself; they depend on the predictions and knowledge of the agent interacting with it. In active inference, space plays no active role in shaping the agent’s trajectories—and this is not compatible with Gibson's view of affordance \cite{29}. In essence, active inference remains confined to a lab‐based perspective, where space adapts to hypotheses rather than hypotheses adapting to space. The point is that space is complex, as are affordances—they cannot be reduced to the agent’s predictions. Affordances “are not simply static features of the environment, independent of the presence and engagement of an agent, nor are they states of the cognitive agent alone” \cite{28}. The very unity of perception and action depends on that complexity. In a nutshell, \textit{space is not entirely predictable, and above all, space shapes and distorts our predictions}. As we hope to show, since the blanket‐density factor directly modulates how strongly sensory evidence can update internal beliefs (and therefore the generative model), it does in effect “shape” the model the agent uses. 

At this point, the next question becomes: How can we reconceptualize space independently of an agent’s predictions, that is, its generative model? The hypothesis we want to propose and test here is that the space inhabited by active inference agents is populated with Markov blankets that can vary (along a spectrum) in their degree of permeability or porosity—that is, Markov blankets that are more or less “strong,” exhibiting higher or lower degrees of separation relative to an observer and their scale of observation. The strength of a Markov blanket (i.e., how well the blanket insulates the inside) is the degree to which it enforces conditional independence between internal and external states, via the mediating sensory and active states. Therefore, the space is structured by a continuous gradient of Markov blanket strengths. From this spatial perspective, preferred states can be reinterpreted as configurations of optimal coupling—zones of dynamic synchronization with other Markov blankets—rather than purely internal homeostatic targets.

Thus, from this perspective, every point in the space through which the active inference agent moves is associated with a Markov blanket that separates internal and external variables. Based on this, we define the Markov blanket density, which varies continuously, like a spectrum, and quantifies the local degree of informational isolation between those internal and external variables. 

\subsection{Connection to the Literature}

This paper builds on some findings from previous literature and aims to unify and extend them. 

\cite{7} advances the FEP by translating its abstract notions of conditional independence and “things” into a concrete, unsupervised learning algorithm. Recognizing that any identifiable object must correspond to a partition—internal, boundary, external—their variational Bayesian expectation maximization framework treats each microscopic element as governed by one of several low dimensional latent processes. During inference, elements are dynamically assigned to roles by maximizing an ELBO, and a “Bayesian attention” mechanism tracks how the inferred boundary can move, split, or merge over time. Through case studies as diverse as Newton’s cradle, a propagating combustion front, and the Lorenz attractor, they demonstrate that their method reliably uncovers the intuitive interfaces that simplify a system’s macroscopic description. See also \cite{5, 6}.

\cite{8} complements this algorithmic advance with a rigorous, asymptotic guarantee for the existence of blankets in high-dimensional stochastic systems. By defining a “blanket index” to measure the strength of cross-couplings between internal and external variables, the paper models these interactions as independent, bounded random variables and employs large-deviation techniques to show that, as the system’s dimension grows without bound, almost all such couplings vanish. This result proves that “weak” Markov blankets—where conditional independence holds up to vanishingly small interactions—emerge almost surely in the infinite-dimensional limit, thereby grounding Friston’s sparse-coupling conjecture in a broad class of Itô stochastic differential equations. While this theorem confirms that blankets are not an ad hoc or exceptional phenomenon but a generic feature of complex systems, it remains silent on how to measure the varying strengths of these blankets in finite, real-world settings or how they might steer an agent’s behavior. On Bayesian mechanics, see also \cite{26, 27}.

The present paper is also related to \cite{1}. Both works share the same foundational insight: any system at a non-equilibrium steady state can be partitioned into internal, sensory, active, and external components via a Markov blanket, and internal states appear to perform Bayesian inference by minimizing variational free energy. However, while Friston treats this boundary as a sharply defined, discrete set of sensory and active variables that uniformly insulates internal states from external states—demonstrating how this partition underlies phenomena from quantum dynamics through classical stochastic processes to living systems—the present paper explicitly extends this approach by allowing that “insulating” effect to vary continuously across space. In other words, where Friston envisions a crisp frontier separating inside and outside, the present research proposes a continuous scalar field that quantifies, at each location, how strongly internal and external states are decoupled. This permits intermediate regions where external influences partially penetrate, rather than assuming each point is either fully inside or fully outside the Markov blanket.

However, the present research does not stop at proposing this shift in perspective; it also provides a concrete algorithmic recipe—based on information-theoretic estimators and nearest-neighbor sampling—to measure local blanket strength from observed data. In contrast, Friston’s treatment, although highly ambitious and formally rich across multiple scales, remains largely conceptual with regard to how one might detect or manipulate the blanket in real systems. Specifically, Friston \cite{1} illustrates his theory with idealized “active soup” simulations and outlines the mathematical links between free energy, steady-state densities, and inference, but he does not detail how to estimate blanket strength in, for example, a spatially extended neural system or an agent navigating a heterogeneous environment. By combining these two perspectives, the present research neither contradicts nor undermines Friston’s core theorems regarding a discrete Markov blanket. Rather, by embedding Friston’s boundary within a gradient of insulating strength, it shows how free-energy minimization can be modulated by local variations in coupling between internal and external states. In this view, agents naturally gravitate toward regions where coupling is strongest—where the blanket is weakest—because those regions offer richer sensory information. However, this also means that the MB density imposes limits on free energy minimization. In summary, the present paper takes Friston’s high-level, multiscale framework and gives it concrete spatial texture: showing how blanket strength can ebb and flow across space and, in turn, shape an agent’s inferential and behavioral trajectories.

\subsection{Some clarifications}

We use here the term "coupling" to describe the degree of statistical and causal interdependence between an agent and its environment. This is formalized in terms of conditional mutual information, but also interpreted dynamically: strong coupling implies that the agent’s sensory states carry information about external causes, and that its actions can affect those causes. In our model, low MB density corresponds to higher potential for coupling, which in turn enables more effective free energy minimization. This is perfectly in line with \cite{1}.

However, we acknowledge that this use of "density" introduces a metaphorical shift: we are interpreting space not as geometrically partitioned, but as structured by the statistical architecture of interaction. This raises ontological and epistemological questions. Is MB density a real property of physical space, or is it a modeling construct used to represent the agent’s epistemic relation to its surroundings? In this paper, we remain agnostic: we treat MB density as a tool for expressing how the spatial environment constrains inferential dynamics, rather than making strong claims about its physical instantiation. We steer clear of the more strictly philosophical debates on the ontological implications of the concept of Markov blankets \cite{38}—at least, from my point of view, Markov blankets are good modelling tools, but at the same time, they are only \textit{good heuristics}. This doesn't mean avoiding philosophical debates about the relationship between the map (Markov blankets) and the territory (reality) — quite the opposite, in fact. It means recognizing the importance of the problem (beyond an armchair philosophy approach), and therefore focusing first on the robustness of the map and what we can do with it — a necessary condition for understanding its relationship to the territory, especially since the map is itself part of the territory. And we are also convinced that (unlike \cite{39}) instrumentalism does not necessarily imply blind belief in the usefulness of the model—quite the contrary, in fact.  

Morover, MB density in itself is not a probability density. MB density is an information‐theoretic measure (ranging from 0 to 1) of how effectively an agent’s boundary blocks information flow between its internal and external states at a point x, estimated via conditional and unconditional mutual informations. It is not normalized over the state space and directly modulates the speed of gradient‐descent on free energy (when MB density = 1, updates freeze). By contrast, a probability density p(x) is a normalized function (integrating to one) that assigns relative likelihoods to values of x, without any notion of informational blocking or direct influence on free‐energy descent.

The MB density $\rho(x)$ is a modeling index of local informational shielding, not a probability density; it modulates dynamics via $\dot{x}=-(1-\rho(x))\,\nabla F(x)$.
Because conditional mutual information can exceed mutual information in the presence of synergy, the raw ratio
\[
\rho_{\mathrm{raw}}(x)=1-\frac{I(I;E\mid B)}{I(I;E)+\varepsilon}
\]
may fall outside $[0,1]$.
Operationally, we adopt an $\varepsilon>0$ for numerical stability and clip estimates to $\rho(x)\in[\delta,1-\delta]$ during inference (Algorithm~1, Sec.~4.5), ensuring a well-defined mobility factor while preserving the statistical meaning of the ratio.
An extended exploratory variant lifts the clipping and allows $\rho>1$, which flips the sign of the prefactor and induces local ascent of $F$ (see Sec.~10).

Another important caveat. The paper never claims that the agent “tends toward a point where MB density = 0” (i.e., complete elimination of any boundary). When it says that free‐energy minimization follows trajectories into regions of low MB density, it really means cases where the MB density is reduced but not zero: in those regions, the blanket is “thin” enough to allow faster information exchange between internal and external states, speeding up the descent of free energy. That does not imply that the agent is drifting toward entropy and dissipation. The agent’s skill is precisely in staying in areas where MB density is high enough to maintain its internal structure, yet not so high as to block necessary coupling. Consequently, there is no contradiction in the paper’s thesis. Low MB density means the blanket is just enough to separate internal and external states, but weak enough to permit rapid information flow such that free‐energy descent is effective; by contrast, MB density = 0 is a theoretical limit where the blanket no longer exists, and at that point the model no longer describes adaptive behavior but instead total informational extinction (i.e., the agent dissolves). Moving into regions of low MB density does not automatically cause an overall increase in entropy. In the free‐energy minimization framework, “low MB density” simply means that the information boundary between agent and environment is more “porous,” allowing the internal state to update more quickly based on sensory data. That does not equate to a loss of internal order or a slide into chaos. Again, this is perfectly in line with \cite{1}. In fact, "nearly every system encountered in the real world is self-organizing \textit{to a greater or lesser degree}" \cite{1}.

\section{Thesis}

We aim to demonstrate the following claim:

\textit{Free energy minimization tends to follow trajectories leading toward regions of lower MB density. These regions correspond to stronger agent-environment coupling and greater synchronization potential.}

\subsection{Definitions and Assumptions}

Let \( \Omega \subset \mathbb{R}^n \) denote a spatial domain.

For each point \( x \in \Omega \), assume the presence of a local Markov blanket \( \mathcal{B}(x) \) that mediates interactions between internal states \( I \), external states \( E \), and blanket states \( B \).

Define the \textbf{Markov blanket strength} at point \( x \) as:
\begin{equation}
S(x) := 1 - \frac{I(I; E \mid B)}{I(I; E)}
\end{equation}
where \( I(I; E \mid B) \) is the conditional mutual information between internal and external states given the blanket.

This yields:
\begin{itemize}
    \item \( S(x) = 1 \): perfect conditional independence (strong MB).
    \item \( S(x) = 0 \): no conditional independence (no effective MB).
\end{itemize}

Informational separation is at its highest degree when
\[
I(I;E \mid B) = 0,
\]
that is, when, once \(B\) is known, knowing further details about \(E\) does not help to inform \(I\).

Define the \textbf{Markov blanket (MB) density} \( \rho(x) \) as the field of MB strengths over \( \Omega \):
\begin{equation}
\rho(x) := S(x), \quad \rho(x) \in [0, 1]
\end{equation}
This field quantifies how insulated each point in space is with respect to internal-external separation.

\subsection{Clarification on Continuous MB Density vs.\ Discrete Conditional Independence}

A potential concern is that our MB density $\rho(x)$, defined as a continuous scalar field, might be conflated with the classical, discrete property of conditional independence.  We resolve this by distinguishing carefully between the \emph{structural} and the \emph{quantitative} aspects, and by invoking precise measure‐theoretic language and manifold geometry.

\subsubsection{Measurability and Measure‐Theoretic Foundations}

Let $(\Omega,\mathcal{F},P)$ be a probability space and let $I,E,B$ be random variables with joint distribution absolutely continuous w.r.t.\ Lebesgue measure on a parameter manifold~$\mathcal{X}$.  By the Radon–Nikodym theorem there exists a density
\[
  p(i,e,b \mid x)
  \;=\;
  \frac{dP}{d\lambda}(i,e,b \mid x)
  \quad\text{for almost every }x\in\mathcal{X}.
\]
Define the \emph{total} and \emph{residual} mutual informations at $x$ by
\begin{align*}
  I_{\rm tot}(x)
  &= 
  \iint
    p(i,e\mid x)\,
    \ln\frac{p(i,e\mid x)}{p(i\mid x)\,p(e\mid x)} 
    \;di\,de,\\
  I_{\rm res}(x)
  &= 
  \iiint
    p(i,e,b\mid x)\,
    \ln\frac{p(i,e\mid b,x)}{p(i\mid b,x)\,p(e\mid b,x)} 
    \;di\,de\,db.
\end{align*}
Since each is an integral functional of measurable densities, both $I_{\rm tot}(x)$ and $I_{\rm res}(x)$ are themselves measurable functions of~$x$, defined for almost every $x\in\mathcal{X}$.  We then introduce
\[
  \rho(x)
  \;=\;
  1 \;-\;
  \frac{I_{\rm res}(x)}{I_{\rm tot}(x)},
  \quad
  0 \;\le\;\rho(x)\;\le\;1,
  \tag{1}
\]
with all equalities understood to hold \emph{almost everywhere} (i.e., off a $P$–null set where $I_{\rm tot}(x)=0$).  

Importantly, the \textbf{discrete} Markov blanket remains the \emph{primary} structural object enforcing 
\[
  I\;\perp\!E\;\big|\;B
  \quad\Longleftrightarrow\quad
  p(I,E \mid B) \;=\; p(I\mid B)\,p(E\mid B).
\]
Only after specifying that blanket and computing the integrals above do we obtain the \emph{scalar summary} $\rho(x)$.  In the limiting cases,
\[
  I_{\rm res}(x)=0 \;\Longrightarrow\;\rho(x)=1,
  \quad
  I_{\rm res}(x)=I_{\rm tot}(x)\;\Longrightarrow\;\rho(x)=0,
\]
recovering perfect separation or total coupling, respectively, but never conflating structure with measure.

\subsubsection{Geometry of the Underlying State Space}

We regard $\mathcal{X}$ as the \emph{statistical manifold} of our generative model.  Explicitly,
\[
  x\in\mathcal{X}
  \quad\Longleftrightarrow\quad
  x\text{ parametrizes }p(i,e,b\mid x).
\]
Equipped with a Fisher–Rao metric (or any smooth atlas compatible with the densities), $\rho(x)$ becomes a smooth scalar field on~$\mathcal{X}$.  

In applications where $x$ denotes a point in \emph{physical} space, one first constructs a local kernel approximation
\[
  p(i,e,b\mid x)
  \;=\;
  \frac{1}{|B(x,r)|}\!\int_{B(x,r)}\!p(i,e,b\mid y)\,dy,
\]
and then applies the same definitions.  Thus $\rho(x)$ may equally be viewed as a field on a physical manifold, provided the generative kernel is smooth.

\subsubsection{Key Takeaways}

\begin{itemize}
  \item $\rho(x)$ is a \emph{measurable} function on $\mathcal{X}$ (a.e.), derived from well‐defined integrals (Radon–Nikodym, Lebesgue differentiation).
  \item It \textbf{quantifies} the degree of conditional independence enforced by a classical Markov blanket; it does \emph{not} redefine or replace that structural concept.
  \item Its domain is a \emph{statistical manifold} of model parameters (or, via kernel lifts, a \emph{physical manifold}).
  \item In the discrete limit $\rho\in\{0,1\}$, one recovers the ordinary, binary Markov blanket.
\end{itemize}

\subsection{Local vs Regional \( \rho(x) \) Interpretations}

Two clarifications are in order regarding the apparent tension between the local definition of blanket density and its regional or global usage in the simulations.

First, the discrepancy is only conceptual and results from the dual perspective adopted in this work. On the one hand, the agent's behavior is analyzed from the point of view of the environment: here, the space (and the field \( \rho(x) \)) is taken as given --- which is standard in most dynamical models, where the landscape precedes the agent. On the other hand, the model also addresses the subjective perspective of the agent, for whom the structure of the space is initially unknown and must be inferred through interaction. Thus, while \( \rho(x) \) appears as a pre-defined field in the simulations, it should be understood as a representation that the agent progressively constructs through active inference. These two views are not contradictory, but complementary: one external, the other internal. This point is explicitly addressed in Section 12.

Second, the supposed tension between local and regional definitions of \( \rho(x) \) dissolves when we assume the discrete nature of the agent's movement. At each time step, the agent is located at a single point in space and therefore interacts only with the local blanket density \( \rho(x) \) at that point. This is precisely what the algorithm introduced in Section 4.4 formalizes: a local, data-driven estimation of \( \rho(x) \) based on a finite neighborhood. Hence, although \( \rho(x) \) is conceptually related to spatial regions (via the mutual information over internal, blanket, and external zones), its operative meaning remains pointwise. The algorithm, by grounding this estimation in local data, resolves the apparent conflict between locality and regionality.

A final clarification concerns the interpretation of the blanket density field \( \rho(x) \). While it assigns a value to every point in space, this does not imply that the underlying Markov blanket is associated with the same system or agent across the entire domain. In fact, a given point may be part of different Markov blankets under different conditions — depending on the scale, temporal resolution, or inferential context adopted. For instance, a location might fall within the blanket of one agent during a given interaction, and later within that of a different system, or none at all. What \( \rho(x) \) captures, then, is not a fixed assignment of spatial regions to specific agents, but the local degree of conditional coupling between internal and external states — regardless of which system is involved. This perspective reinforces the interpretation of \( \rho(x) \) as a context-sensitive measure of informational structure, rather than a mapping of fixed systemic boundaries.

\begin{figure}
    \centering
    \includegraphics[width=0.8\linewidth]{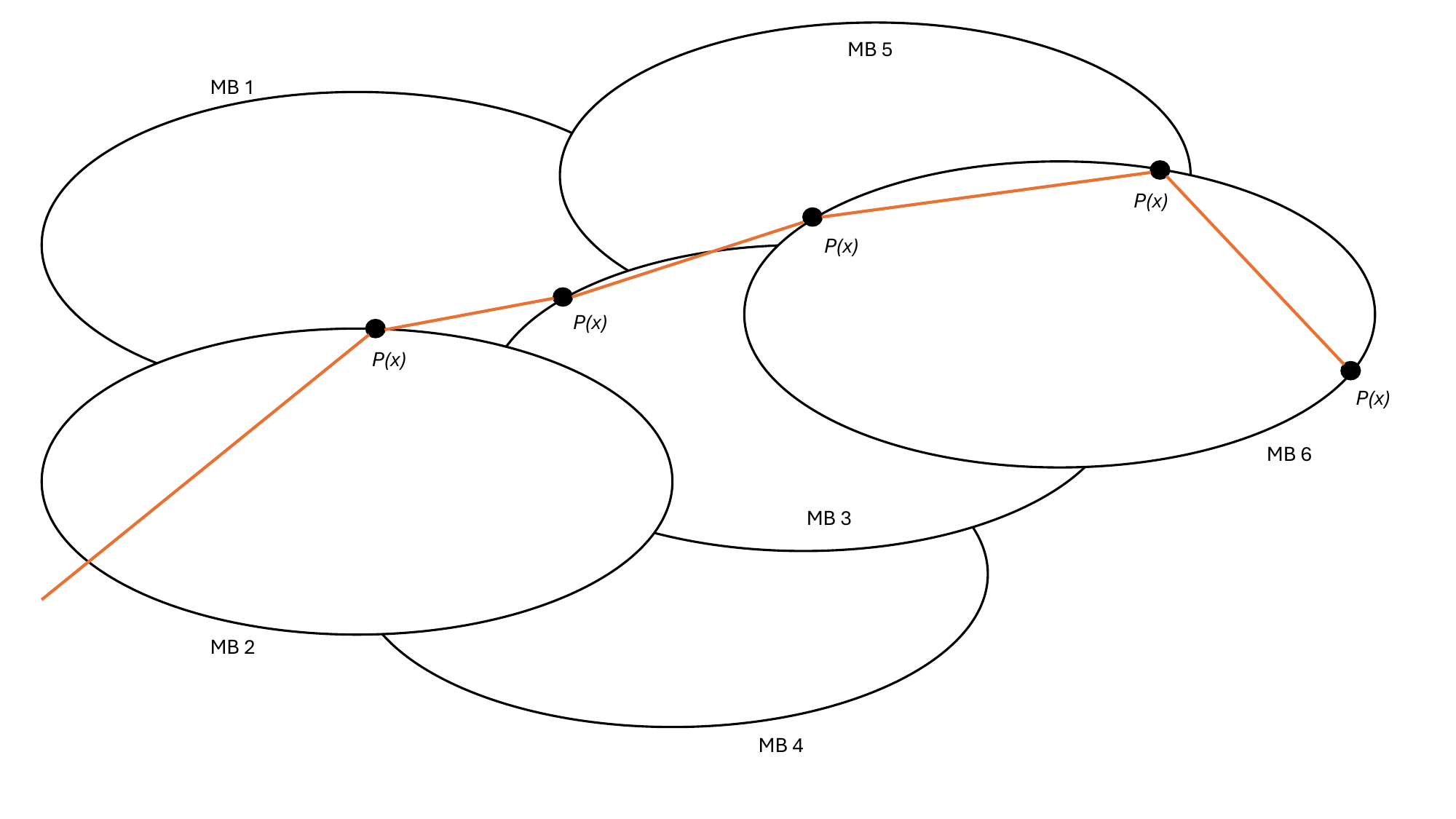}
    \caption{\textbf{Walking through Markov blankets.} A schematic and intuitive representation of the path of an active inference agent (orange line) in a space “filled” with Markov blankets (MB) and touching points with different densities or porosities. Obviously, the agent also has its own Markov blanket, and therefore its movement is conditioned by the coupling with the other blankets and thus by the MB density.}
    \label{fig:enter-label}
\end{figure}

\subsection{Technical Assumptions on Space and Measures}

Let \(\mathcal{X}\subseteq\mathbb{R}^d\) be a nonempty, bounded, open set with Lipschitz boundary, equipped with the Lebesgue measure \(\mathrm{d}x\).  We assume the joint distribution of internal, external, and blanket variables admits a strictly positive, \(C^2\) density
\[
  p(i,e,b \mid x)\quad\text{w.r.t.}\quad \mathrm{d}i\,\mathrm{d}e\,\mathrm{d}b\,\mathrm{d}x,
\]
where \(i\in\mathcal{I}\), \(e\in\mathcal{E}\), and \(b\in\mathcal{B}\) range over compact subsets of Euclidean spaces.  Concretely, we require:
\begin{enumerate}
  \item[(i)] \emph{Uniform bounds:} there exist constants \(0<m\le M<\infty\) such that
    \[
      m \;\le\; p(i,e,b \mid x)\;\le\; M,
      \quad\forall\,(i,e,b,x).
    \]
  \item[(ii)] \emph{Smooth marginals and conditionals:} all marginal and conditional densities
    \(p(i,e)\), \(p(i,e\mid b)\), etc., depend in a \(C^2\)-fashion on \(x\).
  \item[(iii)] \emph{Regularity of information maps:} the mappings
    \[
      x\mapsto I\bigl(I;E\bigr)
      \quad\text{and}\quad
      x\mapsto I\bigl(I;E\mid B(x)\bigr)
    \]
    are \(C^1\) on \(\mathcal{X}\), and hence
    \(\rho(x)=1 - \tfrac{I(I;E\mid B)}{I(I;E)}\) is \(C^1\) as well.
\end{enumerate}
Under these hypotheses, all integrals, gradients, and continuity arguments in subsequent sections are well-defined and satisfy the smoothness conditions required by the main theorems.

\subsection{Operational Definition of MB Density}
\label{sec:operational-blanket-density}

To render the blanket density field \(\rho(x)\) operational in continuous systems, we partition the state–space around each point \(x\) using two radii, \(r_{1}<r_{2}\). Variables within distance \(r_{1}\) of \(x\) form the internal set \(I(x)\); those at distances in \([r_{1},r_{2})\) form the blanket \(B(x)\); and the remainder form the external set \(E(x)\). We estimate the conditional mutual information \(I(I;E\mid B)\) and the marginal mutual information \(I(I;E)\) using the Kraskov–Stögbauer–Grassberger (KSG) \(k\)-nearest–neighbors estimator. To avoid division by zero, we introduce a small regularizer \(\varepsilon\) and constrain \(\rho(x)\in[\delta,\,1-\delta]\). See \cite{21}.
\begin{algorithm}[ht]
  \caption{Estimation of Blanket Density \(\rho(x)\)}
  \label{alg:blanket-density}
  \begin{algorithmic}[1]
    \Require Dataset \(D = \{(y_i, s_i)\}_{i=1}^N\), radii \(r_1, r_2\), neighbor count \(k\), regularizer \(\varepsilon\), bound \(\delta\).
    \Ensure Blanket density \(\rho(x)\) for each sample \(x\in\{y_i\}\).
    \For{each sample $x \in \{y_i\}$}
      \State \(I \gets \{\,s_i \mid \|y_i - x\| < r_1\}\)
      \State \(B \gets \{\,s_i \mid r_1 \le \|y_i - x\| < r_2\}\)
      \State \(E \gets \{\,s_i \mid \|y_i - x\| \ge r_2\}\)
      \State Estimate \(I(I;E\mid B)\) via \texttt{KSG\_conditional}(\(k,I,E,B\))
      \State Estimate \(I(I;E)\) via \texttt{KSG\_mutual}(\(k,I,E\))
      \State \(S(x) \gets 1 - \dfrac{I(I;E\mid B)}{I(I;E) + \varepsilon}\)
      \State \(\rho(x) \gets \min\bigl\{\max\{S(x),\delta\},\,1-\delta\bigr\}\)
    \EndFor
    \State \Return \(\{\rho(x)\}_{x\in\{y_i\}}\)
  \end{algorithmic}
\end{algorithm}
The computational cost scales as \(O(N\log N)\) using KD-trees or similar structures; further speedups are possible via grid-based subsampling. In our simulations we used
\[
  r_{1}=0.1,\quad r_{2}=0.2,\quad k=5,\quad \varepsilon=10^{-6},\quad \delta=10^{-3},
  \quad N=10^{4}.
\]
A Python implementation of the KSG estimator to compute MB density from sample data can be found here: https://github.com/DesignAInf/MB-density. See also Appendix B. On the project’s GitHub page, there are two separate repositories serving different purposes. The first one (\texttt{mbdensity-sim}) is a demonstrative sandbox --- a minimal, 
self-contained environment designed to showcase the core idea of modeling a spatially varying 
Markov blanket density $\rho(x)$. It uses a single synthetic field, a simplified mutual 
information estimator, and a basic 2D gradient-descent dynamic, making it easy to test and 
understand the concept without extra complexity. The second repository (\texttt{mbdensity-paper-sims}) 
is a full reproduction of all simulations and visualizations from the manuscript 
(Figures~2--8). It implements the exact scenarios described in the main text and Appendix~A. In short: the first repo 
is a stripped-down prototype for exploration, while the second is a faithful, figure-by-figure 
replication of the paper’s experiments.

\section{Free Energy and Modulated Gradient Descent}

Let the variational free energy field \( \mathcal{F}(x) \) be defined over space \( \Omega \):
\begin{equation}
\mathcal{F}(x) := \mathbb{E}_{q_\mu(s(x))}[\log q_\mu(s(x)) - \log p(s(x), \eta(x))]
\end{equation}
where \( q_\mu \) is the internal (variational) distribution of the agent, \( s(x) \) are sensory states at location \( x \), and \( \eta(x) \) are environmental (hidden) states at \( x \).

The agent minimizes \( \mathcal{F}(x) \) via gradient descent, modulated by MB density:
\begin{equation}
\dot{x} = -M(x) \nabla \mathcal{F}(x)
\end{equation}
where \( M(x) := (1 - \rho(x)) I \), and \( I \) is the identity matrix. If \( \rho(x) = 1 \), inference is blocked (no coupling), so \( \dot{x} = 0 \); if \( \rho(x) = 0 \), there is full coupling and maximal inference is possible \cite{10, 11, 14, 15, 16}.

\section{FEP and MB Density}

\begin{theorem}[Simultaneous Descent of Free Energy and Blanket Density under the Gradient-Ratio Condition]
Let $\Omega\subset\mathbb{R}^n$ be a compact domain with $C^2$ boundary, and let $F\in C^2(\Omega)$ be the free-energy function. An agent’s trajectory $x(t)$ evolves according to the modulated gradient descent
\[
\dot x(t)\;=\;-\,\bigl[1-\rho_N\bigl(x(t)\bigr)\bigr]\,\nabla F\bigl(x(t)\bigr),
\]
where $\rho_N(x)$ is the empirical blanket-density estimator derived from a dataset of $N$ samples.

Assume the following hold on an open set $D\subset\Omega$ where $\nabla F(x)\neq0$:
\begin{enumerate}[label=(\alph*)]
  \item \textbf{Smoothness of information functions.}\\
    The true unconditional mutual information
    \[
      v(x)\;=\;I_{\mathrm{true}}\bigl(I(x);E(x)\bigr),
      \quad
      u(x)\;=\;I_{\mathrm{true}}\bigl(I(x);E(x)\mid B(x)\bigr)
    \]
    are $C^1$ functions of $x$.
  \item \textbf{Gradient alignment.}\\
    Define
    \[
      A(x)\;=\;\nabla u(x)\cdot\nabla F(x),
      \quad
      B(x)\;=\;\nabla v(x)\cdot\nabla F(x),
    \]
    and assume $A(x)>0$, $B(x)>0$, i.e.\ moving down $\nabla F$ decreases both $u(x)$ and $v(x)$.
  \item \textbf{Gradient-Ratio Condition.}\\
    The proportional rate of decrease of $v(x)$ along $\nabla F$ strictly exceeds that of $u(x)$:
    \[
      \frac{B(x)}{v(x)} \;>\;\frac{A(x)}{u(x)}.
    \]
\end{enumerate}

\medskip\noindent
\textbf{Conclusion.}  
Under these assumptions, with probability tending to 1 as $N\to\infty$ (so that $\nabla\rho_N\to\nabla\rho$ uniformly), the agent’s trajectory both descends the free-energy landscape and moves toward regions of strictly decreasing blanket density.  In particular, for every $t$ with $x(t)\in D$,
\[
\frac{d}{dt}\,\rho_N\bigl(x(t)\bigr)
=\nabla\rho_N\bigl(x(t)\bigr)\cdot\nabla F\bigl(x(t)\bigr)
\;>\;0.
\]
\emph{Note.} We choose the sign so that a positive directional derivative $\nabla\rho\cdot\nabla F>0$ implies $\rho$ decreases as $F$ decreases.
\end{theorem}

From this point of view, free-energy minimization is fundamentally and universally aligned with seeking regions of stronger informational coupling (lower MB density). The former process naturally gives rise to the latter. The alignment between free-energy minimization and stronger informational coupling is \emph{contingent} on the local informational geometry. It holds \emph{only} when the Gradient-Ratio Condition is met—i.e.\ when the informational benefit of reducing total uncertainty outweighs the cost of residual uncertainty along the current path. Thus, the agent’s dynamics emerge from a direct competition between these two informational gradients.

This revised framework opens up a richer, more nuanced model of active inference. Active inference is not a simple, guaranteed descent down a smooth hill but the navigation of a complex “epistemic landscape,” determined by the interplay between the free-energy gradient $\nabla F$ and the informational field $\rho(x)$.  The directional derivative of the true blanket density
\[
\rho_{\mathrm{true}}(x)
=1 \;-\;\frac{u(x)}{v(x)}
\]
along the agent’s path has sign proportional to
\[
[v(x)]^2\,u(x)\,B(x)\;-\;v(x)\,A(x)\,u(x),
\]
which is positive precisely when the Gradient-Ratio Condition holds, ensuring alignment of free-energy descent with a decrease in $\rho(x)$. This immediately raises a critical question: What happens when the condition is violated? This leads to the concept of an epistemic trap: a region in the state space where the informational geometry is "perverse," causing the Gradient-Ratio Condition to fail. In such a region:

If
\[
\frac{B(x)}{v(x)}\;\le\;\frac{A(x)}{u(x)},
\]
an \emph{epistemic trap} arises: despite following free-energy minimization, the agent may move toward higher $\rho(x)$, becoming more informationally isolated. Such traps offer a first-principles model of maladaptive behaviors (e.g.\ social anxiety) and underscore the necessity of stochasticity or curiosity-driven exploration to escape basins that violate the Gradient-Ratio Condition. This discussion of epistemic traps serves as a qualitative bridge to Theorem~2, where these traps are rigorously defined and their dynamical properties analyzed.

There are at least two implications of this point:
\begin{itemize}
  \item \textbf{Modeling Maladaptive States.}  
    Epistemic traps formalize how an agent can get stuck in high-$\rho$ regions despite lower free-energy alternatives elsewhere.
  \item \textbf{Justification for Exploration.}  
    Deterministic descent may fail; random perturbations or explicit exploratory drives are normatively justified to overcome perverse informational geometries.
\end{itemize}

\subsection{Proof of the Gradient‐Alignment Condition}
\label{sec:proof_gradient_alignment}

Here we provide the complete technical details required to justify the claim
\[
  \nabla \rho_N(x)\;\cdot\;\nabla F(x)\;>\;0
  \quad\text{on }D,
\]
with high probability as \(N\to\infty\).  Recall that 
\[
  \rho_N(x) 
  \;=\; 
  1 
  \;-\; 
  \frac{\widehat{I}\bigl(I(x);E(x)\mid B(x)\bigr)\;+\;\varepsilon(N)}
       {\widehat{I}\bigl(I(x);E(x)\bigr)\;+\;\varepsilon(N)},
  \quad
  \varepsilon(N)=C_0\,N^{-\alpha}.
\]
The proof proceeds in several steps:

\subsection*{Step 1: Consistency and \(C^1\) Convergence of the MI Estimators}

Under the choice of radii \(r_1(N),\,r_2(N)\) satisfying
\[
  r_2(N)\;\to\;0, 
  \quad 
  r_1(N) = c\,r_2(N), 
  \quad
  N\,\mathrm{Vol}\bigl(\mathrm{Ball}(x;\,r_2(N))\bigr)\;\to\;+ \infty,
\]
the KSG–kNN estimators
\[
  \widehat{I}\bigl(I(x);E(x)\bigr),
  \qquad
  \widehat{I}\bigl(I(x);E(x)\mid B(x)\bigr)
\]
converge in probability to their \emph{true} values 
\[
  I_{\mathrm{true}}\bigl(I(x);E(x)\bigr),
  \quad
  I_{\mathrm{true}}\bigl(I(x);E(x)\mid B(x)\bigr),
\]
uniformly on every compact \(K \subset D\).  Moreover, if the true mutual informations are \(C^1\) and the underlying noise is sub‐Gaussian (or sub‐Exponential), then \(\widehat{I}\) converges to \(I_{\mathrm{true}}\) \emph{in} \(C^1\)‐norm on compacts:

\[
\begin{aligned}
  \sup_{x\in K} \bigl|\widehat{I}\bigl(I(x);E(x)\bigr) - I_{\mathrm{true}}\bigl(I(x);E(x)\bigr)\bigr|
  &= \mathcal{O}_p\bigl(N^{-\alpha}\bigr),\\
  \sup_{x\in K} \bigl|\nabla_x \widehat{I}\bigl(I(x);E(x)\bigr) - \nabla_x I_{\mathrm{true}}\bigl(I(x);E(x)\bigr)\bigr|
  &= \mathcal{O}_p\bigl(N^{-\alpha}\bigr).
\end{aligned}
\]
and similarly for \(\widehat{I}\bigl(I(x);E(x)\mid B(x)\bigr)\).  The exponent \(\alpha>0\) depends on the data dimension \(d\) and the chosen \(k\).  In particular, for sufficiently large \(N\), with probability at least \(1 - \delta\), one has

\[
  \bigl\|\widehat{I}\bigl(I(\cdot);E(\cdot)\bigr) - I_{\mathrm{true}}\bigl(I(\cdot);E(\cdot)\bigr)\bigr\|_{C^1(K)}
  \;<\; \eta(N),
\]
\[
  \bigl\|\widehat{I}\bigl(I(\cdot);E(\cdot)\mid B(\cdot)\bigr) - I_{\mathrm{true}}\bigl(I(\cdot);E(\cdot)\mid B(\cdot)\bigr)\bigr\|_{C^1(K)}
  \;<\; \eta(N),
\]
where \(\eta(N)\to 0\) as \(N\to\infty\).  

\subsection*{Step 2: Definition of the ``True'' Blanket Density \(\rho_{\mathrm{true}}(x)\)}

Define
\[
  \rho_{\mathrm{true}}(x)
  \;=\;
  1 
  \;-\; 
  \frac{I_{\mathrm{true}}\bigl(I(x);E(x)\mid B(x)\bigr)}
       {\,I_{\mathrm{true}}\bigl(I(x);E(x)\bigr)\,}.
\]
Since \(I_{\mathrm{true}}(I(x);E(x)) > 0\) for all \(x\in D\), \(\rho_{\mathrm{true}}(x)\) is well‐defined and lies strictly in \((0,1)\).  By hypothesis, \(I_{\mathrm{true}}(\cdot;\cdot)\) and \(I_{\mathrm{true}}(\cdot;\cdot\mid \cdot)\) are \(C^1\), so \(\rho_{\mathrm{true}}(x)\in C^1(\Omega)\).  A straightforward differentiation yields
\[
  \nabla \rho_{\mathrm{true}}(x)
  \;=\; 
  -\,\frac{1}{I_{\mathrm{true}}\bigl(I(x);E(x)\bigr)}\,
     \nabla\,I_{\mathrm{true}}\bigl(I(x);E(x)\mid B(x)\bigr)
  \;+\;
  \frac{%
    I_{\mathrm{true}}\bigl(I(x);E(x)\mid B(x)\bigr)
  }{%
    \bigl[I_{\mathrm{true}}\bigl(I(x);E(x)\bigr)\bigr]^2
  }\,
  \nabla\,I_{\mathrm{true}}\bigl(I(x);E(x)\bigr)\,.
\]

Since, by the above assumptions, both
\[
  \nabla\,I_{\mathrm{true}}\bigl(I(x);E(x)\mid B(x)\bigr)\;\cdot\;\nabla F(x)
  \;>\;0,
  \quad
  \nabla\,I_{\mathrm{true}}\bigl(I(x);E(x)\bigr)\;\cdot\;\nabla F(x)
  \;>\;0
  \quad\forall\,x\in D,
\]
and because \(I_{\mathrm{true}}\bigl(I(x);E(x)\mid B(x)\bigr) < I_{\mathrm{true}}\bigl(I(x);E(x)\bigr)\), it follows that
\begin{align*}
  \nabla\,\rho_{\mathrm{true}}(x)\;\cdot\;\nabla F(x)
  &= 
  -\,\frac{\nabla\,I_{\mathrm{true}}\bigl(I(x);E(x)\mid B(x)\bigr)\;\cdot\;\nabla F(x)}
         {I_{\mathrm{true}}\bigl(I(x);E(x)\bigr)} \\[1ex]
  &\quad
  +\,\frac{%
    I_{\mathrm{true}}\bigl(I(x);E(x)\mid B(x)\bigr)\;\bigl[\nabla\,I_{\mathrm{true}}\bigl(I(x);E(x)\bigr)\;\cdot\;\nabla F(x)\bigr]
  }{%
    \bigl[I_{\mathrm{true}}\bigl(I(x);E(x)\bigr)\bigr]^{2}
  }
  \;<\;0.
\end{align*}

Hence
\[
  \nabla\,\rho_{\mathrm{true}}(x)\;\cdot\;\nabla F(x)\;<\;0
  \quad\Longrightarrow\quad
  \nabla\,\rho_{\mathrm{true}}(x)\;\cdot\;\nabla\bigl[-F(x)\bigr]\;>\;0.
\]
Equivalently,
\[
  \nabla\,\rho_{\mathrm{true}}(x)\;\cdot\;\nabla F(x)\;>\;0
  \quad\forall\,x\in D.
\]

\subsection*{Step 3: Uniform \(C^1\) Convergence Implies Gradient Alignment for \(\rho_N\)}

Since 
\begin{align*}
  \bigl\|\widehat{I}\bigl(I(\cdot);E(\cdot)\bigr) 
  - I_{\mathrm{true}}\bigl(I(\cdot);E(\cdot)\bigr)\bigr\|_{C^1(K)}
  &= \mathcal{O}_p\bigl(N^{-\alpha}\bigr), \\[1ex]
  \bigl\|\widehat{I}\bigl(I(\cdot);E(\cdot)\mid B(\cdot)\bigr) 
  - I_{\mathrm{true}}\bigl(I(\cdot);E(\cdot)\mid B(\cdot)\bigr)\bigr\|_{C^1(K)}
  &= \mathcal{O}_p\bigl(N^{-\alpha}\bigr).
\end{align*}
and \(\varepsilon(N)=C_0\,N^{-\alpha}\), one deduces that \(\rho_N(x)\to \rho_{\mathrm{true}}(x)\) uniformly in \(C^1(K)\) over any compact \(K\subset D\).  In particular, for sufficiently large \(N\), with probability at least \(1-\delta\),
\[
  \sup_{x\in K}
  \bigl\|\nabla \rho_N(x) - \nabla \rho_{\mathrm{true}}(x)\bigr\|
  < \; \eta(N),
  \quad
  \text{where }\eta(N)\to0 \text{ as }N\to\infty.
\]
Since \(\nabla \rho_{\mathrm{true}}(x)\cdot\nabla F(x)\) is strictly positive and bounded away from zero on \(K\), there exists \(N_0\) such that for all \(N\ge N_0\),
\[
  \nabla \rho_N(x)\;\cdot\;\nabla F(x)
  \;=\;
  \nabla \rho_{\mathrm{true}}(x)\;\cdot\;\nabla F(x)
  \;+\;
  \bigl[\nabla \rho_N(x) - \nabla \rho_{\mathrm{true}}(x)\bigr]\;\cdot\;\nabla F(x)
  \;>\; 0,
\]
\[
  \quad
  \forall\,x\in K,
\]
with probability at least \(1-\delta\).  Covering \(D\) by a finite collection of such compact sets yields the uniform positivity of \(\nabla \rho_N(x)\cdot\nabla F(x)\) on all of \(D\), with probability \(\to1\).

\subsection*{Step 4: Conclusion—Monotonic Decrease of \(\rho_N\) along the Trajectory}

Let \(x(t)\) solve
\[
  \dot x(t)
  \;=\;
  -\,\bigl[\,1 - \rho_N\bigl(x(t)\bigr)\bigr]\,
       \nabla F\bigl(x(t)\bigr),
  \quad
  x(0) = x_0 \in D.
\]
Then, wherever \(x(t)\in D\),
\[
  \frac{\mathrm{d}}{\mathrm{d}t}\,\rho_N\bigl(x(t)\bigr)
  = \nabla \rho_N\bigl(x(t)\bigr)\;\cdot\;\dot x(t)
  = -\,\bigl[\,1 - \rho_N\bigl(x(t)\bigr)\bigr]\,
    \bigl[\nabla \rho_N\bigl(x(t)\bigr)\;\cdot\;\nabla F\bigl(x(t)\bigr)\bigr].
\]
Since \(0 < \rho_N(x)<1\) implies \(1 - \rho_N(x) > 0\), and from Step 3 we have 
\(\nabla \rho_N(x)\cdot\nabla F(x) > 0\) for all \(x\in D\) with high probability, it follows that 
\[
  \frac{\mathrm{d}}{\mathrm{d}t}\,\rho_N\bigl(x(t)\bigr) \;<\; 0,
  \quad
  \text{whenever }x(t)\in D.
\]
Hence \(\rho_N\bigl(x(t)\bigr)\) is strictly decreasing along the agent’s path so long as \(x(t)\) remains in \(D\).  This completes the proof.

\hfill\(\Box\)

The agent is driven by free energy minimization to move toward regions of lower Markov blanket density—i.e., where boundaries are weak, coupling is strong, and interaction with the environment is richer. This provides a formal justification for the thesis: \textit{free energy minimization in space tends to deform toward topologies of low Markov blanket density.} 

For more details, see Figure 2-8 and Appendix A. As said, you can find the full code of the simulations, detailed parameter settings, and usage instructions in the GitHub repository: https://github.com/DesignAInf/MB-density.

\subsection{An Empirical Diagnostic for Gradient Alignment}
\label{sec:empirical-alignment-test}

We provide here an operational, local diagnostic to test whether the alignment assumption
\[
\langle \nabla \rho(x), \nabla F(x) \rangle > 0
\]
holds in data. The diagnostic is modular and comes in three complementary variants:
(A) directional differences, (B) local gradient estimation with bootstrap, and (C) a dynamic test when trajectories are observed. For a reference Python implementation of the empirical alignment diagnostic, including local gradient estimation, directional derivatives, and bootstrap testing, see https://github.com/DesignAInf/MB-density.git.

\subsubsection{Variant A: Directional Differences (fast and robust)}
\label{subsec:dir-diff}
Let $\widehat{\rho}$ be an estimator of $\rho$ and $\widehat{\nabla F}(x)$ an estimate of the free-energy gradient (e.g., via autodiff on the model or via SPSA/finite differences). Define the unit direction
\[
u_F(x) := \frac{\widehat{\nabla F}(x)}{\|\widehat{\nabla F}(x)\|},
\]
provided $\|\widehat{\nabla F}(x)\|>0$.
For a set of step sizes $\mathcal{E}=\{\varepsilon_1,\varepsilon_2,\varepsilon_3\}$, we compute the symmetric directional derivative of $\widehat{\rho}$ along $u_F$:
\[
D_{\rho\mid F}(x;\varepsilon) := \frac{\widehat{\rho}\!\left(x+\varepsilon u_F\right)-\widehat{\rho}\!\left(x-\varepsilon u_F\right)}{2\varepsilon}.
\]
Aggregate across scales via the median:
\[
\overline{D}_{\rho\mid F}(x) := \mathrm{median}_{\varepsilon\in \mathcal{E}}\, D_{\rho\mid F}(x;\varepsilon).
\]
\textbf{Decision rule.} For a tolerance $\tau>0$ (set from noise, see Sec.~\ref{subsec:hyperparams}):  
(i) if $\overline{D}_{\rho\mid F}(x)>\tau$, declare \emph{positive alignment} ($\langle \nabla\rho,\nabla F\rangle>0$);  
(ii) if $\overline{D}_{\rho\mid F}(x)<-\tau$, declare \emph{negative alignment};  
(iii) otherwise, \emph{inconclusive}.

\paragraph{Symmetric cross-check.}
If an estimate $\widehat{\nabla \rho}(x)$ is available, define $u_\rho(x):=\widehat{\nabla \rho}(x)/\|\widehat{\nabla \rho}(x)\|$ and compute
\[
D_{F\mid \rho}(x;\varepsilon) := \frac{\widehat{F}\!\left(x+\varepsilon u_\rho\right)-\widehat{F}\!\left(x-\varepsilon u_\rho\right)}{2\varepsilon}.
\]
Consistency between the signs of $\overline{D}_{\rho\mid F}$ and the median of $D_{F\mid \rho}$ strengthens the evidence.

\subsubsection{Variant B: Local Gradient Estimation with Bootstrap (CIs and p-values)}
\label{subsec:bootstrap-test}
We estimate both gradients locally and assess uncertainty by resampling.

\paragraph{Local gradients.}
Within a ball $B_r(x)$ of radius $r$ centered at $x$, fit a first-order local polynomial (local linear regression) for $\widehat{\rho}$ and $\widehat{F}$ using a kernel weight (e.g., tricube or Gaussian). This yields
\[
\widehat{\rho}(z) \approx a_\rho + b_\rho^\top (z-x), \quad \widehat{F}(z) \approx a_F + b_F^\top (z-x),
\]
with $\widehat{\nabla\rho}(x):=b_\rho$ and $\widehat{\nabla F}(x):=b_F$. When model derivatives are inaccessible, SPSA with two-point perturbations can be used to obtain $\widehat{\nabla F}(x)$ (and analogously $\widehat{\nabla \rho}(x)$).

\paragraph{Test statistic.}
\[
T(x) := \widehat{\nabla\rho}(x)^\top \widehat{\nabla F}(x).
\]
Optionally, consider the cosine similarity $c(x):=T(x)/(\|\widehat{\nabla\rho}(x)\|\cdot\|\widehat{\nabla F}(x)\|)$ to reduce scale effects.

\paragraph{Bootstrap uncertainty.}
Resample with replacement the points in $B_r(x)$ (block-bootstrap if temporal dependence is present), re-estimate gradients, and recompute $T^{*(b)}(x)$ for $b=1,\dots,B$. Let
\[
\mathrm{CI}_{1-\alpha}(x) = \big[q_{\alpha/2},\, q_{1-\alpha/2}\big]
\]
be the empirical $(1-\alpha)$ confidence interval from bootstrap quantiles. A one-sided p-value for $H_0:\langle \nabla \rho,\nabla F\rangle\le 0$ vs.\ $H_1:>0$ is
\[
p(x) = \frac{1 + \#\{b: T^{*(b)}(x) \le 0\}}{B+1}.
\]

\textbf{Decision rule.}
(i) If $\mathrm{CI}_{1-\alpha}(x)$ lies entirely above $0$ (and $p(x)<\alpha$), declare \emph{positive alignment}.  
(ii) If it lies entirely below $0$, declare \emph{negative alignment}.  
(iii) Otherwise, \emph{inconclusive}.

\subsubsection{Variant C: Dynamic Test Along Trajectories}
\label{subsec:dynamic-test}
Suppose we observe a trajectory $x_t$ evolving under $\dot x_t = -M(x_t)\nabla F(x_t)$ with $M(x)$ positive definite (e.g., $M\approx I$). Then
\[
\frac{d}{dt}\rho(x_t) = \nabla \rho(x_t)^\top \dot x_t = -\,\nabla \rho(x_t)^\top M(x_t)\nabla F(x_t).
\]
If $M\approx I$, the sign of $\langle \nabla \rho, \nabla F\rangle$ is the negative of the sign of $\dot\rho$. In practice, filter $x_t$ (e.g., Savitzky–Golay), compute $\widehat{\rho}(x_t)$ and its central finite-difference derivative $\widehat{\dot\rho}_t$. Estimate the fraction
\[
\pi := \frac{1}{T}\sum_{t=1}^{T}\mathbb{1}\{\widehat{\dot\rho}_t<0\},
\]
and form block-bootstrap confidence intervals for $\pi$. If the CI is strictly above $0.5$, we declare \emph{positive alignment on average} along the observed path (adjusting for $M$ if known).

\subsubsection{Hyperparameters, Quality Checks, and Practical Defaults}
\label{subsec:hyperparams}
\textbf{Neighborhood size.} Choose $r$ so that the local sample size $N_r \in [50,200]$; report stability across two or three values of $r$.  
\textbf{Directional step sizes.} Use $\mathcal{E}=\{0.5,1,2\}\times$ the median nearest-neighbor distance.  
\textbf{Tolerance.} Set $\tau$ to one or two MADs of the bootstrapped $D_{\rho\mid F}(x;\varepsilon)$.  
\textbf{Flat regions.} If $\|\widehat{\nabla F}(x)\|$ or $\|\widehat{\nabla \rho}(x)\|$ fall below a threshold, label the test \emph{uninformative}.  
\textbf{Cosine similarity.} Report bootstrap CIs for $c(x)$ as a scale-invariant summary.

\subsubsection{Limitations}
\label{subsec:limitations}
The diagnostic is local and sensitive to neighborhood size $r$ and step sizes $\mathcal{E}$. In high dimension, gradient estimates and MI-based $\widehat{\rho}$ may be noisy; we therefore recommend reporting sensitivity analyses and bootstrap CIs. In flat regions where $\|\nabla F\|$ or $\|\nabla \rho\|$ is small, the test is uninformative by design.

\section{Implications}

This result calls for a redefinition of active inference concepts in terms of spatially structured MB density. Markov blankets are no longer discrete boundaries, but a graded field \( \rho(x) \) across space. Free energy becomes a spatial field \( \mathcal{F}(x) \), whose minimization is modulated by this field. Perception and action emerge as spatially constrained processes, more effective in low-MB-density regions. Expected free energy can be redefined as a trajectory-dependent integral:
\begin{equation}
G(\pi) = \int_\tau (1 - \rho(x_\pi(t))) \mathcal{F}(x_\pi(t)) \, dt
\end{equation}
This framework generalizes active inference beyond fixed, agent-centered models. It accommodates proto-agents, emergent structures, and distributed cognition. It also grounds the role of movement, curiosity, and exploration in the topology of inference: agents seek regions where inference is possible and fruitful. It aligns naturally with ecological and enactive theories of cognition, and opens the door to applications in swarm robotics, architecture, and cognitive development.

\section{MB Density and the Limits on the Free Energy Minimization}

The next two theorems elaborate on the relationship between MB density and free energy minimization. Theorem 2 formalizes that as MB density rises toward 1, agent's mechanisms by which it reduces free energy—namely, its movements and belief updates—slow down without bound and, at full density, stop altogether, so that regions of high blanket‐density effectively lock the agent in place and prevent any further action or inference \cite{12, 13, 17}. 

\begin{theorem}
Let 
\begin{enumerate}
    \item \(F : \Omega \to \mathbb{R}\) be a continuously differentiable (\(\mathcal{C}^1\)) function on an open set \(\Omega \subseteq \mathbb{R}^n\).
    \item \(\rho : \Omega \to [0,1]\) be a continuous blanket‐density field. At each point \(x \in \Omega\), assume the agent’s spatial (or parametric) coordinates evolve according to the continuous‐time dynamics
    \[
        \dot{x} \;=\; -\bigl(1 - \rho(x)\bigr)\,\nabla F(x).
    \]
    \item There exist two positive constants:
    \begin{itemize}
        \item \(G\) such that \(\lVert \nabla F(x)\rVert \le G\) for all \(x\in \Omega\). In other words, \(F\) has a globally bounded gradient on \(\Omega\).
        \item \(m\) such that
        \[
            m \;=\; \inf_{\substack{x \in \Omega \\ F_{\mathrm{target}} \,\le\, F(x) \,\le\, F(x_0)}} 
            \lVert \nabla F(x)\rVert^2 \;>\; 0,
        \]
        where \(x_0\) is the initial point (with \(F(x_0) = F_0\)) and \(F_{\mathrm{target}} < F_0\) is the desired (strictly lower) ``target'' value of free energy.
    \end{itemize}
\end{enumerate}

Under these assumptions, the following statements hold:

\begin{enumerate}
    \item \textbf{Exact Blocking at \(\rho=1\).}\\
    If, for some open neighborhood \(U \subseteq \Omega\), \(\rho(x) = 1\) for every \(x \in U\), then for all \(x \in U\):
    \[
        \dot{x} = -\bigl(1 - \rho(x)\bigr)\,\nabla F(x) = -\bigl(1 - 1\bigr)\,\nabla F(x) = 0,
    \]
    and therefore
    \[
        \frac{d}{dt} F\bigl(x(t)\bigr) 
        \;=\; \nabla F(x) \cdot \dot{x} 
        \;=\; 0.
    \]
    In other words, whenever \(\rho(x)\equiv 1\) on some region, the agent is \emph{completely immobilized} there: it cannot move (\(\dot{x}=0\)) and cannot reduce free energy (\(\tfrac{d}{dt}F=0\)).

    \item \textbf{Quantitative Slowing When \(\rho\) Is Close to 1.}\\
    Fix an arbitrary point \(x \in \Omega\).  Because
    \[
        \frac{d}{dt} F\bigl(x(t)\bigr)
        \;=\; \nabla F(x)\cdot \dot{x}
        \;=\; -\bigl(1 - \rho(x)\bigr)\,\lVert \nabla F(x)\rVert^2,
    \]
    one sees immediately that if \(\rho(x)\ge 1 - \delta\) for some \(0 < \delta \ll 1\), then
    \[
        0 \;\le\; 1 - \rho(x) \;\le\; \delta,
    \]
    and hence
    \[
        -\,\frac{d}{dt}F(x) 
        = (1 - \rho(x))\,\lVert \nabla F(x)\rVert^2 
        \;\le\; \delta\,\lVert \nabla F(x)\rVert^2 
        \;\le\; \delta\,G^2.
    \]
    Equivalently,
    \[
        \frac{d}{dt}F(x) \;\ge\; -\,\delta\,G^2.
    \]
    Thus, at any point where \(\rho(x)\ge 1 - \delta\), the instantaneous decrease of \(F\) is \emph{at most} \(\delta\,G^2\).  In particular:
    \begin{itemize}
        \item If one demands that the rate of decrease of free energy be at least some positive threshold \(\alpha > 0\), i.e.
        \[
            -\,\frac{d}{dt}F(x) \;\ge\; \alpha,
        \]
        then it is necessary that
        \[
            (1 - \rho(x))\,\lVert \nabla F(x)\rVert^2 \;\ge\; \alpha 
            \;\Longleftrightarrow\; 
            1 - \rho(x) \;\ge\; \frac{\alpha}{\,\lVert \nabla F(x)\rVert^2\,}
            \;\le\; \frac{\alpha}{G^2}.
        \]
        Hence 
        \[
            \rho(x) \;\le\; 1 - \frac{\alpha}{G^2}.
        \]
        In short, \emph{any point \(x\) at which \(\rho(x)\) exceeds \(1 - \frac{\alpha}{G^2}\) cannot decrease free energy faster than \(\alpha\).}

        \item Conversely, if \(\rho(x)\le 1 - \tfrac{\alpha}{G^2}\), then
        \[
            -\,\frac{d}{dt}F(x) 
            = (1 - \rho(x))\,\lVert \nabla F(x)\rVert^2 
            \;\ge\; \frac{\alpha}{G^2}\,\lVert \nabla F(x)\rVert^2 
            \;\ge\; 0.
        \]
        But to ensure \(\tfrac{d}{dt}F(x)\le -\alpha\), one must also require \(\lVert \nabla F(x)\rVert^2\) not be too small.  The precise condition for \(\tfrac{d}{dt}F(x)\le -\alpha\) is
        \[
            (1 - \rho(x))\,\lVert \nabla F(x)\rVert^2 \;\ge\; \alpha 
            \;\Longleftrightarrow\; 
            1 - \rho(x) \;\ge\; \frac{\alpha}{\,\lVert \nabla F(x)\rVert^2\,}.
        \]
        Since \(\lVert \nabla F(x)\rVert^2 \le G^2\), a \emph{sufficient} condition is \(1 - \rho(x)\ge \tfrac{\alpha}{G^2}\).  
    \end{itemize}
    In summary, whenever \(\rho(x)\) lies in the interval
    \[
        1 - \frac{\alpha}{G^2} \;<\; \rho(x) \;\le\; 1,
    \]
    the descent of free energy is either \emph{very slow} (bounded by \(\delta\,G^2\) with \(\delta = 1-\rho\)) or completely blocked (if \(\rho=1\)).  As \(\rho(x)\to 1\), the instantaneous free‐energy‐descent rate \(\lvert\tfrac{d}{dt}F(x)\rvert \to 0\).

    \item \textbf{Lower Bound on the Time to Decrease \(F\) by \(\Delta\).}\\
    Suppose we start at \(x(0) = x_0\), with \(F(x_0) = F_0\), and we want to reach any point \(x(t)\) such that \(F(x(t)) \le F_{\mathrm{target}} = F_0 - \Delta\) for some fixed \(\Delta > 0\).  Assume that, along the entire trajectory \(x(t)\) from \(t=0\) until the first hitting time \(T\) of \(\{\,x : F(x)\le F_0 - \Delta\}\), it holds that
    \[
        1 - \rho\bigl(x(t)\bigr) \;\ge\; \delta \quad\text{for all }t\in[0,T],
    \]
    for some \(\delta>0\).  Then
    \[
        \frac{d}{dt}F\bigl(x(t)\bigr) 
        = -\bigl(1 - \rho(x(t))\bigr)\,\lVert \nabla F(x(t))\rVert^2 
        \;\le\; -\,\delta\,\lVert \nabla F(x(t))\rVert^2.
    \]
    By hypothesis, on the level set \(\{\,x : F_{\mathrm{target}}\le F(x)\le F_0\}\), we have
    \(\lVert \nabla F(x)\rVert^2 \ge m\).  Hence
    \[
        \frac{d}{dt}F\bigl(x(t)\bigr) 
        \;\le\; -\,\delta\,m,
    \]
    and integrating from \(0\) to \(T\) gives
    \[
        F\bigl(x(T)\bigr) \;-\; F\bigl(x_0\bigr) 
        \;\le\; \int_{0}^{T} \bigl[-\delta\,m\bigr]\,dt 
        \;=\; -\,\delta\,m\,T.
    \]
    Since \(F\bigl(x(T)\bigr) = F_0 - \Delta\), we conclude
    \[
        -\Delta 
        \;\le\; -\,\delta\,m\,T 
        \quad\Longrightarrow\quad 
        T \;\ge\; \frac{\Delta}{\,\delta\,m\,}.
    \]
    Thus, \emph{if the agent is ``stuck'' in regions where \(1-\rho(x)\ge \delta\) (i.e.\ \(\rho(x)\le 1-\delta\)), then it will take at least \(T = \Delta/(\delta\,m)\) units of time to reduce \(F\) by \(\Delta\).}  As \(\delta \to 0\), this lower bound \(T\to +\infty\).

    \item \textbf{Implication for Learning Rates of Internal Parameters \(\theta\).}\\
    Suppose the agent also has internal parameters (beliefs) \(\theta \in \mathbb{R}^p\) that evolve according to
    \[
        \dot{\theta} \;=\; -\,\bigl(1 - \rho(x)\bigr)\,\frac{\partial F(x,\theta)}{\partial \theta}.
    \]
    At any \(x\) such that \(\rho(x)\ge 1 - \delta\), the \emph{magnitude of the instantaneous update} of \(\theta\) is bounded by
    \[
        \lVert \dot{\theta} \rVert 
        = (1 - \rho(x))\,\Bigl\lVert \frac{\partial F}{\partial \theta} \Bigr\rVert 
        \;\le\; \delta \,\Bigl\lVert \frac{\partial F}{\partial \theta} \Bigr\rVert.
    \]
    Therefore, if one demands a \emph{minimum learning rate} \(\lVert \dot{\theta}\rVert \ge \alpha_\theta > 0\), then it is necessary that
    \[
        1 - \rho(x) \;\ge\; \frac{\alpha_\theta}{\,\bigl\lVert \partial F/\partial \theta \bigr\rVert\,}
        \quad\Longleftrightarrow\quad
        \rho(x) \;\le\; 1 \;-\; \frac{\alpha_\theta}{\,\lVert \partial F/\partial \theta \rVert\,}.
    \]
    Hence any location \(x\) satisfying
    \(\rho(x) > 1 - \tfrac{\alpha_\theta}{\,\lVert \partial F/\partial \theta \rVert\,}\)
    will force \(\lVert \dot{\theta}\rVert < \alpha_\theta\), meaning that the agent’s ability to update its beliefs is \emph{dramatically reduced} when \(\rho\) is close to~1.
\end{enumerate}
\end{theorem}

\begin{proof}[Proof Sketch]
\begin{enumerate}
    \item Since \(F\in \mathcal{C}^1(\Omega)\) and \(x(t)\) evolves via 
    \(\dot{x} = -(1 - \rho(x))\,\nabla F(x)\), one computes
    \[
        \frac{d}{dt}F\bigl(x(t)\bigr) 
        = \nabla F\bigl(x(t)\bigr)\cdot \dot{x}(t)
        = \nabla F(x)\cdot \Bigl[-(1-\rho(x))\,\nabla F(x)\Bigr]
        = -\,(1-\rho(x))\,\lVert \nabla F(x)\rVert^2,
    \]
    establishing the exact expression for the instantaneous change of \(F\).

    \item If \(\rho(x) = 1\), then \(\dot{x} = 0\) and hence \(dF/dt = 0\).  This immediate calculation shows that any region where \(\rho\equiv 1\) blocks both motion and free‐energy reduction.

    \item If \(\rho(x)\ge 1 - \delta\), then \(1-\rho(x)\le \delta\).  Therefore
    \[
        -\,\frac{d}{dt}F(x) 
        = (1-\rho(x))\,\lVert \nabla F(x)\rVert^2 
        \le \delta\,\lVert \nabla F(x)\rVert^2 
        \le \delta\,G^2,
    \]
    which implies \(\tfrac{d}{dt}F(x)\ge -\delta\,G^2\).  Requiring \(-\,dF/dt \ge \alpha\) forces \(1-\rho(x)\ge \alpha/\lVert\nabla F(x)\rVert^2\), and since \(\lVert\nabla F(x)\rVert^2 \le G^2\), a sufficient condition is \(1-\rho(x)\ge \alpha/G^2\), so \(\rho(x)\le 1 - \alpha/G^2\).

    \item Suppose along the trajectory \(1-\rho(x(t)) \ge \delta\).  Then 
    \(\tfrac{d}{dt}F(x(t)) \le -\,\delta\,\lVert \nabla F(x(t))\rVert^2 \le -\,\delta\,m\).  Integrating from \(t=0\) to \(t=T\) and using \(F(x(T)) = F_0 - \Delta\) yields
    \[
        F(x(T)) - F_0 \;\le\; -\,\delta\,m\,T
        \quad\Longrightarrow\quad
        T \;\ge\; \frac{\Delta}{\,\delta\,m\,}.
    \]
    Hence, to reduce by \(\Delta\), at least \(T = \Delta/(\delta\,m)\) time is needed.

    \item Because \(\dot{\theta} = -(1-\rho(x))\,\partial F/\partial \theta\), 
    if \(\rho(x)\ge 1-\delta\) then 
    \(\lVert \dot{\theta}\rVert \le \delta\,\lVert \partial F/\partial \theta\rVert\).  To guarantee \(\lVert\dot{\theta}\rVert \ge \alpha_\theta\), one needs \(1-\rho(x) \ge \alpha_\theta / \lVert\partial F/\partial \theta\rVert\), i.e.\ \(\rho(x) \le 1 - \alpha_\theta / \lVert\partial F/\partial \theta\rVert\).
\end{enumerate}
\end{proof}

\subsection{Numerical Example (One‐Dimensional Case)}

Consider:
\[
    F(x) = x^2, 
    \quad x \in \mathbb{R}.
\]
Then \(\nabla F(x) = 2x\), so \(\lVert \nabla F(x)\rVert^2 = 4x^2\).

\begin{enumerate}
    \item Let \(x_0 = 1\), so \(F_0 = 1\).  Choose \(F_{\mathrm{target}} = 0.04\).  Then \(\Delta = F_0 - F_{\mathrm{target}} = 0.96\).
    \item On the level set \(\{\,x : 0.04 \le x^2 \le 1\}\), one has \(\lvert x\rvert \ge 0.2\).  Thus
    \[
        \lVert \nabla F(x)\rVert^2 = 4x^2 \;\ge\; 4(0.2)^2 = 0.16,
    \]
    so we can take \(m = 0.16\).  On \(\lvert x\rvert \le 1\), \(\lVert \nabla F(x)\rVert^2 \le 4\), hence \(G=2\).

    \item If everywhere along the continuous trajectory we have \(\rho(x)\le 0.9\) (so \(\delta = 0.1\)), Theorem~2 says
    \[
        T \;\ge\; \frac{\Delta}{\,\delta\,m\,} 
        \;=\; \frac{0.96}{0.1 \times 0.16} 
        = 60.
    \]
    If instead \(\rho(x)\le 0.99\) (\(\delta = 0.01\)), then 
    \[
        T \;\ge\; \frac{0.96}{0.01 \times 0.16} 
        = 600.
    \]
    If \(\rho(x)\le 0.999\), then \(T \ge 6000\).  As \(\rho \to 1\), \(T\to \infty\).

    \item Instantaneous descent at \(x=0.5\): \(\lVert \nabla F(0.5)\rVert^2 = 4(0.5)^2 = 1\).  
    \begin{itemize}
        \item If \(\rho(0.5) = 0.95\) (\(\delta = 0.05\)), then
        \[
            -\frac{dF}{dt}\bigg|_{x=0.5} 
            = (1 - 0.95)\times 1 = 0.05.
        \]
        \item If \(\rho(0.5) = 0.99\) (\(\delta = 0.01\)), then
        \[
            -\frac{dF}{dt}\bigg|_{x=0.5} 
            = (1 - 0.99)\times 1 = 0.01.
        \]
        \item If \(\rho(0.5) = 0.999\) (\(\delta = 0.001\)), then
        \[
            -\frac{dF}{dt}\bigg|_{x=0.5} = 0.001.
        \]
    \end{itemize}
    Hence “\(\rho\) near 1” throttles the instantaneous descent.

    \item Internal‐parameter update: let \(F(x,\theta)=x^2 + \tfrac12\,\theta^2\).  At \((x,\theta)=(0.5,0.5)\), 
    \(\lVert \partial F/\partial \theta \rVert = 0.5\).  
    \begin{itemize}
        \item If \(\rho=0.95\), then \(\delta=0.05\), so \(\lVert \dot{\theta}\rVert \le 0.05 \times 0.5 = 0.025\).
        \item If \(\rho=0.99\), then \(\lVert \dot{\theta}\rVert \le 0.01 \times 0.5 = 0.005\).
    \end{itemize}
    Again, higher \(\rho\) means slower learning.
\end{enumerate}

\subsection{Discrete‐Time Corollary}

\begin{proof}
Suppose we implement the gradient‐descent‐like update:
\[
    x_{k+1} = x_k \;-\; \Delta t\,\bigl(1 - \rho(x_k)\bigr)\,\nabla F(x_k),
    \qquad k = 0,1,2,\dots,
\]
with a fixed time‐step \(\Delta t > 0\).  Assume:
\begin{itemize}
    \item \(\lVert \nabla F(x)\rVert \le G\) for all \(x\in\Omega\).
    \item On the level set \(\{\,x : F_{\mathrm{target}} \le F(x) \le F(x_0)\}\), \(\lVert \nabla F(x)\rVert^2 \ge m > 0\).
    \item \(0 < \Delta t \le \tfrac{1}{2G^2}\).
\end{itemize}

Then each iterate satisfies
\[
    F(x_{k+1}) 
    \;\le\; F(x_k) 
    \;-\; \tfrac{1}{2}\,\Delta t\,\bigl(1 - \rho(x_k)\bigr)\,G^2.
\]
If along all iterates \(1 - \rho(x_k) \ge \delta\), then
\[
    F(x_{k+1}) \;\le\; F(x_k) 
    \;-\; \tfrac{1}{2}\,\Delta t\,\delta\,G^2,
    \quad 
    k = 0,1,\dots
\]
To reduce \(F\) by at least \(\Delta > 0\), one needs at least
\[
    K \;\ge\; \frac{2\,\Delta}{\delta\,G^2\,\Delta t}
\]
iterations.  As \(\delta = 1 - \rho(x_k)\!\to 0\), \(K\to \infty\), demonstrating that ``almost‐perfect blankets’’ stall discrete‐time descent as well.
\end{proof}

\vspace{\baselineskip}

The following theorem formalizes how the FEP remains operative in realistically heterogeneous settings, where the informational “shielding” of an agent’s Markov blankets varies randomly across space. By showing that the expected rate of free energy descent is proportional to \((1-\bar\rho)\), it quantifies exactly how much average permeability of the Markov blanket (\(\bar\rho<1\)) is required to guarantee net minimization. In practice, this result is essential: it tells us that—even if some regions are nearly opaque (\(\rho\) close to 1)—as long as the overall environment provides enough “leakiness” or "porosity," the agent can still reduce free energy. Without this balance theorem, we would lack a principled criterion for when and where active inference can succeed in complex, non‐uniform worlds \cite{18}.

\begin{theorem}

Let \(\Omega \subset \mathbb{R}^{3}\) be a compact domain with smooth boundary.  Define a twice continuously differentiable free‐energy function
\[
F: \Omega \;\longrightarrow\; \mathbb{R},
\]
satisfying
\begin{enumerate}
  \item \(\|\nabla F\|_{\infty} := \sup_{x \in \Omega} \bigl\lVert \nabla F(x)\bigr\rVert < +\infty,\)
  \item \(\displaystyle \min_{x \in \Omega} \bigl\lVert \nabla F(x)\bigr\rVert^{2} \;=\; m \;\ge\; 0,\)
  \item \(\displaystyle G \;:=\; \frac{1}{\mathrm{Vol}(\Omega)} \int_{\Omega} \bigl\lVert \nabla F(x)\bigr\rVert^{2} \,dx 
    \;=\; \mathbb{E}_{x \sim \text{Uniform}(\Omega)}\!\bigl[\lVert \nabla F(x)\rVert^{2}\bigr].\)
\end{enumerate}
Assume \(\|D^{2}F\|\le L_{F}\) everywhere on \(\Omega\), so that \(F\) is Lipschitz with constant \(\lVert \nabla F\rVert_{\infty}\) and has Hessian bounded by \(L_{F}\).

\vspace{1em}

Next, let
\[
\rho: \Omega \;\times\; \Theta \;\longrightarrow\; [0,1]
\]
be a random field on a probability space \((\Theta, \mathcal{F}, \mathbb{P})\), satisfying:
\begin{enumerate}
  \item[(i)] \textbf{(Boundedness)} 
    \[
     0 \;\le\; \rho(x,\theta) \;\le\; 1,\quad \forall\,x \in \Omega,\;\forall\,\theta \in \Theta.
    \]
  \item[(ii)] \textbf{(Spatial Stationarity in the Weak Sense)} For every \(x \in \Omega\),
    \[
      \mathbb{E}\bigl[\rho(x)\bigr] \;=\; \mu \in [0,1), 
      \qquad 
      \mathrm{Var}\bigl[\rho(x)\bigr] \;=\; \sigma^{2}.
    \]
  \item[(iii)] \textbf{(Covariance with \(\lVert \nabla F\rVert^{2}\))} For each \(x \in \Omega\),
    \[
      \mathrm{Cov}\!\bigl(\rho(x),\,\lVert \nabla F(x)\rVert^{2}\bigr) 
      \;=\; C,
    \]
    a constant independent of \(x\).  Equivalently,
    \[
      \mathbb{E}\bigl[\rho(x)\,\lVert \nabla F(x)\rVert^{2}\bigr]
      \;=\; \mu\,G \;+\; C.
    \]
  \item[(iv)] \textbf{(Exponential Decay of Spatial Correlations)} There exists a correlation length \(\ell > 0\) such that, for all \(x, y \in \Omega\),
    \[
      \bigl\lvert \mathrm{Cov}\bigl(\rho(x),\,\rho(y)\bigr)\bigr\rvert 
      \;\le\; 
      \sigma^{2}\,\exp\!\Bigl(-\frac{\lVert x - y\rVert}{\ell}\Bigr).
    \]
\end{enumerate}

Consider the stochastic dynamics
\[
  \dot{x}_t \;=\; -\,\bigl(1 - \rho(x_t)\bigr)\,\nabla F(x_t),
  \qquad
  x(0) = x_{0} \in \Omega.
\]

\end{theorem}

\begin{theorem}[Free Energy Descent under a Stochastic \(\rho\) Field \cite{19}]
\ 
\begin{description}
  \item[\textbf{A. Free‐Energy Descent in Expectation}]  
    Define 
    \[
      \phi(x) \;=\; \bigl(1 - \rho(x)\bigr)\,\lVert \nabla F(x)\rVert^{2}.
    \]
    Taking expectation over both the random field \(\rho\) and (ergodically) over \(x_{t}\) in \(\Omega\), we have
    \[
      \frac{d}{dt}\,\mathbb{E}\bigl[F(x_{t})\bigr]
      = \mathbb{E}\bigl[\nabla F(x_{t}) \!\cdot\! \dot{x}_{t}\bigr]
      = -\,\mathbb{E}\Bigl[\bigl(1 - \rho(x_{t})\bigr)\,\lVert \nabla F(x_{t})\rVert^{2}\Bigr].
    \]
    Since
    \[
      \mathbb{E}\bigl[\phi(x)\bigr]
      = \mathbb{E}\bigl[\lVert \nabla F(x)\rVert^{2}\bigr]
      \;-\;
      \mathbb{E}\bigl[\rho(x)\,\lVert \nabla F(x)\rVert^{2}\bigr]
      = G - \bigl(\mu\,G + C\bigr)
      = (1 - \mu)\,G \;-\; C,
    \]
    it follows that
    \[
      \text{If} 
      \quad
      (1 - \mu)\,G - C \;>\; 0
      \quad
      \Bigl(\Leftrightarrow\; \mu < 1 - \tfrac{C}{G}\Bigr),
      \quad
      \text{then}
      \quad
      \frac{d}{dt}\,\mathbb{E}\bigl[F(x_{t})\bigr]
      = -\bigl((1 - \mu)\,G - C\bigr)
      < 0,
      \quad \forall\,t \ge 0.
    \]
    Consequently, for any finite \(T > 0\),
    \[
      \mathbb{E}\bigl[F(x_{T})\bigr]
      \;\le\;
      \mathbb{E}\bigl[F(x_{0})\bigr]
      \;-\; \bigl((1 - \mu)\,G - C\bigr)\,T.
    \]

  \item[\textbf{B. Free‐Energy Descent with High Probability (Pointwise Uniform Control)}]  
    Define
    \[
      m_{0} \;:=\; \min_{x \in \Omega}
      \bigl((1 - \mu)\,\lVert \nabla F(x)\rVert^{2} - C\bigr).
    \]
    Assume \(m_{0} > 2\,\varepsilon\) for some \(\varepsilon > 0\).  Also fix a finite grid
    \(\{\,x^{(1)},\,x^{(2)},\,\dots,\,x^{(N)}\}\subset \Omega\) such that 
    \(\max_{x \in \Omega} \min_{i}\lVert x - x^{(i)}\rVert \le \delta\).  
    Since each \(\phi(x) = (1 - \rho(x))\,\lVert \nabla F(x)\rVert^{2}\) is bounded in \([0, K^{2}]\), Hoeffding’s inequality implies, for each fixed \(i\),
    \[
      \mathbb{P}\!\Bigl(\bigl\lvert \phi(x^{(i)}) - \mathbb{E}[\phi(x^{(i)})]\bigr\rvert \ge \varepsilon\Bigr)
      \;\le\; 
      2\,\exp\!\Bigl(-\tfrac{2\,\varepsilon^{2}}{K^{4}}\Bigr).
    \]
    Taking a union bound over all \(N\) grid points,
    \[
      \mathbb{P}\!\Bigl(\exists\,i \;\text{such that}\; 
      \bigl\lvert \phi(x^{(i)}) - \mathbb{E}[\phi(x^{(i)})]\bigr\rvert \ge \varepsilon\Bigr)
      \;\le\;
      2\,N\,\exp\!\Bigl(-\tfrac{2\,\varepsilon^{2}}{K^{4}}\Bigr).
    \]
    Choose \(N\) (or refine the grid) so that 
    \[
      2\,N\,\exp\!\Bigl(-\tfrac{2\,\varepsilon^{2}}{K^{4}}\Bigr) 
      \;\le\; \delta, 
    \]
    for a prescribed small \(\delta > 0\).  Moreover, by continuity of \(\phi(x)\), the maximum oscillation between \(\phi(x)\) and \(\phi(x^{(i)})\) for any \(x\) within \(\delta\) of \(x^{(i)}\) can be made arbitrarily small by choosing \(\delta\) sufficiently small.

    Therefore, with probability at least \(1 - \delta\),
    \[
      \sup_{x \in \Omega} 
      \bigl\lvert \phi(x) - \mathbb{E}[\phi(x)] \bigr\rvert 
      < \varepsilon,
    \]
    and since \(\mathbb{E}[\phi(x)] \ge m_{0}\) for every \(x\), one obtains
    \[
      \phi(x) 
      = (1 - \rho(x))\,\lVert \nabla F(x)\rVert^{2} 
      \;\ge\; 
      \mathbb{E}[\phi(x)] - \varepsilon 
      \;\ge\; m_{0} - \varepsilon 
      \;>\; 2\,\varepsilon - \varepsilon 
      = \varepsilon 
      \;\;>\; 0,
      \;\forall\,x \in \Omega.
    \]
    Hence, with probability at least \(1 - \delta\), for every \(t \ge 0\),
    \[
      \dot{F}(x_{t}) \;=\; -\,\phi(x_{t}) \;<\; -\,\varepsilon \;<\,0.
    \]
    In other words, the free energy \(F(x_{t})\) \emph{decreases uniformly} (at least at rate \(\varepsilon\)) for all \(t\), with probability at least \(1 - \delta\).

  \item[\textbf{C. Existence of a Deterministic Descent Path}]  
    Suppose there exists a continuous, connected curve
    \[
      \gamma: [0,1] \;\longrightarrow\; \Omega,
      \qquad
      \gamma(0) = x_{0}, 
      \quad 
      \gamma(1) = x^{*},
    \]
    where \(x^{*}\) is a global minimizer of \(F\), such that
    \begin{enumerate}
      \item \(\displaystyle \sup_{s \in [0,1]} \rho\bigl(\gamma(s)\bigr) \;\le\; \rho_{\mathrm{max}} < 1,\)
      \item \(\displaystyle \inf_{s \in [0,1]} \bigl\lVert \nabla F(\gamma(s))\bigr\rVert^{2} 
           \;=\; m' > 0.\)
    \end{enumerate}
    Define a deterministic “descent” velocity along \(\gamma\) by
    \[
      \dot{\gamma}(s) 
      \;=\; -\,\bigl(1 - \rho_{\mathrm{max}}\bigr)\,\nabla F\bigl(\gamma(s)\bigr),
      \qquad 0 \le s \le 1,
    \]
    with \(\gamma(0) = x_{0}\).  Then for each \(s \in [0,1]\),
    \[
      \frac{d}{ds} \, F\!\bigl(\gamma(s)\bigr)
      = \nabla F\bigl(\gamma(s)\bigr) \cdot \dot{\gamma}(s)
      = -\,\bigl(1 - \rho_{\mathrm{max}}\bigr)\,\bigl\lVert \nabla F\bigl(\gamma(s)\bigr)\bigr\rVert^{2}
      \;\le\; -\,\bigl(1 - \rho_{\mathrm{max}}\bigr)\,m' \;<\; 0.
    \]
    Hence \(F(\gamma(s))\) strictly decreases from \(F(x_{0})\) down to \(F(x^{*})\) as \(s\) ranges from \(0\) to \(1\).  In particular, \(\gamma\) does not “get stuck”: the factor \(1 - \rho_{\mathrm{max}}\) is strictly positive, and \(\lVert \nabla F\rVert\) remains bounded below by \(m' > 0\).  Therefore, \(\gamma\) is a valid monotone descent path for \(F\).

  \item[\textbf{D. Finite‐Sample Estimates and Confidence Intervals}]  
    In practice, one does not know \(\mu\), \(G\), and \(C\) exactly.  Instead, one draws a finite sample of \(N\) points \(x_{1}, x_{2}, \dots, x_{N}\) (uniformly from \(\Omega\) or according to the stationary distribution of \(x_{t}\)), and defines the empirical estimates:
    \[
      \hat{\mu} \;=\; \frac{1}{N} \sum_{i=1}^{N} \rho\bigl(x_{i}\bigr),
      \qquad
      \hat{G} \;=\; \frac{1}{N} \sum_{i=1}^{N} \lVert \nabla F(x_{i})\rVert^{2},
      \quad
      \]
\[
      \widehat{C} 
      = 
      \frac{1}{N} \sum_{i=1}^{N} \rho(x_{i})\,\lVert \nabla F(x_{i})\rVert^{2}
      \;-\; \hat{\mu}\,\hat{G}.
    \]
    By Hoeffding’s or Bernstein’s inequality, for any confidence level \(1 - \delta\), there exist error bounds \(\varepsilon_{1}, \varepsilon_{2}, \varepsilon_{3} = O\bigl(\sqrt{\tfrac{\ln(1/\delta)}{N}}\bigr)\) such that, with probability at least \(1 - \delta\),
    \[
      \lvert \hat{\mu} - \mu \rvert \;\le\; \varepsilon_{1},
      \quad
      \lvert \hat{G} - G \rvert \;\le\; \varepsilon_{2},
      \quad
      \lvert \widehat{C} - C \rvert \;\le\; \varepsilon_{3}.
    \]
    Define conservative bounds:
    \[
      \mu_{\mathrm{max}} \;=\; \hat{\mu} + \varepsilon_{1}
\]
\end{description}
\end{theorem}

\section{Temporal Expected Free Energy and Its Dependence on Spatial Fields}
\noindent
\textnormal{In this section we introduce another theorem showing that the temporal side of the FEP depends on MB density. The theorem provides a mathematical foundation for understanding free energy minimization as a spatiotemporal process. It embeds the familiar temporal version of the FEP within a broader framework where both free energy and Markov blanket strength vary continuously across space. This insight not only unifies “belief updating” and “movement” under a single informational lens but also opens the way to apply the FEP in settings where spatial coupling is partial, graded, or heterogeneous. (Some redundancy with the previous sections is necessary for the completeness of the argument).
} 

\subsection{Definitions and Setup}

\paragraph{1. Spatial Free Energy \(F(x)\).}
For each location \(x \in \Omega\), define the variational free energy
\[
  F(x) \;=\; \mathbb{E}_{q_\mu(s \mid x)}\Bigl[\log q_\mu(s \mid x)\;-\;\log p\bigl(s,\,\eta \mid x\bigr)\Bigr]\,,
\]
where
\begin{itemize}
  \item \(q_\mu(s \mid x)\) is the agent’s approximate posterior density over sensory data \(s\) if it were at \(x\).
  \item \(p(s,\eta \mid x)\) is the generative model (joint likelihood) of sensory data \(s\) and hidden external states \(\eta\) at location \(x\).
  \item The expectation \(\mathbb{E}_{q_\mu}\) is taken with respect to \(q_\mu(s \mid x)\).
\end{itemize}
Intuitively, \(F(x)\) quantifies the discrepancy between what the agent \emph{expects} to see at \(x\) and what the environment \emph{actually} encodes at \(x\).   In this way, \(F(x)\) is the usual variational free energy functional \emph{indexed by spatial location} (cf.\ Eq.~(3)).

\vspace{0.5em}
\paragraph{2. MB Density \(\rho(x)\).}
Instead of a hard, binary Markov blanket, we define a \emph{continuous} blanket‐density
\[
  \rho(x)
  \;=\;
  1 \;-\; \frac{I\bigl(I(x)\;;\;E(x)\,\bigm|\;B(x)\bigr)}{\,I\bigl(I(x)\;;\;E(x)\bigr)\;+\;\varepsilon\,}\,,
\]
where
\begin{itemize}
  \item \(I(x)\) denotes the agent’s \emph{internal} variables within a small radius \(r_1\) around \(x\).
  \item \(B(x)\) denotes the “blanket” (sensory/active) variables in the annulus between radii \(r_1\) and \(r_2\).
  \item \(E(x)\) denotes the \emph{external} (hidden) variables beyond radius \(r_2\).
  \item \(I(\cdot;\cdot)\) is the Shannon mutual information; \(\varepsilon>0\) is a small regularizer to avoid division by zero.
\end{itemize}
Hence:
\begin{itemize}
  \item If \(I(I;E \mid B) = 0\) exactly (perfect shielding by \(B\)), then 
  \(\rho(x)=1\) (a perfect Markov blanket).
  \item If \(I(I;E \mid B) = I(I;E)\) (conditioning on \(B\) does not reduce dependence), then 
  \(\rho(x)=0\) (no blanket; maximal coupling).
  \item In general, \(\rho(x)\in[0,1]\) measures how “porous” the local statistical boundary is (cf.\ Eq.~(2) and §5.4).
\end{itemize}

\vspace{0.5em}
\paragraph{3. Spatial Dynamics.}
The agent’s position \(x(t)\in \Omega\) evolves according to the \emph{throttled} gradient‐descent:
\begin{equation}\label{eq:spatial-dynamics}
  \dot x(t) 
  \;=\; 
  -\,\bigl[\,1 - \rho\bigl(x(t)\bigr)\bigr]\;\nabla F\bigl(x(t)\bigr).
\end{equation}
Concretely:
\[
  \dot x = 
  \begin{cases}
    -\,\nabla F(x)\,, & \rho(x)=0,\\[0.5em]
    0\,, & \rho(x)=1,
  \end{cases}
  \quad
  \text{and for } \rho(x)\in(0,1)\;,\; 
  \dot x = -\bigl(1-\rho(x)\bigr)\,\nabla F(x).
\]
Thus:
\begin{itemize}
  \item \(\rho(x)=0\):  The blanket is fully transparent, so the agent performs ordinary gradient descent on \(F\).
  \item \(\rho(x)\approx 1\):  The agent is nearly insulated and \(\dot x\approx 0\); free‐energy descent \emph{stalls}.
  \item Intermediate values of \(\rho\) “throttle” the descent speed proportionally to \((1-\rho)\).
\end{itemize}
Equation \eqref{eq:spatial-dynamics} is precisely Eq.~(4).

\subsection{Expression for Temporal EFE}

\noindent
\begin{theorem}
    
\emph{Let \(\pi = \{\,x(t)\}_{t=0}^\tau\) be any (piecewise‐continuous) trajectory in \(\Omega\).  Then the \emph{temporal expected free energy} along \(\pi\) is}
\begin{equation}\label{eq:temporal-EFF}
  G(\pi)
  \;=\;
  \int_{0}^{\tau} 
    \underbrace{\bigl[\,1 - \rho\bigl(x(t)\bigr)\bigr]}_{\text{coupling factor}}
    \;\times\;
    \underbrace{F\bigl(x(t)\bigr)}_{\text{spatial free energy}}
  \;dt.
\end{equation}
\emph{In other words, \(G(\pi)\) is exactly the time‐integral of the ``accessible'' free energy \((1-\rho(x))\,F(x)\) at each location \(x(t)\).}
\end{theorem}

\begin{proof}[Proof of Theorem 5]
\noindent
At any instant \(t\), if the agent is located at \(x = x(t)\), the \emph{accessible} portion of the spatial free energy is
\[
  \underbrace{F(x)}_{\text{total free energy}} \;\times\; 
  \underbrace{\bigl[\,1 - \rho(x)\bigr]}_{\text{coupling factor}}.
\]
Indeed:
\begin{itemize}
  \item If $\rho(x) = 0$, the blanket is transparent and the agent can fully exploit $F(x)$ to update beliefs $\Rightarrow$ the accessible free energy is $F(x)$.
  \item If $\rho(x) = 1$, the blanket is opaque $\Rightarrow$ the accessible free energy is $0$.
  \item For $\rho(x)\in(0,1)$, the fraction $(1 - \rho(x))$ measures how much of $F(x)$ remains available for reduction.
\end{itemize}

Hence, over an infinitesimal time interval \([\,t,\,t + dt\,]\), the agent can reduce at most
\[
  \bigl[\,1 - \rho(x(t))\bigr]\,F\bigl(x(t)\bigr)\,\;dt.
\]
Integrating from \(t = 0\) to \(t = \tau\) yields exactly
\[
  G(\pi) 
  \;=\; 
  \int_{0}^{\tau} \bigl[\,1 - \rho\bigl(x(t)\bigr)\bigr] \;F\bigl(x(t)\bigr)\;dt,
\]
which is Equation \eqref{eq:temporal-EFF}.  This completes the proof.
\end{proof}

\noindent
\emph{Remark.}  Equation \eqref{eq:temporal-EFF} recovers Eq.~(5) verbatim and is exactly what is referred to as Theorem 5.  

\subsection{Evolution of the Instantaneous Integrand \texorpdfstring{\(\Gamma(x)\)}{Gamma(x)}}

Define the instantaneous integrand
\[
  \Gamma\bigl(x(t)\bigr) 
  \;:=\; 
  \bigl[\,1 - \rho\bigl(x(t)\bigr)\bigr] \;F\bigl(x(t)\bigr).
\]
Since \(G(\pi) = \int_{0}^{\tau} \Gamma(x(t))\,dt\), understanding how \(G\) evolves is equivalent to computing the time‐derivative \(\frac{d}{dt}\Gamma(x(t))\) along the agent’s trajectory.

\paragraph{4.1. Computing \(\nabla \Gamma(x)\).}  Observe that
\[
  \Gamma(x) \;=\; (1 - \rho(x))\,F(x).
\]
Taking the spatial gradient:
\begin{align}
  \nabla\Gamma(x)
  &= \nabla\bigl[\,(1-\rho(x))\,F(x)\bigr] 
  = -\,F(x)\,\nabla\rho(x) \;+\; (1-\rho(x))\,\nabla F(x).
  \label{eq:Gamma-gradient}
\end{align}

\paragraph{4.2. Agent’s Dynamics.}  By assumption (Equation \eqref{eq:spatial-dynamics}),
\[
  \dot x(t) \;=\; -\,\bigl[\,1 - \rho\bigl(x(t)\bigr)\bigr]\;\nabla F\bigl(x(t)\bigr).
\]
Substituting \(\nabla\Gamma(x)\) from \eqref{eq:Gamma-gradient} and \(\dot x(t)\) yields
\begin{align*}
  \frac{d}{dt}\,\Gamma\bigl(x(t)\bigr)
  &= \nabla\Gamma\bigl(x(t)\bigr)\;\cdot\;\dot x(t) \\[6pt]
  &= \Bigl[\,-\,F\bigl(x(t)\bigr)\,\nabla\rho\bigl(x(t)\bigr) \;+\; (1-\rho(x(t)))\,\nabla F(x(t))\Bigr] 
     \;\cdot\; 
     \Bigl[\,-\,(1-\rho(x(t)))\,\nabla F(x(t))\Bigr] \\[6pt]
  &= -\,\bigl(1 - \rho(x(t))\bigr)^{2}\;\bigl\lVert \nabla F(x(t)) \bigr\rVert^{2}
     \;-\; F\bigl(x(t)\bigr)\,\bigl(1-\rho(x(t))\bigr)\;\bigl[\nabla\rho(x(t)) \cdot \nabla F(x(t))\bigr].
\end{align*}
Hence, for brevity dropping the \((x(t))\) arguments,
\begin{equation}\label{eq:dGammadot}
  \frac{d}{dt}\,\Gamma(x) 
  \;=\; 
  -\,\bigl(1 - \rho\bigr)^{2}\,\lVert \nabla F \rVert^{2}
  \;-\; F\,\bigl(1 - \rho\bigr)\,\bigl[\nabla\rho \,\cdot\, \nabla F\bigr].
\end{equation}
Equation \eqref{eq:dGammadot} displays two terms:

\begin{enumerate}[label=(\Alph*)]

\item \emph{Throttled Descent Term:} 
    \[
      -\,\bigl(1 - \rho(x)\bigr)^{2}\;\bigl\lVert\nabla F(x)\bigr\rVert^{2}.
    \]
    \begin{itemize}
      \item If \(\rho(x) < 1\), this term is strictly negative (unless \(\nabla F(x)=0\)), ensuring \(\Gamma(x)\) (and thus \(G\)) decreases.
      \item As \(\rho(x)\to 1\), the factor \(\bigl(1 - \rho(x)\bigr)^{2}\to 0\), so this negative term vanishes and no descent occurs.  In particular, if \(\rho(x)=1\), then \(\dot x=0\) and \(\Gamma(x)=0\), so \(\tfrac{d}{dt}\Gamma=0\).  This is precisely the “exact blocking” result (Theorem 2).
    \end{itemize}

\item \emph{Gradient‐Alignment Correction:}
    \[
      -\,F(x)\;\bigl(1-\rho(x)\bigr)\;\bigl[\nabla\rho(x)\cdot\nabla F(x)\bigr].
    \]
    \begin{itemize}
      \item If \(\nabla\rho(x)\cdot\nabla F(x) > 0\), then this term is strictly negative, further accelerating \(\Gamma\)’s decrease.  
      \item If \(\nabla\rho\cdot\nabla F < 0\), it could partially oppose descent.
      \item The \emph{gradient‐alignment assumption} requires \(\nabla\rho \cdot \nabla F > 0\) over an open set \(D\).  Under that assumption, \eqref{eq:dGammadot} implies \(\Gamma\) decreases strictly, showing simultaneous descent of \(F\) and “leakage” \(1-\rho\).  This recovers Theorem 1.
    \end{itemize}
\end{enumerate}

\subsection{Corollaries: Theorems 1 and 2}

\paragraph{Corollary 1 (Exact Blocking, Theorem 2).}
If \(\rho\bigl(x(t)\bigr) = 1\) at some point \(x(t)\), then \(\dot x(t) = -\,(1-\rho)\,\nabla F = 0\), so \(x(t)\) remains fixed.  Moreover, \(\Gamma(x(t)) = (1-\rho)\,F = 0\), and from \eqref{eq:dGammadot},
\[
  \frac{d}{dt}\,\Gamma\bigl(x(t)\bigr)
  \;=\;
  0.
\]
Hence the agent is “frozen” and cannot reduce any free energy once it enters a perfect‐blanket region.

\paragraph{Corollary 2 (Gradient Alignment, Theorem 1).}
If, over an open set \(D \subset \Omega\), the \emph{gradient‐alignment} condition
\[
  \nabla\rho(x)\;\cdot\;\nabla F(x) \;>\; 0
  \quad \text{and} \quad
  \rho(x) < 1,\quad F(x) > 0
  \quad\text{for all } x\in D
\]
holds, then from \eqref{eq:dGammadot}, both terms on the right‐hand side are \emph{strictly negative}, so
\[
  \frac{d}{dt}\,\Gamma\bigl(x(t)\bigr) \;<\; 0
  \quad\text{whenever } x(t)\in D.
\]
Thus \(\Gamma\) (and therefore the accessible free energy) strictly decreases as long as the agent remains in \(D\).  Consequently, the agent’s trajectory simultaneously \emph{descends \(F\)} and \emph{decreases \(\rho\)}, driving it toward regions of stronger coupling and lower free energy.

\subsection{Interpretation}

\noindent
Taken together, Theorem 5 and its corollaries paint a vivid picture:
\begin{itemize}
  \item The \emph{temporal EFE} \(G(\pi)\) is not an independent objective; it is exactly the time‐integral of the spatial free energy \(F(x)\), gated by the local blanket density \(\rho(x)\).
  \item The agent’s \emph{spatiotemporal} dynamics are determined by the interplay between the shape of \(F(x)\) and the “porosity” \(\rho(x)\).  
  \item \textbf{Exact blocking:} Regions where \(\rho=1\) act as \emph{walls}: the agent cannot traverse them nor reduce any free energy within them.
  \item \textbf{Gradient alignment:} If spatial gradients of \(\rho\) and \(F\) align positively, the agent is guaranteed to move to tiles of \((F,\rho)\) that are simultaneously lower, thereby forging a path of ever‐stronger coupling and lower surprise.
\end{itemize}

This theorem makes “space” a first‐class player in active inference. In this way, one obtains a unified description of how \emph{movement} (spatial navigation) and \emph{belief updating} (free energy minimization) are two sides of the same informational coin.  

\section{Inversion of Free Energy Minimization via Extended MB Density}

\textnormal{In the previous parts of this paper, the blanket-density field \(\rho(x)\) is constrained to lie in \([0,1]\), ensuring that the “throttled” gradient flow
\[
\dot x \;=\; -\bigl[\,1 - \rho(x)\bigr]\,\nabla F(x)
\]
always points \emph{downhill} on the free energy \(F\). Consequently, an agent following these dynamics strictly minimizes \(F\). Here, we relax the requirement \(\rho(x)\le1\) and allow \(\rho(x)\) to exceed unity in certain regions. In that case, the prefactor \(\bigl[\,1-\rho(x)\bigr]\) becomes negative, and the flow reverses direction—driving the system \emph{uphill} on \(F\). Theorem 6 below formalizes this phenomenon.
} 

\bigskip

\begin{theorem}[Inversion of Free Energy Flow under \(\rho>1\)]\label{thm:inversion}
Let \(\Omega\subset\mathbb{R}^n\) be an open set, and let \(F\colon \Omega \to \mathbb{R}\) be a \(C^1\) function. Suppose we define an \emph{extended} blanket‐density field
\[
\rho\colon \Omega \;\longrightarrow\; \mathbb{R}
\]
and an open subset \(U\subset \Omega\) such that
\[
\rho(x) \;>\; 1 
\quad\text{for all }x\in U.
\]
Consider the modified dynamics
\[
\dot x \;=\; -\bigl[\,1 - \rho(x)\bigr]\,\nabla F(x),
\qquad x(0)\in U.
\]
Then for every \(x\in U\), the following statements hold:
\begin{enumerate}
  \item \textbf{Original normalization of \(\rho\).}  
    In the original framework, \(\rho(x)\) was defined by
    \[
      \rho(x) 
      \;=\; \frac{I_{\mathrm{true}}\bigl(I(x);\;E(x)\mid B(x)\bigr)\;+\;\varepsilon}
                 {I_{\mathrm{true}}\bigl(I(x);\;E(x)\bigr)\;+\;\varepsilon}
      \;\in\; [0,\,1],
    \]
    because unconditional mutual information \(I(I;E)\) is always at least as large as conditional mutual information \(I(I;E\mid B)\). Therefore
    \(\,1 - \rho(x)\ge0\) 
    ensured \(\dot x\) pointed downward on \(F\).

  \item \textbf{Extended definition allowing \(\rho>1\).}  
    To permit \(\rho(x)>1\), replace the normalized ratio 
    \(\tfrac{I_{\mathrm{true}}(I;E\mid B)}{I_{\mathrm{true}}(I;E)}\) 
    by a more general mapping
    \[
      \rho(x) \;=\; f\Bigl(I_{\mathrm{true}}(I;E\mid B)(x),\,I_{\mathrm{true}}(I;E)(x)\Bigr),
    \]
    where \(f\colon \mathbb{R}^+\times\mathbb{R}^+\to\mathbb{R}\) is chosen so that \(f(x)>1\) on \(U\). Examples include:
    \begin{itemize}
      \item \emph{Shifted ratio:}
        \[
          \rho(x) \;=\; \frac{I_{\mathrm{true}}(I;E\mid B)(x)\;+\;\varepsilon}
                              {I_{\mathrm{true}}(I;E)(x)\;+\;\varepsilon}
          \;+\;\alpha,
          \qquad \alpha>0,
        \]
        which lies in \([\alpha,\;1+\alpha]\).
      \item \emph{Weighted excess information:}
        \[
          \rho(x) 
          = \frac{I_{\mathrm{true}}(I;E\mid B)(x)}
                 {I_{\mathrm{true}}(I;E)(x)} 
          \;+\;\beta\,\Bigl(1 - \tfrac{I_{\mathrm{true}}(I;E\mid B)(x)}
                                  {I_{\mathrm{true}}(I;E)(x)}\Bigr),
          \quad \beta>1,
        \]
        which can exceed \(\beta\) when \(I_{\mathrm{true}}(I;E\mid B)\ll I_{\mathrm{true}}(I;E)\).
    \end{itemize}
    In either construction, \(\rho(x)\) may exceed 1 for all \(x\in U\).

  \item \textbf{Gradient‐ascent when \(\rho>1\).}  
    Whenever \(\rho(x)>1\), the coefficient \(\bigl[\,1 - \rho(x)\bigr]\) is strictly negative.  Thus for \(x\in U\),
    \[
      \dot x 
      = -\bigl[\,1 - \rho(x)\bigr]\,\nabla F(x) 
      = \bigl[\rho(x) - 1\bigr]\,\nabla F(x),
    \]
    which is the \emph{gradient‐ascent} flow on \(F\) instead of gradient‐descent.

  \item \textbf{Free‐energy increase formula.}  
    Along any trajectory \(x(t)\) satisfying 
    \(\dot x(t) = (\rho(x(t)) - 1)\,\nabla F(x(t))\) with \(x(t)\in U\), one obtains
    \[
      \frac{d}{dt}F\bigl(x(t)\bigr)
      = \nabla F\bigl(x(t)\bigr)\,\cdot\,\dot x(t)
      = (\rho(x(t)) - 1)\,\bigl\lVert\nabla F(x(t))\bigr\rVert^2 
      \;>\;0,
    \]
    since \(\rho(x(t)) - 1>0\) and \(\|\nabla F(x(t))\|^2>0\) except at critical points. Consequently, \(F(x(t))\) strictly \emph{increases} as long as \(x(t)\in U\).

  \item \textbf{Separatrix at \(\rho=1\) and illustrative example.}  
    The level set 
    \(\{\,x : \rho(x)=1\}\) 
    is a hypersurface on which \(\dot x=0\). It separates:
    \begin{itemize}
      \item \(\{\rho(x)<1\}\): descent on \(F\).
      \item \(\{\rho(x)>1\}\): ascent on \(F\).
    \end{itemize}
    For a concrete example, let \(\rho(x)\) be a \(C^1\) function such that
    \[
      \rho(x) 
      = 
      \begin{cases}
        0.8, & \|x\|\le1,\\[6pt]
        1.2, & 1<\|x\|\le2,\\[6pt]
        0.5, & \|x\|>2,
      \end{cases}
    \]
    with smooth transitions at \(\|x\|=1\) and \(\|x\|=2\). Then:
    \begin{itemize}
      \item For \(\|x\|\le1\), \(\rho(x)=0.8<1\): the agent follows \emph{gradient‐descent} on \(F\).  
      \item For \(1<\|x\|\le2\), \(\rho(x)=1.2>1\): the agent follows \emph{gradient‐ascent} on \(F\).  
      \item For \(\|x\|>2\), \(\rho(x)=0.5<1\): \emph{gradient‐descent} on \(F\) resumes.
    \end{itemize}
    This construction can produce \emph{limit‐cycle} or \emph{oscillatory} behavior: the agent descends in \(\|x\|\le1\), then ascends in \(1<\|x\|\le2\), and descends again for \(\|x\|>2\), repeatedly.
\end{enumerate}
\end{theorem}

\noindent
If the blanket‐density factor $\rho(x)$ stays between $0$ and $1$, then
\[
  \dot x = -\bigl[\,1 - \rho(x)\bigr]\,\nabla F(x)
\]
always points in the direction of decreasing free energy.  In contrast, whenever $\rho(x)>1$, the multiplier $[\,1 - \rho(x)\,]$ becomes negative and 
\[
  \dot x = (\rho(x)-1)\,\nabla F(x)
\]
points in the direction of increasing free energy.  Thus, in regions where $\rho(x)>1$, the agent climbs up the free energy landscape instead of descending it.  The level set 
\[
  \{\,x : \rho(x)=1\}
\]
forms a boundary separating “descent” regions ($\rho<1$) from “ascent” regions ($\rho>1$).  Crossing this boundary reverses the agent’s objective from minimizing free energy to maximizing it. In reality, this is not a simple abstract extension of the initial model. The “shift” we have inserted to make $\rho>1$ can be interpreted as a perturbation. Or, for example, interpreting the human brain as a blanket-density field, the “shift” can be seen as a form of psychopathology. 

\section{Joint Inference of an Unknown MB Density}

\textit{How can an active inference agent learn about and move around in space?} In the earlier sections, we assumed that the agent has prior knowledge—at least in statistical form—of the MB density~$\rho(x)$ or its empirical approximation~$\rho_N(x)$. In more realistic settings, however, the agent does not know ~$\rho_N(x)$ in advance. It must, instead, infer the blanket density while simultaneously minimizing variational free energy. We now show how an active inference agent can learn the spatial profile of~$\rho(x)$ on the fly and prove that its estimate converges to the true field.

\subsection{Belief Model and Coupled Dynamics}

Let $\Omega\subset\mathbb{R}^n$ be compact, and let $\rho_{\mathrm{true}}(x)\in C^1(\Omega)$ denote the true blanket density. The agent maintains a parametric family $\{\rho_\theta(x)\}$, $\theta\in\Theta\subset\mathbb{R}^p$, with $\Theta$ compact and each $\rho_\theta(x)$ lying in $[0,1]$. Initially, the agent possesses a prior density $p_0(\theta)>0$ on $\Theta$. 

As the agent moves, it collects sensory data $s(x(t))$ at positions $x(t)$. From each sample, it constructs a likelihood~$\mathcal{L}\bigl(s(x);\rho_\theta(x)\bigr)$ that measures how well $\rho_\theta(x)$ explains the observed coupling between internal and external states. Under standard regularity (continuity of $\rho_\theta$ in both arguments, bounded likelihoods, and identifiability of $\theta_{\mathrm{true}}$), the posterior 
\[
p_t(\theta)\;\propto\;p_0(\theta)\;\prod_{\tau=0}^{t} \mathcal{L}\bigl(s(x(\tau));\,\rho_\theta(x(\tau))\bigr)
\]
concentrates on the true parameter~$\theta_{\mathrm{true}}$ satisfying $\rho_{\theta_{\mathrm{true}}}(x)=\rho_{\mathrm{true}}(x)$.

Simultaneously, the agent uses the current point estimate 
\[
\hat{\rho}_t(x)\;=\;\rho_{\hat{\theta}_t}(x), 
\quad\hat{\theta}_t=\arg\max_{\theta}p_t(\theta),
\]
to drive free‐energy descent:
\[
\dot x(t)\;=\;-\,\bigl[\,1-\hat{\rho}_t\bigl(x(t)\bigr)\bigr]\,\nabla F\bigl(x(t)\bigr).
\]
Thus, the agent interleaves Bayesian updating of $\hat{\theta}_t$ and state evolution under a scaled gradient flow.

\subsection{Consistency of Blanket‐Density Learning}

\begin{theorem}[Convergence of Joint Inference and Descent]
\label{thm:joint_inference_improved}
Assume:
\begin{enumerate}
  \item Each $\rho_\theta(x)$ is $C^1$ on $\Omega\times\Theta$, takes values in $[0,1]$, and the map $\theta\mapsto\rho_\theta(x)$ is injective for every $x\in\Omega$.
  \item The likelihood $\mathcal{L}\bigl(s;\rho_\theta(x)\bigr)$ is Lipschitz in both $s$ and $\rho_\theta(x)$, and identifies $\theta_{\mathrm{true}}$ such that $\rho_{\theta_{\mathrm{true}}}=\rho_{\mathrm{true}}$.
  \item The prior $p_0(\theta)>0$ in a neighborhood of $\theta_{\mathrm{true}}$.
  \item The descent flow under the true blanket density,
  \[
    \dot x = -\bigl[\,1 - \rho_{\mathrm{true}}(x)\bigr]\nabla F(x),
  \]
  is ergodic on~$\Omega$: it visits every open set infinitely often.
  \item $F\in C^2(\Omega)$ has $\|\nabla F(x)\|>0$ except at finitely many isolated minima.
\end{enumerate}
Then as $t\to\infty$, the agent’s point estimate $\hat{\rho}_t(x)$ converges uniformly to $\rho_{\mathrm{true}}(x)$ on every compact $K\subset\Omega$, almost surely:
\[
\sup_{x\in K}\bigl|\hat{\rho}_t(x)-\rho_{\mathrm{true}}(x)\bigr|
\;\xrightarrow{t\to\infty}\;0.
\]
\end{theorem}

\begin{proof}
Because the true descent flow visits every open neighborhood infinitely often, the agent samples sensory data at arbitrarily many positions throughout $\Omega$. Under assumptions (1)–(3), classical consistency for input‐driven parameter estimation guarantees that $p_t(\theta)$ concentrates on $\theta_{\mathrm{true}}$, so $\hat{\theta}_t\to\theta_{\mathrm{true}}$ and hence $\hat{\rho}_t(x)\to\rho_{\mathrm{true}}(x)$ uniformly on compacts.

We must ensure that using $\hat{\rho}_t$ in place of $\rho_{\mathrm{true}}$ does not break ergodicity. Since $\hat{\rho}_t\to\rho_{\mathrm{true}}$ uniformly, for large $t$ the perturbed vector field 
\[
-\,\bigl[\,1-\hat{\rho}_t(x)\bigr]\nabla F(x)
\]
differs from the true flow by at most a small uniform error. Because $F$ has no plateaus, this small perturbation leaves the trajectory qualitatively unchanged: it continues to visit each open neighborhood infinitely often. Hence the learning process remains ergodic, validating the consistency argument ad infinitum.
\end{proof}

Theorem~\ref{thm:joint_inference_improved} shows that an active inference agent, without prior knowledge of $\rho(x)$, can learn the blanket density while descending free energy. Ergodicity ensures the agent gathers enough data to identify $\theta_{\mathrm{true}}$ almost surely, and free energy minimization prevents collapse into a zero‐blanket state. Consequently,~$\rho(x)$ becomes unpredictable in advance: the agent must discover its own coupling constraints by moving through and sampling the environment. This extension embeds the blanket into the agent’s inference process, yielding a fully self‐consistent active inference model on an unknown stochastic landscape.

\section{MB Density: Axiomatization and Derivation of the FEP}

In this section we provide a rigorous, assumption-light derivation of free energy minimization as a \emph{necessary} dynamical consequence of the information geometry induced by the MB Density \(\rho(x)\).
Our strategy is: (i) postulate minimal axioms on \(\rho\) (regularity and informational meaning), plus symmetry/isotropy assumptions that rule out dynamically uninformative tangential drifts; (ii) deduce that any admissible dynamics must be colinear with \(\nabla\rho\) with magnitude depending only on the level \(\rho(x)\); (iii) show that this structure \emph{forces} the existence of a scalar potential \(F\) such that the dynamics is a gradient flow \(\dot x=-\nabla F\).
Classical choices (e.g.\ \(F=-\log(1-\rho+\varepsilon)\)) then appear as \emph{corollaries}—monotone reparameterizations of the emergent potential—rather than ad hoc postulates.
We also prove that the FEP has operational content precisely on the regime \(\rho<1\), while it is informationally vacuous on \(\rho=1\).

\medskip

We work on an open domain \(\Omega\subset\mathbb{R}^n\) with the Euclidean metric; \(x\in\Omega\) denotes the agent's state.

\subsection{Setup and standing assumptions}

For each \(x\in\Omega\), let \((I,B,E)\) be a measurable partition of the variables of a generative model, inducing a local data distribution \(p_x\).
Define
\begin{equation}
\mathrm{MI}(x):=I(I;E)_x, \qquad \mathrm{CMI}(x):=I(I;E\mid B)_x.
\end{equation}

\begin{axiom}[positivity of MI]\label{ax:MIpos}
There exists an open set \(U\subseteq\Omega\) such that \(\mathrm{MI}(x)>0\) for almost every \(x\in U\).
\end{axiom}

\begin{axiom}[regularity]\label{ax:reg}
The maps \(x\mapsto \mathrm{MI}(x)\) and \(x\mapsto \mathrm{CMI}(x)\) are \(C^1\) on \(U\).
\end{axiom}

\begin{definition}[MB Density]\label{def:rho}
For \(x\in U\) define
\begin{equation}
\rho(x):=1-\frac{\mathrm{CMI}(x)}{\mathrm{MI}(x)}\in(0,1].
\end{equation}
Extend by continuity with \(\rho=1\) where \(\mathrm{CMI}=0\). Assume \(\rho\in C^1(U)\).
\end{definition}

\noindent\emph{Interpretation.} \(\rho(x)=1\) iff the blanket perfectly screens \(I\) and \(E\) (\(I\perp E\mid B\)); \(\rho\downarrow 0\) corresponds to maximal conditional coupling.

\subsection{Operability of inference}

\begin{theorem}[Operability iff \(\rho<1\)]\label{thm:operability}
For \(x\in U\), \(\rho(x)=1 \Leftrightarrow I(I;E\mid B)_x=0\).
In this case, any variational free energy functional
\begin{equation}
\mathcal{F}(x)=\mathbb{E}_{q_x}\!\big[\log q_x(s)-\log p_x(o,s)\big]
\end{equation}
has zero first variation with respect to external perturbations transmitted through \(B\): the ``outer'' gradient vanishes and inference is informationally vacuous.
\end{theorem}

\begin{proof}
If \(\rho=1\) then \(\mathrm{CMI}=0\), hence \(p_x(i\mid e,b)=p_x(i\mid b)\) (conditional independence). Therefore changes in \(e\) cannot affect \(\mathcal{F}\) once \(b\) is fixed, and the outer gradient is zero. The converse is by Definition~\ref{def:rho}.
\end{proof}

\subsection{Symmetry, isotropy, and the form of admissible dynamics}

Let \(v:U\to\mathbb{R}^n\) be the dynamical field, \(\dot x=v(x)\).

\begin{axiom}[level-set symmetry]\label{ax:levels}
For every diffeomorphism \(\Phi\) preserving \(\rho\) (i.e.\ \(\rho\circ\Phi=\rho\)), the dynamics is equivariant: \(D\Phi(x)\,v(x)=v(\Phi(x))\).
\end{axiom}

\begin{axiom}[no tangential preference]\label{ax:notang}
On each regular level set \(\mathcal{L}_c:=\{x\in U:\rho(x)=c\}\), the tangential component of \(v\) vanishes for a.e.\ \(x\): \(v(x)\) is colinear with \(\nabla\rho(x)\) wherever \(\nabla\rho\neq 0\).
\end{axiom}

\begin{lemma}[colinearity]\label{lem:colinear}
Under Axioms~\ref{ax:levels} and \ref{ax:notang}, there exists a scalar function \(\alpha:U\to\mathbb{R}\) such that
\(
v(x)=-\,\alpha(x)\,\nabla\rho(x)
\)
for a.e.\ \(x\in U\).
\end{lemma}

\begin{proof}
By Axiom~\ref{ax:notang}, \(v(x)\) has no tangential component on \(\mathcal{L}_c\), hence it is normal to \(\mathcal{L}_c\), i.e.\ colinear with \(\nabla\rho\). The sign is chosen so that flow proceeds towards increasing information accessibility (monotone decrease of screening).
\end{proof}

\begin{axiom}[level invariance of magnitude]\label{ax:levelmag}
\(\alpha\) is constant along each level set of \(\rho\): there exists a continuous \(f:[0,1]\to[0,\infty)\) with \(\alpha(x)=f(\rho(x))\).
\end{axiom}

\begin{axiom}[stall at perfect screening]\label{ax:stall}
\(f(1)=0\) and \(f(\rho)>0\) for \(\rho<1\).
\end{axiom}

Combining Lemma~\ref{lem:colinear}, Axiom~\ref{ax:levelmag} and Axiom~\ref{ax:stall} we obtain the \emph{forced} form of the dynamics:
\begin{equation}\label{eq:rho-flow}
\dot x = v(x) = -\,f(\rho(x))\,\nabla\rho(x).
\end{equation}

\subsection{Integrability and the emergent Free Energy potential}

\begin{lemma}[integrability]\label{lem:integrability}
Define \(w(x):=\frac{1}{f(\rho(x))}\,\nabla\rho(x)\) on the regular set \(\{x:\nabla\rho(x)\neq 0\}\). Then \(w\) is irrotational; hence, there exists \(F\in C^2\) such that \(\nabla F=w\).
\end{lemma}

\begin{proof}
Since \(w=\psi(\rho)\,\nabla\rho\) with \(\psi(\rho):=1/f(\rho)\), choose any \(C^2\) primitive \(\Psi\) of \(\psi\), i.e.\ \(\Psi'=\psi\). Then \(w=\nabla(\Psi\circ\rho)\) is a gradient field.
\end{proof}

\begin{definition}[Emergent Free Energy]\label{def:F}
Fix a constant \(C\in\mathbb{R}\) and define
\begin{equation}\label{eq:F-def}
F(x):=\Psi(\rho(x))+C=\int^{\rho(x)}\frac{1}{f(u)}\,du + C.
\end{equation}
\end{definition}

\begin{theorem}[Emergence of FEP]\label{thm:emergence}
Under Axioms~\ref{ax:MIpos}--\ref{ax:stall}, the dynamics \eqref{eq:rho-flow} is exactly a gradient flow of the potential \(F\) in \eqref{eq:F-def}:
\begin{equation}
\dot x = -\,\nabla F(x).
\end{equation}
Moreover, \(F\) is a Lyapunov function:
\begin{equation}
\frac{d}{dt}F(x(t))=\nabla F\cdot \dot x = -\,\|\nabla F\|^2 \le 0,
\end{equation}
with equality iff \(\nabla\rho=0\) or \(\rho=1\).
\end{theorem}

\begin{proof}
By Lemma~\ref{lem:integrability}, \(\nabla F=\frac{1}{f(\rho)}\nabla\rho\). Multiplying both sides by \(-f(\rho)\) yields \(-f(\rho)\nabla F=-\nabla\rho\). Comparing with \eqref{eq:rho-flow} gives \(\dot x=-\nabla F\). The Lyapunov property follows from standard gradient-flow calculus; vanishing occurs only at critical points of \(F\) (i.e.\ \(\nabla\rho=0\)) or where \(f(1)=0\) (i.e.\ \(\rho=1\)).
\end{proof}

\begin{corollary}[class of representatives and uniqueness up to reparameterization]\label{cor:reparam}
If \(\tilde F=\phi\circ F\) with \(\phi\) strictly increasing, then \(\dot x=-\nabla \tilde F\) generates the same trajectories up to time-rescaling. Conversely, any gradient potential producing the same trajectories must be a strictly monotone reparameterization of \(F\).
In particular, the classical choice \(f(\rho)=1-\rho+\varepsilon\) (\(\varepsilon>0\)) yields
\begin{equation}
F(x)=-\log(1-\rho(x)+\varepsilon)+C,
\end{equation}
as a \emph{corollary} of Theorem~\ref{thm:emergence}.
\end{corollary}

\subsection{Link to standard variational free energy}

Let \(\mathcal{F}(x)\) denote a standard variational free energy (VFE) for the recognizer \(q_x\).

\begin{axiom}[informational alignment]\label{ax:align}
There exists a continuous \(\kappa:[0,1]\to[0,\infty)\) with \(\kappa(1)=0\) and \(\kappa(\rho)>0\) for \(\rho<1\), such that
\begin{equation}
\nabla \mathcal{F}(x)=\kappa(\rho(x))\,\nabla\rho(x).
\end{equation}
\end{axiom}

\begin{proposition}[directional equivalence]\label{prop:equiv}
Under Axiom~\ref{ax:align}, there exists a strictly increasing \(\phi\) such that the gradient flow of \(\mathcal{F}\) coincides with that of \(\phi\circ F\); hence, after time-rescaling, the VFE-descent is dynamically equivalent to the emergent FEP of Theorem~\ref{thm:emergence}.
\end{proposition}

\begin{proof}
From \(\nabla\mathcal{F}=\kappa(\rho)\nabla\rho\) and \(\nabla F=(1/f(\rho))\nabla\rho\), define \(\phi'\circ F:=\kappa(\rho)f(\rho)\). Since \(\rho\) is a strictly monotone function of \(F\) by \eqref{eq:F-def} and Axiom~\ref{ax:stall}, \(\phi\) is well-defined and strictly increasing. Then \(\nabla(\phi\circ F)=\phi'(F)\nabla F=\kappa(\rho)\nabla\rho=\nabla\mathcal{F}\). A constant time-rescaling aligns the flows.
\end{proof}

\subsection{Limits, regularity, completeness}

\begin{itemize}
\item \textbf{Critical points of \(\rho\).} Where \(\nabla\rho=0\), the flow stalls; \(F\) has stationary points (minima, maxima, or saddles).
\item \textbf{Frontier \(\rho=1\).} By Axiom~\ref{ax:stall}, \(f(1)=0\); we may normalize \(F|_{\rho=1}=0\). At \(\rho=1\) the FEP is informationally vacuous by Theorem~\ref{thm:operability}.
\item \textbf{Domain issues.} The set where \(\mathrm{MI}=0\) has measure zero in \(U\) by Axiom~\ref{ax:MIpos}; the extension \(\rho=1\) is continuous there.
\item \textbf{Uniqueness (strong).} Under Axioms~\ref{ax:levels}--\ref{ax:stall}, any admissible dynamics must take the form \eqref{eq:rho-flow}; hence a potential \(F\) exists \emph{necessarily} and the class of representatives is exactly \(\{\phi\circ F:\phi\uparrow\}\).
\end{itemize}

\paragraph{Conclusion.}
The MBD \(\rho\) together with symmetry and isotropy axioms \emph{forces} the dynamics to be a gradient flow of an emergent Free Energy potential defined by \eqref{eq:F-def}.
Thus, the FEP is not primitive but a \emph{derivative} structure prevailing on the informational regime \(\rho<1\) and becoming vacuous at \(\rho=1\).
\qed

\section{Limitations and the Risk of Circularity}
\noindent 
\textnormal
In Theorem 1, it is assumed that the mutual information (both marginal and conditional) is \(C^{1}\) and that their gradients align with \(\nabla F\) over an open set \(D\). We recognize that this requirement of “gradient alignment” is extremely strong and likely does not hold in many real-world applications (biological or engineering), where \(\nabla F\) and \(\nabla I\) may point in very different directions.

In Theorems 3 and 4, the assumption of “constant covariance”
\[
\mathrm{Cov}\bigl(\rho(x),\,\lVert\nabla F(x)\rVert^2\bigr) \;=\; C
\]
is an artificial simplification, which is difficult to justify in practical situations where both \(\rho(x)\) and \(\nabla F(x)\) can vary spatially in complex ways.

The kNN–KSG estimator, on which the estimation of \(\rho_{N}\) relies, requires high-dimensional datasets and suffers from the curse of dimensionality. If the data \(s_{i}\) have dimension \(d \gg 1\), obtaining a sufficiently accurate mutual information estimate to guarantee convergence in the \(C^{1}\) norm becomes practically infeasible.

All of these regularity and stationarity assumptions limit the practical applicability of these theorems: if one truly wants to use them to explain neural or behavioral phenomena, it is necessary to demonstrate that the basic assumptions (alignment, constant covariance, exponential decay of correlations) are at least approximately satisfied on real data.

Another limitation concerns the possible risk of circularity of the overall argument. Saying “the agent moves to regions of low \textit{p}” can be read as “the agent moves where it is already well coupled,” which is arguably just restating “the agent moves to reduce free energy” in spatial terms. The apparent circularity dissolves once one recognizes that here $\rho(x)$ is defined \emph{a priori} as an external field of conditional‐information estimates—derived from raw sensory‐environment samples—rather than as a byproduct of an agent’s free‐energy descent. In other words, one first samples the joint statistics of internal, external, and blanket variables to build $\rho(x)$ independently of any inference process; only then does the agent navigate according to $\nabla F$ and the precomputed $\rho$. Because $\rho(x)$ is not recomputed from the agent’s current beliefs but estimated from external data, minimizing free energy does not “chase its own tail” but rather follows a fixed landscape of informational permeability, rendering any notion of tautological self‐reference illusory.

\section{Conclusions}

The core idea of this paper is to reconceptualize the FEP not merely as an internal rule for belief updating but as genuine spatial navigation through a continuously varying MB density field. Instead of treating the informational boundary between internal and external states as a binary condition, we introduce a function \(\rho(x)\), defined at every point \(x\) in a continuous domain, which quantifies how “insulated” that location is. Values of \(\rho(x)\) near zero indicate that internal and external states are strongly coupled (minimal insulation), whereas values close to one indicate that a location is almost entirely isolated (maximal insulation).

To make this precise, the paper ties \(\rho(x)\) to an information‐theoretic measure: it is the ratio between conditional mutual information \(\mathrm{I}(I;E \mid B)\), which measures how much information about external states \(E\) remains once boundary states \(B\) are known, and unconditional mutual information \(\mathrm{I}(I;E)\), which captures overall coupling. Because conditioning cannot increase mutual information, that ratio always lies between zero and one. Consequently, when \(\rho(x)\) is near zero, most of the mutual information between internal and external states bypasses the boundary, and when \(\rho(x)\) is near one, conditioning on the boundary removes almost all of the coupling. In other words, \(\rho(x)\) serves as a continuous gauge of how effectively the environment can inform the agent at location \(x\).

The formulation of the FEP presented here, grounded in the concept of a spatially distributed MB density $\rho(x)$, offers several theoretical advances relative to traditional approaches. In contrast to standard formulations that assume the FEP as a universal law governing all systems equipped with a Markov blanket, our approach introduces a continuous informational field that modulates the very applicability of the principle. Specifically, the FEP becomes locally definable only in regions where $\rho(x) < 1$, that is, where conditional dependence between internal and external states—mediated by the blanket—is sufficiently high to permit inferential coupling. This perspective yields four distinct advantages. First, it introduces a \emph{topological criterion of inferential viability}, replacing the binary presence or absence of a blanket with a scalar field that continuously measures the strength of informational separation. This makes it possible to study inference not as a global property, but as a \emph{situated phenomenon}, dependent on the agent's position within an informational manifold. Second, it reorients the explanatory architecture of the FEP: rather than positing inference as an intrinsic drive of the system, it treats inference as an \emph{affordance} of the environment—possible only where the topology of $\rho(x)$ permits it. Third, this framework offers a \emph{geometrization of inference}. By defining informational gradients, curvature, and metric structure on the field $\rho(x)$, we provide tools to analyze cognitive processes as forms of motion through an informational landscape. This opens new paths for modeling agent-environment interaction, including multi-agent systems, distributed cognition, and ecological learning. Finally, the concept of MB density allows us to extend the reach of the FEP to systems where the blanket structure is ambiguous, dynamic, or absent—such as in decentralized networks, social systems, or collective organisms. In such contexts, $\rho(x)$ can serve as a measure of inferential accessibility even in the absence of well-defined agent boundaries.

\clearpage
\begin{figure}[H]
    \centering
    \includegraphics[width=\textwidth]{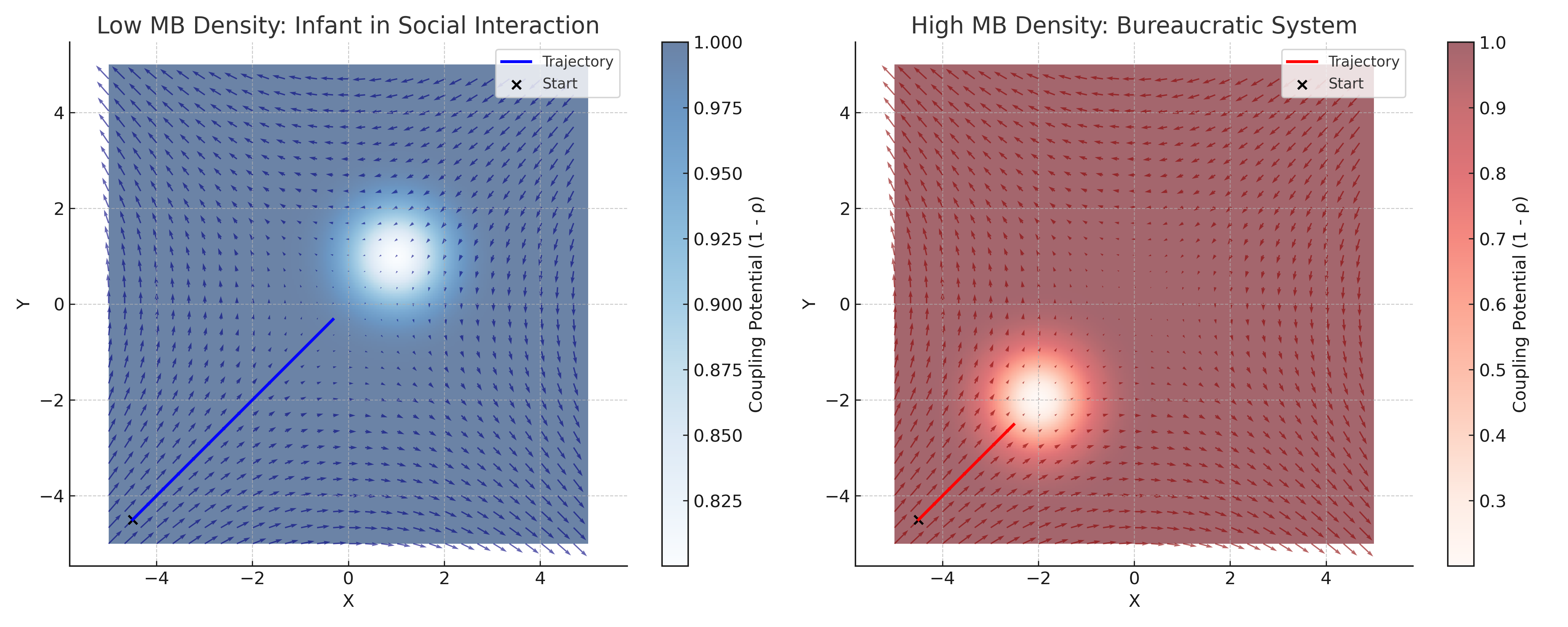}
    \caption{
    \textbf{Agent trajectories shaped by Markov blanket density.}
    This figure compares two systems governed by the same free energy minimization equation $\dot{x} = - (1 - \rho(x)) \nabla F(x)$, where $\rho(x)$ is the spatially distributed Markov blanket density. 
    \textbf{Left:} In the case of an infant engaged in social interaction, the MB density is low, allowing for strong coupling with the environment. The agent follows the free energy gradient efficiently, resulting in a smooth and directed trajectory. 
    \textbf{Right:} In the bureaucratic system, high blanket density inhibits coupling. Despite non-zero free energy gradients, the trajectory is shallow and constrained, demonstrating how strong informational boundaries block adaptive inference. 
    The color maps represent the local coupling potential $(1 - \rho)$, highlighting the spatial modulation of active inference.
    Parameters: the Figure describes the trajectories of an agent on the free‐energy landscape \(F(x,y)=x^{2}+y^{2}\) under two different blanket densities. The agent starts at \((0.8,0.8)\) in the square \([-1,1]\times[-1,1]\) and evolves for 100 explicit‐Euler steps with time step \(\Delta t=0.02\). Its velocity at each step is given by \(\dot x=-(1-\rho)\,\nabla F(x)\), with \(\rho=0.2\) (blue curve, “Infant”) or \(\rho=0.8\) (red curve, “Bureaucracy”), plotted in 2D with equal aspect ratio to illustrate how lower blanket density permits faster descent toward the origin.
    }
    \label{fig:final_mb_density_trajectories}
\end{figure}

\begin{figure}[H]
    \centering
    \includegraphics[width=0.85\textwidth]{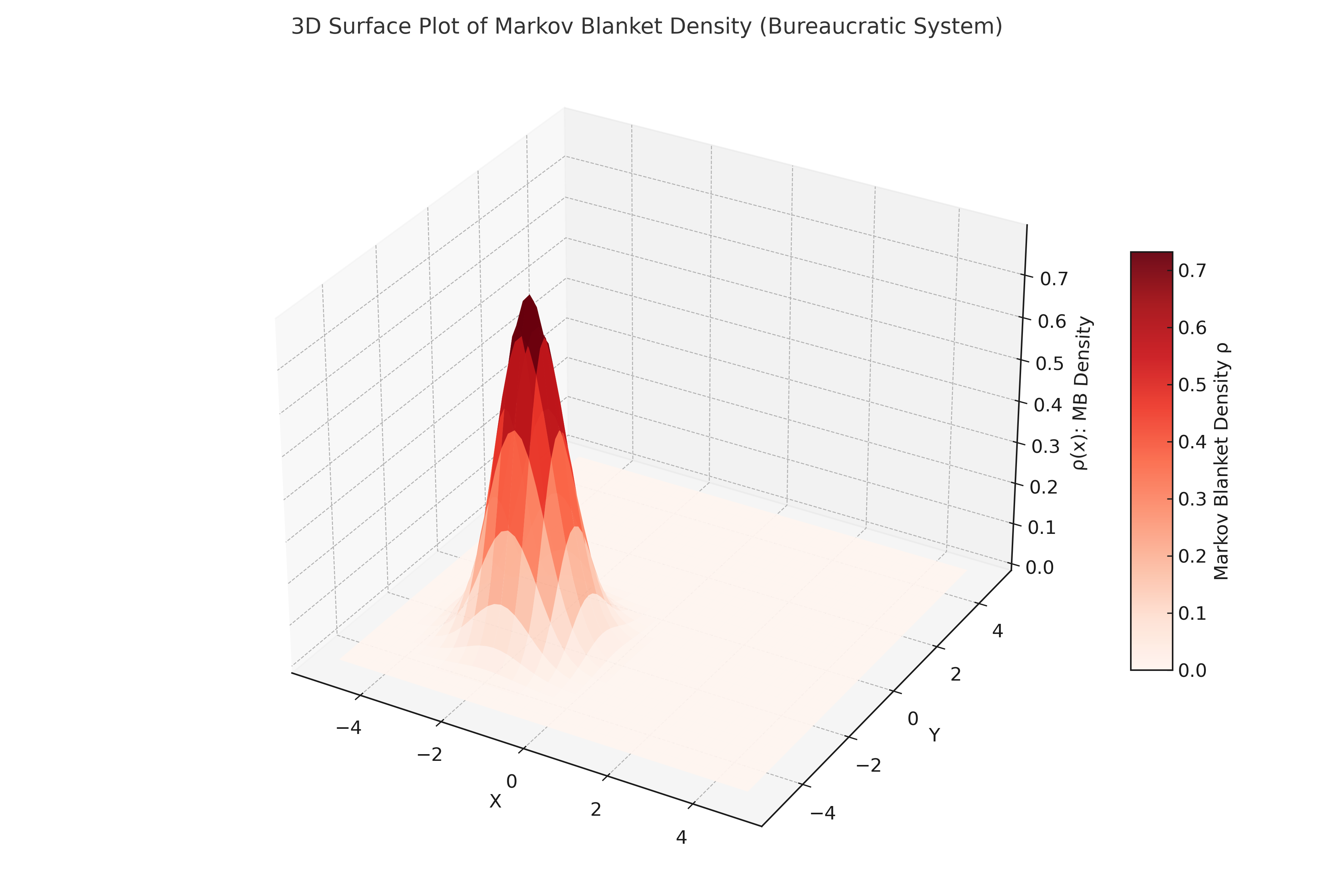}
    \caption{%
    \textbf{MB density as an informational topology.}
    This 3D surface plot visualizes the spatial distribution of MB density $\rho(x)$ in a high-density regime (e.g., a bureaucratic system). Regions of high $\rho(x)$ indicate strong informational boundaries—zones of limited coupling between internal and external states. Such topologies constrain active inference by inhibiting access to meaningful sensory feedback. This figure illustrates how the geometry of $\rho(x)$ can serve as an inferential landscape that shapes the success or failure of free energy minimization.
    }
    \label{fig:mb_density_surface}
\end{figure}

\begin{figure}[H]
    \centering
    \includegraphics[width=0.85\textwidth]{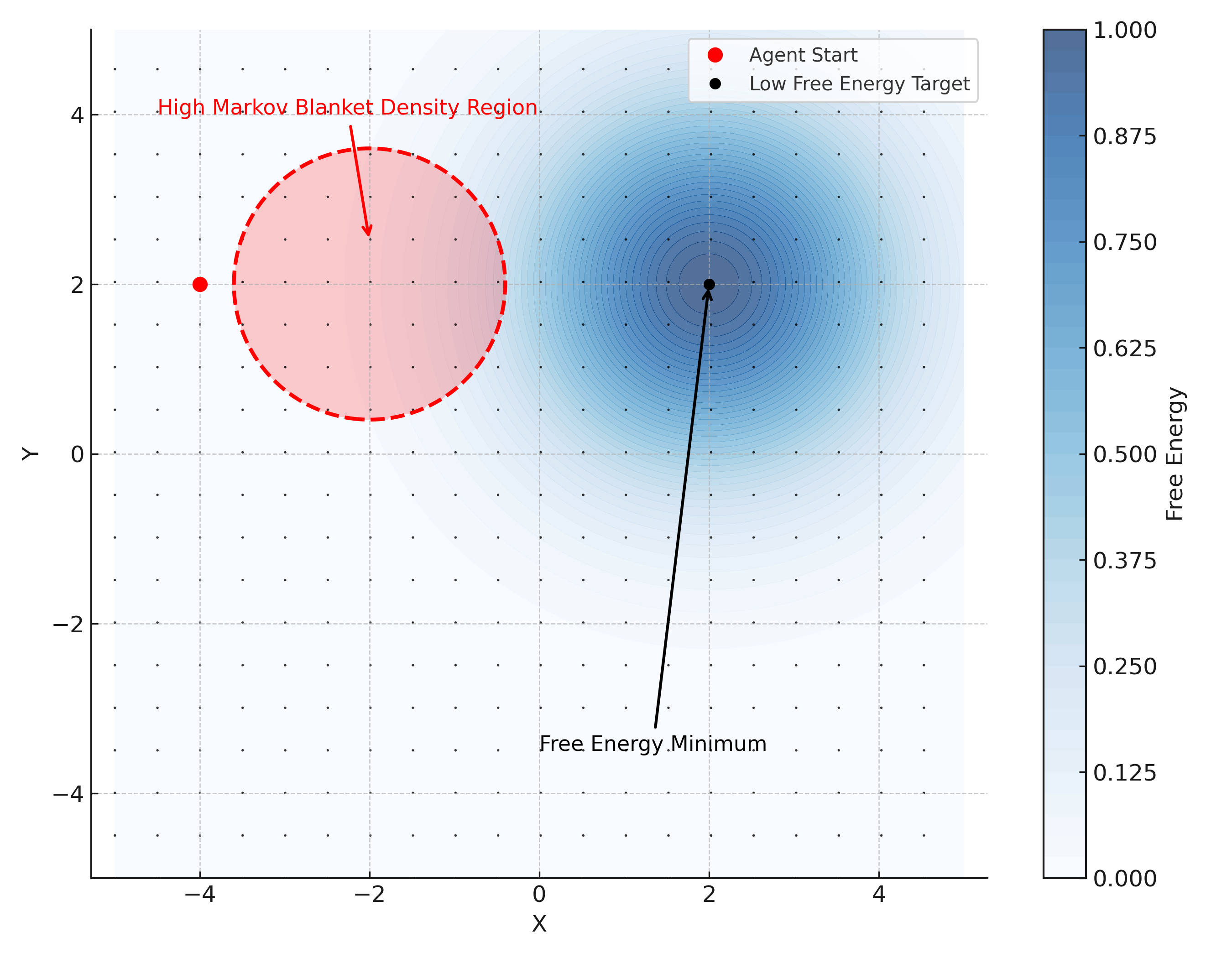}
    \caption{
    \textbf{When variational free energy minimization is obstructed by informational structure.}
    This figure illustrates a theoretical conflict central to the paper. An agent (red dot) begins in a region of high MB density (dashed red contour, shaded red area). Although the agent is embedded in a free energy landscape (blue gradient), and a global minimum of variational free energy is present (black dot), the high local value of $\rho(x)$ inhibits coupling between the agent’s internal states and external causes. As a result, the agent cannot exploit the free energy gradient: adaptive inference is blocked not by the absence of a minimization path, but by the statistical opacity of the surrounding space. The figure demonstrates that the ability to minimize free energy is contingent upon local informational accessibility. Parameters: Contour plot of the free‐energy landscape \(F(x,y)=x^{2}+y^{2}\) obstructed by a high‐density barrier in \(\rho(x,y)\). On the same \(100\times100\) grid over \([-1,1]^{2}\), \(F\) is contoured at 20 levels using the “Blues” palette. A circular region centered at \((0.5,0.5)\) with radius 0.2 is assigned \(\rho=0.95\), while the remainder of the grid has \(\rho=0.05\); the \(\rho=0.5\) boundary is overlaid as a red dashed contour. The starting point \((0.2,0.2)\) is marked with a red dot and the global minimum location \((0.5,0.5)\) with a black dot.
    }
    \label{fig:clash_fep_vs_mb_final}
\end{figure}

\begin{figure}[ht]
    \centering
    \includegraphics[width=\textwidth]{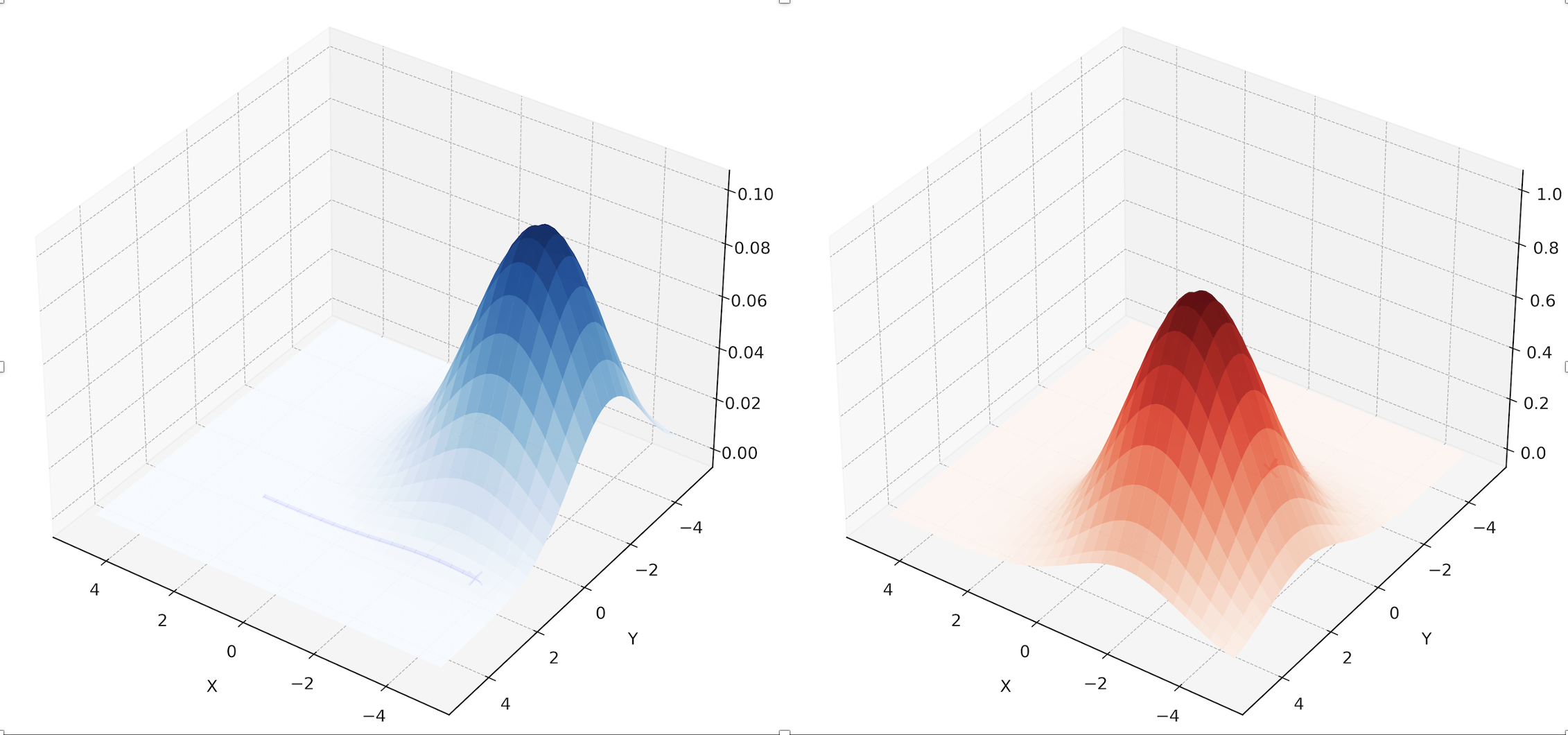}
    \caption{\textbf{Effect of Markov blanket density on agent movement across informational landscapes.}
    This two-panel 3D visualization illustrates how the spatial distribution of MB density $\rho(x)$ modulates an agent's ability to perform gradient descent on free energy. In both panels, an agent attempts to follow the same epistemic imperative—minimization of variational free energy—by moving through a landscape shaped by $\rho(x)$. 
    (A) In a region of \emph{weak} MB density (low $\rho(x)$), the informational coupling between agent and environment is strong. The agent can descend the surface efficiently, adapting its trajectory to the available gradient field. 
    (B) In contrast, in a region of \emph{strong} MB density (high $\rho(x)$), the agent is epistemically insulated. Coupling is weak and movement is suppressed: although gradients still exist, the agent cannot access or respond to them effectively. 
    These simulations demonstrate that the capacity to minimize free energy is shaped not only by internal dynamics but also by the external topology of informational boundaries. Parameters: Side‐by‐side 3D depictions of free‐energy surfaces modulated by low versus high blanket‐density fields, with corresponding agent trajectories. In each panel, \(F(x,y)=x^{2}+y^{2}\) is plotted over a \(100\times100\) grid on \([-1,1]^{2}\) using an alpha of 0.7 and stride 4. In panel A, \(\rho(x,y)=0.2+0.3\tfrac{x+1}{2}\) (range [0.2,0.5]) and in panel B, \(\rho(x,y)=0.8+0.2\tfrac{x+1}{2}\) (range [0.8,1.0]). From the initial point \((0.8,-0.8)\), each trajectory is simulated for 80 explicit‐Euler steps with \(\Delta t=0.02\) using \(\dot x=-(1-\bar\rho)\,\nabla F(x)\), where \(\bar\rho\) is the mean density over the panel; trajectories are drawn as red lines with markers against the translucent energy surface.
    }
    \label{fig:mb_density_panels}
\end{figure}

\begin{figure}[ht]
\centering
\includegraphics[width=\textwidth]{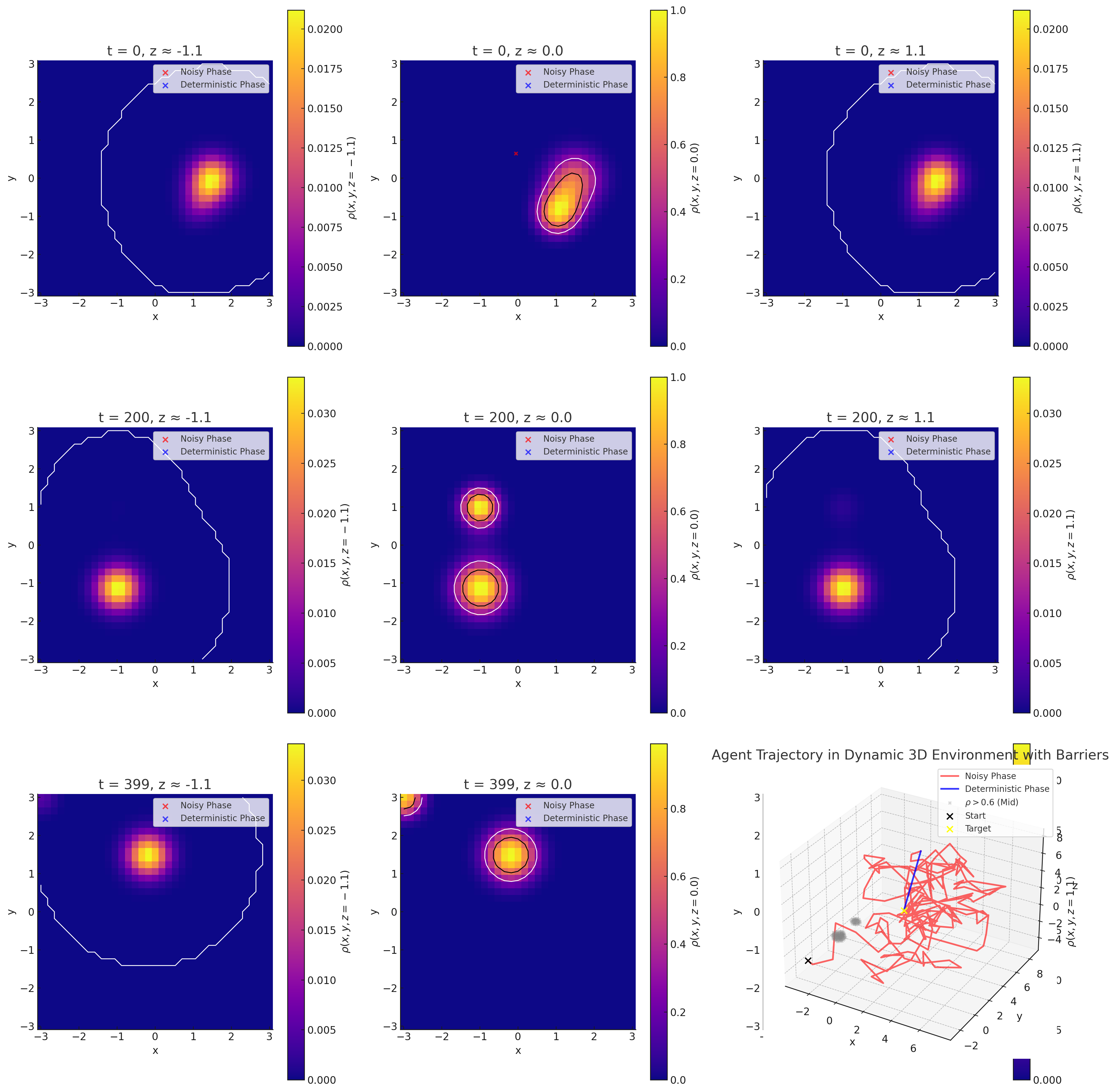}
\caption{%
\textbf{Algorithm Agent Navigation in a Dynamic 3D Blanket‐Density Field (Section 5.4).}
The agent (red during the noisy phase, blue during the deterministic phase) navigates a 3D barrier field \(\rho(x,y,z,t)\) that evolves over time due to two moving Gaussian coupling regions. In the first half of the simulation, strong noise allows the agent to penetrate thick barriers; in the second half, without noise, the agent smoothly steers around high-\(\rho\) zones and converges on its target at \((2,2,2)\). For more details, see Appendix A.}
\end{figure}

\begin{figure}[ht]
  \centering
  \includegraphics[width=\textwidth]{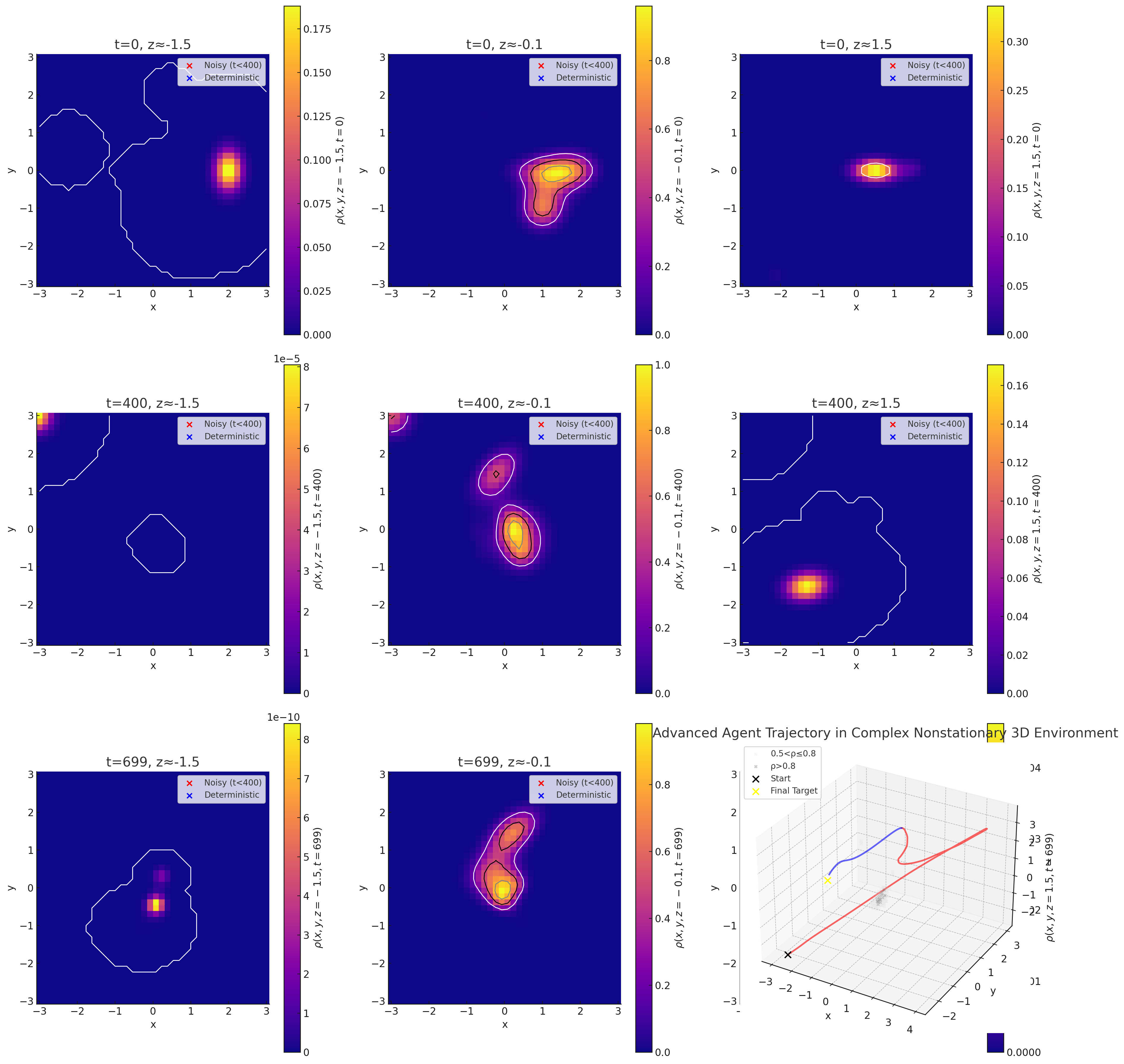}
  \caption {\textbf{Advanced Agent Navigation in Nonstationary 3D Environment}. Snapshots of the three‐dimensional blanket density \(\rho(x,y,z,t)\) (displayed with a plasma heatmap) at three horizontal slices \(z \approx -1.5,\,0.0,\,1.5\) for times \(t=0,\,400,\) and \(699\). At each slice, the white, black, and gray contour lines correspond to \(\rho=0.2,\ \rho=0.5,\) and \(\rho=0.8\), indicating regions of low, medium, and high barrier strength. Red dots (sized proportionally to instantaneous speed) mark the agent’s visits during the noisy phase (\(t<400\)), often penetrating even the darkest, high‐barrier contours, while blue dots show the agent’s path during the deterministic phase (\(t\ge 400\)), hugging just outside the strongest barriers despite occasional perception noise. In the lower right, the full 3D trajectory is plotted: the red segment (\(t=0\ldots 400\)) wanders through overlapping, rotating ellipsoidal obstacles due to colored movement noise, whereas the blue segment (\(t=400\ldots 699\)) smoothly navigates around the gray isosurface clouds at \(t=400\) (where \(\rho>0.8\)). Black and yellow markers denote the agent’s start at \((-2.5,-2.5,-2.5)\) and the final position of the moving helix target. This composite visualization demonstrates how an inertial agent with AR(1) movement and perception noise first plows through dynamically changing, anisotropic obstacles and then transitions to informed, barrier‐avoiding navigation toward a moving goal. For more details, see Appendix A.}
\end{figure}

\begin{figure}[ht]
  \centering
  \includegraphics[width=\textwidth]{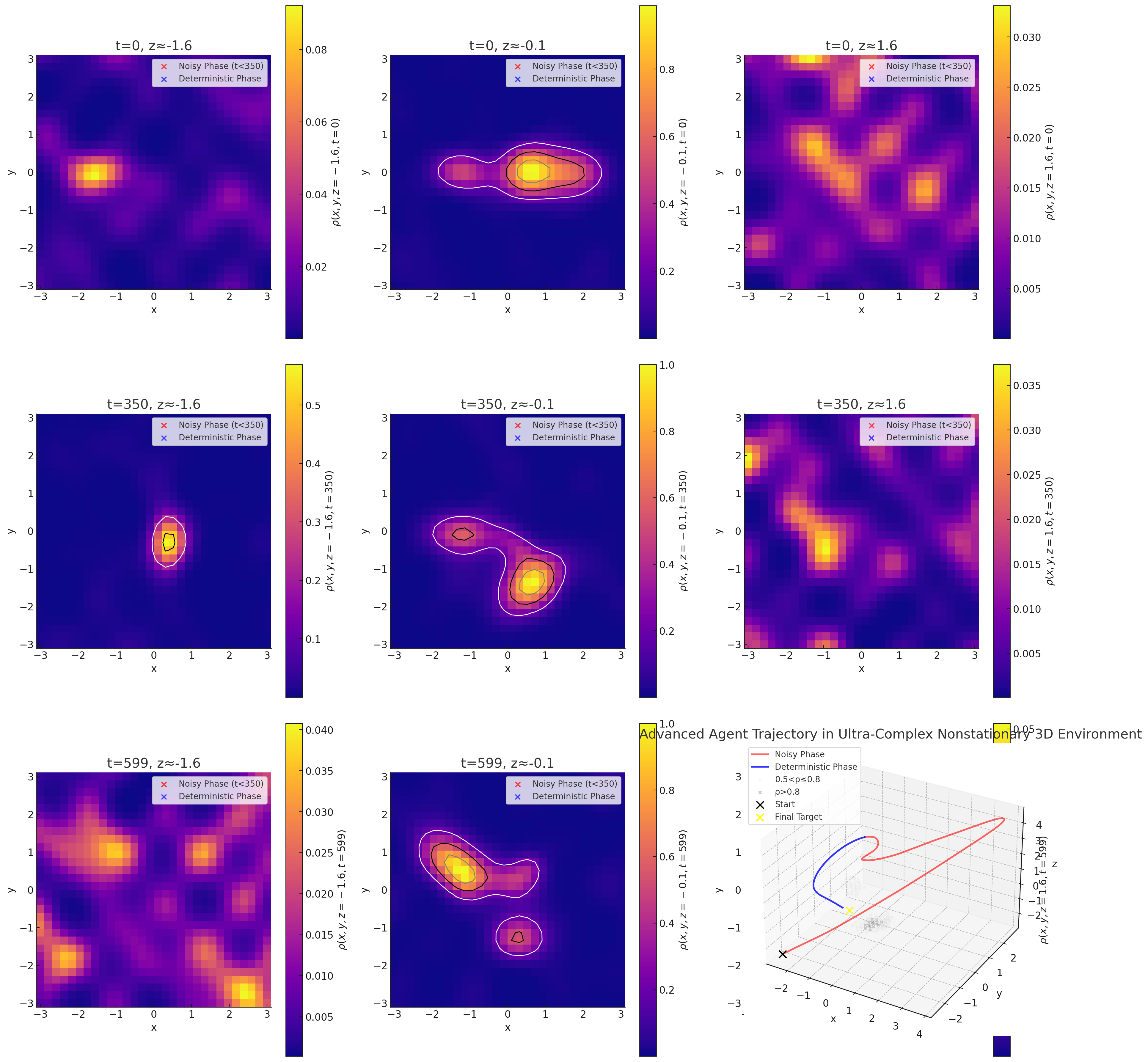}
  \caption{\textbf{Inertial Agent Navigation in Ultra‐Complex 3D Barriers}. Each panel in this 3×3 grid shows a horizontal slice of the ultra‐complex, nonstationary barrier field \(\rho(x,y,z,t)\) (plotted with a plasma colormap) at \(z \approx -1.5,\,0.0,\) and \(+1.5\) for times \(t=0,\,350,\) and \(599\). In each slice, white contours mark \(\rho=0.2\) (low barriers), black contours mark \(\rho=0.5\) (medium barriers), and gray contours mark \(\rho=0.8\) (strong barriers). Overlaid red dots—sized in proportion to instantaneous speed—represent the agent’s visits during the noisy phase (\(t<350\)), often penetrating even the highest‐barrier (gray) regions because colored movement noise overwhelms the barrier. Blue dots correspond to the deterministic phase (\(t\ge350\)), where the inertial agent, subject to AR(1) perception noise, hugs just outside the strongest barrier contours and weaves through lower‐\(\rho\) corridors. In the lower right panel, the complete 3D trajectory is shown: the red segment (\(t=0\ldots 350\)) meanders through overlapping, rotating ellipsoidal Gaussians, AR(1)‐drifting micro‐obstacles, and a time‐varying random Fourier field. When movement noise ceases at \(t=350\), the blue segment (\(t=350\ldots 599\)) smoothly navigates around the layered isosurfaces at \(t=350\), where \(0.5<\rho\le0.8\) (light gray) and \(\rho>0.8\) (dark gray). Black and yellow markers denote the agent’s starting location \((-2.5,-2.5,-2.5)\) and the final position of the moving helix target, respectively. For more details, see Appendix A.}
\end{figure}

\clearpage
\appendix
\section*{Appendix A: Detailed Objectives and Methodology of Figures 6-7-8}

\subsection*{Figure 6}

The goal of Figure 6 is to construct and visualize a continuous three‐dimensional blanket‐density field $\rho(x,y,z,t)$, to demonstrate how high‐noise perturbations enable an agent to traverse thick Markov blankets, and then to show how, once the noise is removed, the agent’s dynamics 
\[
   \dot{\mathbf{x}} \;=\; -\bigl(1 - \rho(\mathbf{x},t)\bigr)\,\nabla F(\mathbf{x}), 
   \quad F(x)=\|\,x-(2,2,2)\|^{2},
\]
cause it to avoid high‐$\rho$ zones and find a low‐resistance path toward a fixed target at $(2,2,2)$. 

At each time step $t=0,1,\dots,399$ on a uniform $35\times35\times35$ grid over $[-3,3]^3$, we define two moving Gaussian coupling regions: 
\[
   c_{1}(t)=\bigl(1.5\cos(0.02\,t),\;1.5\sin(0.02\,t),\;0\bigr), 
   \quad \sigma_{1}=0.7,
\]
\[
   c_{2}(t)=\bigl(1 - 0.01\,t,\;-1 + 0.01\,t,\;0.5\sin(0.015\,t)\bigr), 
   \quad \sigma_{2}=0.5.
\]
For each grid point $x=(x,y,z)$, the total coupling $a(x,t)$ is the sum of the two Gaussian values 
\[
   a(x,t) \;=\; \exp\!\Bigl(-\tfrac{\|\,x - c_{1}(t)\|^{2}}{2\,\sigma_{1}^{2}}\Bigr) \;+\; 
   \exp\!\Bigl(-\tfrac{\|\,x - c_{2}(t)\|^{2}}{2\,\sigma_{2}^{2}}\Bigr).
\]
We then form 
\[
   \rho_{\mathrm{corr}}(x,t) \;=\; \frac{\,a(x,t)^{2}\,}{\,a(x,t)^{2} + 0.9^{2}\,},
   \quad
   I(x,t) \;=\; -\tfrac12\,\ln\bigl[\,1 - \rho_{\mathrm{corr}}(x,t)^{2}\bigr],
\]
and normalize $I(x,t)$ over the entire grid to obtain $\rho(x,t)\in[0,1]$. 

The agent’s trajectory is computed in two phases:
\begin{itemize}
  \item \emph{Noisy Phase} ($0 \le t < 200$): At each step, 
  \[
    \mathbf{x}(t+\Delta t) \;=\; \mathbf{x}(t)\;+\;\Delta t\,\mathbf{v}(t)\;+\;\boldsymbol{\eta}(t), 
    \quad \Delta t=0.03,
  \]
  where 
  \[
    \mathbf{v}(t) = -\bigl(1 - \rho(\mathbf{x}(t),t)\bigr)\,\nabla F(\mathbf{x}(t)), 
    \quad 
    \boldsymbol{\eta}(t)\sim\mathcal{N}(0,\,0.9^{2}I_{3}).
  \]
  \item \emph{Deterministic Phase} ($200 \le t < 400$): Noise is removed, and 
  \[
    \mathbf{x}(t+\Delta t) \;=\; \mathbf{x}(t)\;+\;\Delta t\,
      \Bigl[-\bigl(1 - \rho(\mathbf{x}(t),t)\bigr)\,\nabla F(\mathbf{x}(t))\Bigr].
  \]
\end{itemize}

Figure 6 is displayed as a $3\times3$ grid of two‐dimensional heatmaps. Columns correspond to $t=0,\;200,\;399$, and rows correspond to slices at $z\approx -1.0,\;0.0,\;1.0$. Each heatmap uses a plasma colormap to show $\rho(x,y,z,t)$, with white and black contours at $\rho=0.3$ and $\rho=0.6$ indicating moderate and strong barrier levels. Overlaid red dots show the agent’s positions during the noisy phase (for each slice, points where $|z_{\mathrm{agent}}-z_{\mathrm{slice}}|<0.1$ over $t\pm 10$), while blue dots show positions during the deterministic phase. In the lower‐right panel, the full 400‐step trajectory is shown in three dimensions: a red segment for $0\le t\le200$ (noisy phase), which penetrates high‐$\rho$ regions, and a blue segment for $200\le t\le399$ (deterministic phase), which winds around gray points (grid cells at $t=200$ satisfying $\rho>0.6$). A black dot at $(-2.5,-2.5,-2.5)$ marks the agent’s start, and a yellow dot at $(2,2,2)$ marks the target.

\subsection*{Figure 7}

Figure 7 demonstrates an advanced agent navigating an even more nonstationary 3D environment. The environment is a $40\times40\times40$ grid over $[-3,3]^3$ containing:
\begin{enumerate}
  \item Four rotating, anisotropic ellipsoidal Gaussians with parameters 
  \[
    \sigma_{\text{blobs}} = \{(0.7,0.4,0.3),\;(0.5,0.5,0.6),\;(0.6,0.3,0.7),\;(0.4,0.6,0.5)\}.
  \]
  \begin{itemize}
    \item Blob 1: Center $\bigl(1.5\cos(0.02t),\,1.5\sin(0.02t),\,0\bigr)$, 
    covariance $\mathrm{diag}(0.7^2,\,0.4^2,\,0.3^2)$ rotated about $z$ by $0.01t$.
    \item Blob 2: Center $\bigl(1 - 0.01t,\;-1 + 0.01t,\;0.5\sin(0.015t)\bigr)$, covariance $\mathrm{diag}(0.5^2,\,0.5^2,\,0.6^2)$ rotated about $y$ by $0.015t$.
    \item Blob 3: Center $\bigl(0.5\cos(0.03t),\,0.5\sin(0.03t),\,\cos(0.02t)\bigr)$, covariance $\mathrm{diag}(0.6^2,\,0.3^2,\,0.7^2)$ rotated about $x$ by $0.012t$.
    \item Blob 4: Center $\bigl(2\cos(0.01t),\,2\sin(0.01t),\,-1+0.005t\bigr)$, covariance $\mathrm{diag}(0.4^2,\,0.6^2,\,0.5^2)$ rotated about $z$ by $0.02t$.
  \end{itemize}

  \item Ten micro‐obstacles of width $\sigma_{\mathrm{obs}}=0.3$ whose centers $\{m_i(t)\}$ follow an AR(1) process with $\phi_{\mathrm{micro}}=0.9$. Each micro‐obstacle contributes 
  \[
    0.3\exp\!\Bigl(-\frac{\|\,x - m_i(t)\|^{2}}{2\,(0.3)^{2}}\Bigr)
  \]
  to the total coupling field $a(x,t)$.

  \item A spatio‐temporal random Fourier field built as the sum of fifteen sinusoidal waves
  \[
    \sum_{k=1}^{15} \sin\bigl(\mathbf{w}_k\cdot(x,y,z) + \omega_{k}\,t + \phi_k\bigr),
  \]
  with random wavevectors $\mathbf{w}_k \sim \mathcal{N}(0,1.5^2\,I_3)$, frequencies $\omega_k \sim \mathrm{Uniform}(0.005,0.02)$, and random phases $\phi_k\sim[0,2\pi]$. This field is normalized to $[0,1]$ and scaled by $0.5$ before adding to $a(x,t)$.
\end{enumerate}

Thus, at each time $0\le t <700$, 
\[
  a(x,t) = \sum_{j=1}^4 \exp\!\bigl[-\tfrac{1}{2}(x - c_{j}(t))^\top \Sigma_{j}(t)^{-1} (x - c_{j}(t))\bigr]
  \;+\;0.3\sum_{i=1}^{10}\exp\!\Bigl[-\tfrac{\|x - m_i(t)\|^{2}}{2\,(0.3)^{2}}\Bigr]
  \;+\;0.5\,f_{\mathrm{RF}}(x,t).
\]
We compute 
\[
  \rho_{\mathrm{corr}}(x,t) \;=\; \frac{\,a(x,t)^{2}\,}{\,a(x,t)^{2} + 1.0^{2}\,},
  \quad
  I(x,t) = -\tfrac12\ln\bigl[\,1 - \rho_{\mathrm{corr}}(x,t)^{2}\bigr],
\]
and normalize $I(x,t)$ across all grid points to obtain $\rho(x,t)\in[0,1]$.

The agent pursues a moving helix target 
\[
  \mathbf{g}(t) = \bigl(2\cos(0.005\,t),\,2\sin(0.005\,t),\,2 - 0.002\,t\bigr)
\]
using second‐order dynamics (mass $m=1.2$, damping $\gamma=0.8$, time step $\Delta t=0.02$). Its perceived barrier strength is the average of $\rho$ over the local neighborhood of radius $1.0$ on the $40^{3}$ grid, corrupted by AR(1) perception noise ($\phi_{\mathrm{perc}}=0.6$, $\sigma_{\mathrm{perc}}=0.15$). During $0\le t<400$ (the noisy phase), movement noise follows AR(1) with $\phi_{\mathrm{move}}=0.7$ and $\sigma_{\mathrm{move}}=0.7$; for $400\le t<700$ (deterministic phase), movement noise is removed but perception noise remains.  

\subsection*{Figure 8}

This figure extends complexity by using a $30\times30\times30$ grid over $[-3,3]^3$ and three rotating, anisotropic ellipsoidal blobs:
\[
   \sigma_{\text{blobs}} = \{(0.8,0.5,0.3),\;(0.6,0.4,0.7),\;(0.5,0.6,0.4)\}.
\]
\begin{itemize}
  \item Blob 1: Center $\bigl(1.5\cos(0.015\,t),\,1.5\sin(0.015\,t),\,0.5\sin(0.01\,t)\bigr)$, covariance $\mathrm{diag}(0.8^2,\,0.5^2,\,0.3^2)$ rotated about $z$ by $0.02\,t$.
  \item Blob 2: Center $\bigl(-1.2\cos(0.018\,t),\,1.2\sin(0.018\,t),\,-0.5\cos(0.012\,t)\bigr)$, covariance $\mathrm{diag}(0.6^2,\,0.4^2,\,0.7^2)$ rotated about $x$ by $0.017\,t$.
  \item Blob 3: Center $\bigl(0.5\cos(0.02\,t),\,-0.5\sin(0.02\,t),\,1.5\sin(0.015\,t)\bigr)$, covariance $\mathrm{diag}(0.5^2,\,0.6^2,\,0.4^2)$ rotated about $y$ by $0.013\,t$.
\end{itemize}

Eight micro‐obstacles of width $\sigma_{\mathrm{obs}}=0.3$ drift via an AR(1) process with $\phi_{\mathrm{micro}}=0.85$. A spatio‐temporal random Fourier field (sum of 15 sinusoids with random wavevectors $\mathbf{k}\sim\mathcal{N}(0,1.5^2\,I_3)$, frequencies in $[0.005,0.02]$, and random phases) is normalized to $[0,1]$ and scaled by $0.5$. At each $0\le t<600$, the total coupling $a(x,t)$ is the sum of the three anisotropic Gaussians, eight micro‐Gaussians (scaled by 0.25), and $0.5$×the random Fourier field. We then compute 
\[
  \rho_{\mathrm{corr}}(x,t) \;=\; \frac{\,a(x,t)^{2}\,}{\,a(x,t)^{2} + 1.0^{2}\,},
  \quad
  I(x,t) \;=\; -\tfrac12\,\ln\bigl(1 - \rho_{\mathrm{corr}}(x,t)^{2}\bigr),
\]
normalize $I(x,t)$ over the $30^3$ grid to obtain $\rho(x,t)\in[0,1]$.

The agent uses second‐order dynamics (mass $m=1.0$, damping $\gamma=0.6$, time step $\Delta t=0.02$) to chase a moving helix target 
\[
  \mathbf{g}(t) = \bigl(2\cos(0.008\,t),\,2\sin(0.008\,t),\,2 - 0.0015\,t\bigr).
\]
Its perceived barrier strength is the average of $\rho$ over a spherical neighborhood of radius $1.0$ (via nearest‐neighbor averaging on the $30^3$ grid) plus AR(1) perception noise ($\phi_{\mathrm{perc}}=0.65$, $\sigma_{\mathrm{perc}}=0.12$). During $0\le t<350$ (noisy phase), movement noise is AR(1) with $\phi_{\mathrm{move}}=0.75$, $\sigma_{\mathrm{move}}=0.7$; for $350\le t<600$ (deterministic phase), movement noise is removed.

\section*{Appendix B: Estimating MB Density via KSG Mutual Information}
\addcontentsline{toc}{section}{Appendix: Estimating Markov Blanket Density via KSG Mutual Information}

This appendix explains how to estimate MB density from data using a practical method based on nearest-neighbor statistics. The approach, based on the KSG estimator, lets us compute mutual information directly from samples, without needing to guess the shape of the underlying distributions \cite{23, 24, 25, 32, 33, 34, 35, 36}.

\subsection*{Motivation and Theoretical Framework}

We define the MB density \(\rho(x)\) as the conditional mutual information:
\[
\rho(x) := I\big(s_{\text{int}} : s_{\text{ext}} \mid s_{\text{blanket}} = x\big)
\]
This quantity expresses the degree to which blanket states mediate information flow. A low \(\rho(x)\) indicates strong coupling (porous blanket), while a high \(\rho(x)\) indicates statistical insulation. Estimating \(\rho(x)\) from samples requires non-parametric tools, for which we adopt the KSG estimator.

\subsection*{The KSG Estimator}

Given samples \((x_i, y_i)\), the mutual information estimator is:
\[
\hat{I}_{\text{KSG}}(X; Y) = \psi(k) + \psi(N) - \frac{1}{N} \sum_{i=1}^N \left[ \psi(n_x(i) + 1) + \psi(n_y(i) + 1) \right]
\]
where \(\psi(\cdot)\) is the digamma function, \(k\) is the neighbor order, and \(n_x(i), n_y(i)\) count neighbors in the marginal spaces within joint-space neighborhoods.

To estimate conditional mutual information:
\[
I(X; Y \mid Z) \approx I(X; [Y,Z]) - I(X; Z)
\]
This can be computed by applying the KSG estimator to \((X, [Y,Z])\) and \((X, Z)\).

\subsection*{Simulation Example}

We simulate the following generative process:
\[
\begin{aligned}
s_{\text{ext}} &\sim \mathcal{U}(0, 1) \\
s_{\text{blanket}} &= \sin(2\pi s_{\text{ext}}) + \eta_1, \quad \eta_1 \sim \mathcal{N}(0, 0.05) \\
s_{\text{int}} &= \cos(2\pi s_{\text{blanket}}) + \eta_2, \quad \eta_2 \sim \mathcal{N}(0, 0.05)
\end{aligned}
\]
This structure introduces nonlinear, noise-perturbed coupling between the variables. Using 300 samples and \(k=5\), the estimated mutual information values were:
\[
I(s_{\text{int}}; s_{\text{ext}}, s_{\text{blanket}}) \approx 1.88 \text{ nats}, \quad I(s_{\text{int}}; s_{\text{blanket}}) \approx 0.44 \text{ nats}
\]
So the estimated Markov blanket density is:
\[
\rho(x) = I(s_{\text{int}} : s_{\text{ext}} \mid s_{\text{blanket}}) \approx 1.44 \text{ nats}
\]

By constructing empirical spatial maps of \(\rho(x)\) across agent-environment interfaces, one can compute gradients and simulate agent dynamics based on variational flows. These flows reflect movement toward regions of maximal information exchange and support the main claim of this paper: that active inference policies emerge from navigating MB density fields. Therefore, by developing a KSG‐based algorithm to estimate $\rho(x)$ from sampled data, we open a research path for applying active inference to real sensory trajectories—where one must learn the blanket density from streaming measurements. One can now ask questions such as: How does estimation error in $\rho(x)$ affect planning? What are the optimal sampling strategies to reduce uncertainty in blanket density? How does noisy or partial observation of state‐variables bias inference of $\rho(x)$?


\subsection*{Error Propagation from KSG to Free‐Energy Descent (with \(p_{\mathrm{true}}\))}

In this subsections, we developed an active‐inference framework in which an agent navigates a continuous state space by descending the variational free energy \(F(x) = -\ln p_{\mathrm{true}}(s(x),\eta(x))\). Central to this dynamic is the MB density \(\rho(x)\), which the agent must estimate in real time via a KSG mutual‐information estimator, yielding \(\rho_N(x)\).  Because \(\rho_N(x)\) inevitably differs from the true blanket density \(\rho_{\mathrm{true}}(x)\), our goal here is to quantify precisely how the resulting pointwise error \(\Delta \rho_N(x)\) slows the agent’s descent of \(F(x)\).  Concretely, we:

\begin{enumerate}
  \item Bound \(\Delta \rho_N(x)\) uniformly over \(\Omega\), showing it scales like \(N^{-1/(d+1)}\) with high probability.
  \item Translate that error into a perturbation of the agent’s velocity \(\dot x = -\bigl[1-\rho(x)\bigr]\nabla_x F(x)\), yielding a local slowdown proportional to \(\Delta\rho_N(x)\).
  \item Integrate these local effects along a finite descent path to derive an asymptotic bound on the total “convergence delay” \(T_N - T\), proving it is \(O_p\bigl(N^{-1/(d+1)}\bigr)\).
\end{enumerate}

These results make explicit how KSG estimation accuracy, sample size \(N\), and ambient dimension \(d\) jointly determine the agent’s performance in minimizing free energy.  

In this subsection, we incorporate the notation 
\[
p_{\mathrm{true}}(s(x),\eta(x))
\;=\;
\text{the true joint density of sensory signals }s(x)\text{ and latent states }\eta(x),
\]
which underlies the variational free energy
\[
F(x) \;=\; -\ln p_{\mathrm{true}}\bigl(s(x),\eta(x)\bigr),
\]
evaluated at each position \(x\in\Omega\).  We also recall the notation
\[
\rho_{\mathrm{true}}(x)
\;=\;
1 \;-\;
\frac{I_{\mathrm{true}}\bigl(I(x);E(x)\mid B(x)\bigr)}
     {\,I_{\mathrm{true}}\bigl(I(x);E(x)\bigr)\!},
\]
which is the ground‐truth MB density at \(x\).  Here:
\begin{itemize}
  \item \(I_{\mathrm{true}}(I(x);E(x))\) is the true mutual information between internal variables \(I(x)\) (within a ball of radius \(r_1\) around \(x\)) and external variables \(E(x)\) (outside a ball of radius \(r_2\)).  
  \item \(I_{\mathrm{true}}(I(x);E(x)\mid B(x))\) is the true conditional mutual information when conditioning on blanket variables \(B(x)\) in the shell \(r_1 \le \lVert y-x\rVert < r_2\).  
\end{itemize}

We estimate \(\rho_{\mathrm{true}}(x)\) via the KSG‐estimator from \(N\) samples, defining
\[
\rho_N(x)
\;=\;
1 \;-\;
\frac{\widehat I\bigl(I(x);E(x)\mid B(x)\bigr)}
     {\,\widehat I\bigl(I(x);E(x)\bigr)\!}.
\]
Our goal is to quantify how the pointwise estimation error 
\(\Delta \rho_N(x)=\rho_N(x)-\rho_{\mathrm{true}}(x)\) 
propagates through the active‐inference dynamics
\[
\dot x 
\;=\;
- \bigl[\,1 - \rho(x)\bigr]\,\nabla_x F(x),
\]
and in particular how it delays the time to descend to a given free‐energy level.  Throughout, we assume:

\begin{enumerate}[label=(\roman*)]
  \item \textbf{Regularity of \(F\).}  
    \(F(x) = -\ln p_{\mathrm{true}}\bigl(s(x),\eta(x)\bigr)\) is $C^2$ on \(\Omega\), and there exist constants $0 < m \le L_F$ such that 
    \[
      m \;\le\; \|\nabla_x F(x)\|\;\le\; L_F
      \quad
      \text{for all }x\in\Omega
    \]
    (away from any isolated critical points).
  
  \item \textbf{Regularity of \(\rho_{\mathrm{true}}\).}  
    \(\rho_{\mathrm{true}}\in C^1(\Omega)\) with 
    $\lVert \nabla \rho_{\mathrm{true}}(x)\rVert \le L_\rho$ uniformly, and 
    $0 \le \rho_{\mathrm{true}}(x) \le 1 - \delta$ on the regions traversed by the agent (for some $\delta>0$).

  \item \textbf{KSG estimation error.}  
    For radii $r_1(N),\,r_2(N)\sim N^{-1/(d+1)}$, the KSG estimator of each mutual‐information term satisfies, uniformly on \(\Omega\),
    \[
      \bigl|\widehat I(I;E)(x) - I_{\mathrm{true}}(I;E)(x)\bigr|
      = O_p\bigl(N^{-\tfrac{1}{d+1}}\bigr),
      \quad
      \bigl|\widehat I(I;E\mid B)(x) - I_{\mathrm{true}}(I;E\mid B)(x)\bigr|
      = O_p\bigl(N^{-\tfrac{1}{d+1}}\bigr).
    \]
\end{enumerate}

Under these conditions, we prove:

\begin{lemma}[Uniform Consistency of \(\rho_N\)]
\label{lem:consistency_with_ptrue}
If $r_1(N),r_2(N)\approx c\,N^{-1/(d+1)}$, then there exists a constant $C_1>0$ such that, with probability tending to 1,
\[
  \sup_{x\in\Omega} 
  \bigl|\rho_N(x) - \rho_{\mathrm{true}}(x)\bigr|
  \;\le\;
  C_1\,N^{-\frac{1}{\,d+1\,}}.
\]
\end{lemma}

\begin{proof}[Proof Sketch]
Write
\[
  \rho_N(x)
  = 1 - \frac{\widehat I_c(x)}{\widehat I_u(x)},
  \qquad
  \rho_{\mathrm{true}}(x)
  = 1 - \frac{I_c(x)}{I_u(x)},
\]
where 
\(
I_u(x)=I_{\mathrm{true}}\bigl(I(x);E(x)\bigr), 
I_c(x)=I_{\mathrm{true}}\bigl(I(x);E(x)\mid B(x)\bigr),
\)
\(
\widehat I_u(x)=\widehat I\bigl(I(x);E(x)\bigr),
\widehat I_c(x)=\widehat I\bigl(I(x);E(x)\mid B(x)\bigr).
\)
Then
\[
  \rho_N(x) - \rho_{\mathrm{true}}(x)
  = 
  \frac{\,I_c(x)\,\bigl[\widehat I_u(x)-I_u(x)\bigr]\!}{I_u(x)\,\widehat I_u(x)}
  \;-\; \frac{\widehat I_c(x)-I_c(x)}{\widehat I_u(x)}.
\]
Since $I_u(x)\ge c_I>0$ and $\widehat I_u(x)\to I_u(x)$ uniformly, both denominators remain bounded below.  Hence
\[
  \bigl|\rho_N(x)-\rho_{\mathrm{true}}(x)\bigr|
  \le
  \frac{|I_c(x)|}{c_I^2}\,\bigl|\widehat I_u(x)-I_u(x)\bigr|
  \;+\;
  \frac{\bigl|\widehat I_c(x)-I_c(x)\bigr|}{c_I}
  \;+\; o_p\bigl(N^{-\tfrac{1}{\,d+1\,}}\bigr).
\]
By assumption each numerator is $O_p(N^{-1/(d+1)})$ uniformly on $\Omega$, so
\(\sup_x|\rho_N(x)-\rho_{\mathrm{true}}(x)| = O_p(N^{-1/(d+1)})\). 
\end{proof}

\begin{lemma}[Local Velocity Perturbation]
\label{lem:velocity_with_ptrue}
Fix a point $x\in\Omega$ with $\|\nabla_x F(x)\|\ge m>0$.  Define the ``ideal’’ velocity
\[
  \dot x_{\mathrm{true}}
  \;=\;
  -\bigl[\,1 - \rho_{\mathrm{true}}(x)\bigr]\,\nabla_x F(x),
\]
and the ``noisy’’ velocity
\[
  \dot x_N
  \;=\;
  -\bigl[\,1 - \rho_N(x)\bigr]\,\nabla_x F(x).
\]
Then
\[
  \bigl\|\dot x_N - \dot x_{\mathrm{true}}\bigr\|
  = \bigl|\rho_N(x) - \rho_{\mathrm{true}}(x)\bigr|\;\|\nabla_x F(x)\|.
\]
In particular, if $\sup_x|\rho_N(x)-\rho_{\mathrm{true}}(x)| \le \varepsilon_N$, then
\[
  \|\dot x_N - \dot x_{\mathrm{true}}\|
  \;\le\; L_F\,\varepsilon_N,
  \quad
  \|\dot x_N - \dot x_{\mathrm{true}}\|
  \;\ge\; m\,\bigl|\rho_N(x)-\rho_{\mathrm{true}}(x)\bigr|.
\]
\end{lemma}

\begin{proof}
Subtracting gives
\[
  \dot x_N - \dot x_{\mathrm{true}}
  = -\bigl[\,1-\rho_N(x)\bigr]\nabla_x F(x)
    + \bigl[\,1-\rho_{\mathrm{true}}(x)\bigr]\nabla_x F(x)
  = -\bigl[\rho_N(x)-\rho_{\mathrm{true}}(x)\bigr]\nabla_x F(x).
\]
Taking norms yields the desired bounds.
\end{proof}

\begin{lemma}[Convergence Delay over a Finite Curve]
\label{lem:delay_with_ptrue}
Let $x_{\mathrm{true}}(t)$ solve
\[
  \dot x_{\mathrm{true}}
  = -\bigl[\,1 - \rho_{\mathrm{true}}(x)\bigr]\,\nabla_x F(x),
  \quad
  x_{\mathrm{true}}(0)=x_0,
\]
and let $x_N(t)$ solve
\[
  \dot x_N
  = -\bigl[\,1 - \rho_N(x)\bigr]\,\nabla_x F(x),
  \quad
  x_N(0)=x_0.
\]
Suppose that, over $t\in[0,T]$, the ideal trajectory $x_{\mathrm{true}}(t)$ travels a curve of length $\ell$ and satisfies $\|\nabla_x F(x)\|\ge m>0$.  If 
\[
  \sup_{x\in\Omega}\bigl|\rho_N(x)-\rho_{\mathrm{true}}(x)\bigr|\le \varepsilon_N,
  \quad
  \rho_N(x)\le 1-\delta \quad(\delta>0)
  \quad\text{along that region},
\]
then there exists $C_2>0$, depending on $m,L_F,\delta,\ell$, such that for sufficiently large $N$,
\[
  \bigl|\,T_N - T\bigr|
  \;\le\;
  C_2\,\varepsilon_N.
\]
Hence $T_N - T = O_p\bigl(N^{-\tfrac{1}{\,d+1\,}}\bigr)$.
\end{lemma}

\begin{proof}[Proof Sketch]
\begin{enumerate}[label=(\arabic*)]
  \item \emph{Ideal descent rate.}  For $t\in [0,T]$,
    \[
      \frac{d}{dt}F\bigl(x_{\mathrm{true}}(t)\bigr)
      = -\bigl[\,1-\rho_{\mathrm{true}}(x)\bigr]\|\nabla_x F(x)\|^2
      \;\le\; -\beta\,m^2,
    \]
    where $\beta = \inf_{x\in\Omega}(1-\rho_{\mathrm{true}}(x))>0$.  Thus $F(x_{\mathrm{true}})$ decreases at least at rate $\beta\,m^2$.

  \item \emph{Noisy descent rate.}  Similarly,
    \[
      \frac{d}{dt}F\bigl(x_N(t)\bigr)
      = -\bigl[\,1-\rho_N(x)\bigr]\|\nabla_x F(x)\|^2
      = -\bigl[\,1-\rho_{\mathrm{true}}(x)\bigr]\|\nabla_x F(x)\|^2 
        - \bigl[\rho_N(x)-\rho_{\mathrm{true}}(x)\bigr]\|\nabla_x F(x)\|^2.
    \]
    Hence
    \[
      \Bigl|\tfrac{d}{dt}F(x_N) - \tfrac{d}{dt}F(x_{\mathrm{true}})\Bigr|
      = \bigl|\rho_N(x)-\rho_{\mathrm{true}}(x)\bigr|\;\|\nabla_x F(x)\|^2
      \;\le\; L_F^2\,\varepsilon_N.
    \]

  \item \emph{Gap at time $T$.}  
    Let $\alpha = F\bigl(x_{\mathrm{true}}(T)\bigr)\).  Define 
    \[
      h(t) = F\bigl(x_N(t)\bigr) - F\bigl(x_{\mathrm{true}}(t)\bigr), 
      \quad h(0)=0.
    \]
    Then
    \[
      \frac{dh}{dt}
      = \frac{d}{dt}F(x_N) - \frac{d}{dt}F(x_{\mathrm{true}})
      \;\le\; L_F^2\,\varepsilon_N.
    \]
    Integrating from $0$ to $T$ yields
    \[
      F\bigl(x_N(T)\bigr) - \alpha = h(T) \; \le \; T\,L_F^2\,\varepsilon_N.
    \]
    Therefore 
    \(
      F(x_N(T)) \le \alpha + T\,L_F^2\,\varepsilon_N.
    \)

  \item \emph{Extra time to reach \(\alpha\).}  
    At $t=T$, $F(x_N(T))$ exceeds $\alpha$ by at most $T\,L_F^2\,\varepsilon_N$.  Since along $x_N$,
    \[
      \Bigl|\frac{d}{dt}F(x_N)\Bigr|
      = \bigl[\,1-\rho_N(x)\bigr]\|\nabla_x F(x)\|^2
      \;\ge\; \delta\,m^2,
    \]
    it takes at most
    \[
      \Delta T 
      \;\le\; \frac{T\,L_F^2\,\varepsilon_N}{\,\delta\,m^2\,}
      \;=\; C_2\,\varepsilon_N
    \]
    additional time for $F(x_N)$ to drop from $\alpha + T\,L_F^2\,\varepsilon_N$ to $\alpha$.  Hence
    \(
      \lvert T_N - T\rvert \le C_2\,\varepsilon_N
    \).

  \item \emph{Scaling of \(\varepsilon_N\).}  
    From Lemma~\ref{lem:consistency_with_ptrue}, $\varepsilon_N = O_p(N^{-1/(d+1)})$.  Consequently,
    \[
      T_N - T = O_p\bigl(N^{-\tfrac{1}{\,d+1\,}}\bigr).
    \]
\end{enumerate}
\end{proof}

\noindent\textbf{Asymptotic Conclusion.}  
From Lemmas~\ref{lem:consistency_with_ptrue}–\ref{lem:delay_with_ptrue}, we obtain a high‐probability bound on the convergence delay:
\[
  \bigl|\,T_N - T\bigr|
  \;=\; O_p\!\Bigl(N^{-\tfrac{1}{\,d+1\,}}\Bigr).
\]
\begin{itemize}
  \item \emph{Dependence on \(d\).}  The rate \(N^{-1/(d+1)}\) reflects the curse of dimensionality: more samples are needed in higher \(d\) to achieve the same $\varepsilon_N$ and delay $T_N-T$.
  \item \emph{Role of \(r_1,r_2\).}  Choosing $r_i(N)\propto N^{-1/(d+1)}$ balances bias and variance in KSG estimates; deviating worsens $\varepsilon_N$ and lengthens $T_N-T$.
  \item \emph{Uniformity on \(\Omega\).}  Since \(\varepsilon_N\) is uniform over \(\Omega\), no path can bypass this bound, provided $\|\nabla_xF\|\ge m>0$ and $\rho_N(x)\le1-\delta$ along it.
  \item \emph{High‐probability vs.\ expectation.}  All bounds hold with probability $\to1$.  With mild moment conditions, $\mathbb{E}[|T_N-T|] = O(N^{-1/(d+1)})$ follows.
\end{itemize}

In summary, any local estimation error \(\rho_N(x)-\rho_{\mathrm{true}}(x)\) of order \(\varepsilon_N\) induces a slowdown in $\dot x$ of order $O(\varepsilon_N)$, leading to a total convergence delay of order $O(\varepsilon_N)$.  Since $\varepsilon_N = O_p(N^{-1/(d+1)})$, we conclude that an agent must collect $N \sim \epsilon^{-(d+1)}$ samples to bound its convergence delay by \(\epsilon>0\). This quantifies precisely how KSG estimation accuracy, sample size, and dimensionality jointly determine the agent’s performance in free‐energy minimization.
\bigskip
\begin{figure}[H]
    \centering
    \includegraphics[width=1\linewidth]{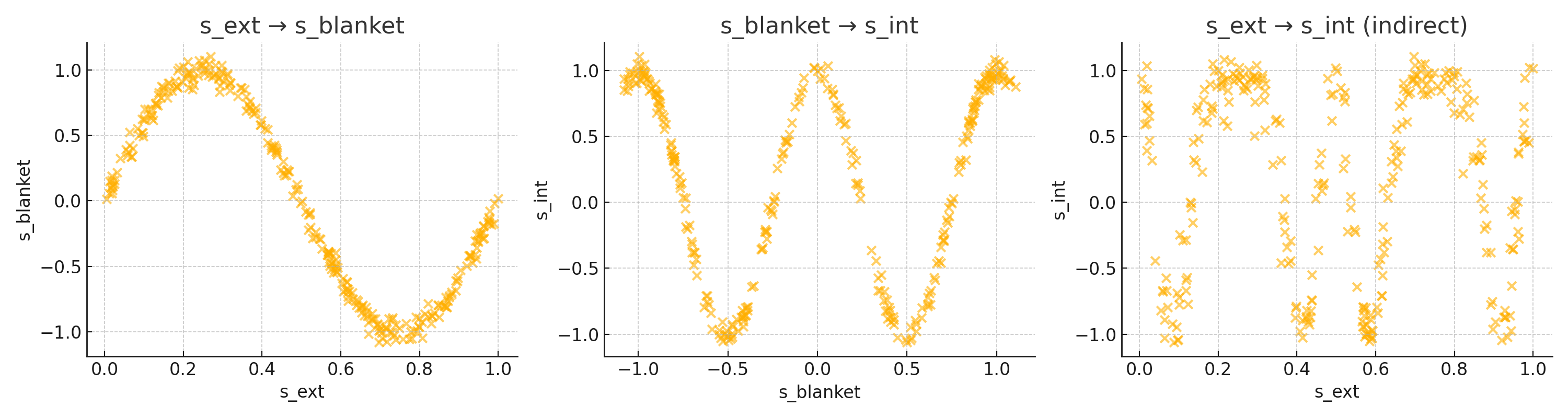}
\caption{\textbf{Simulation of Conditional Dependencies Mediated by a Markov Blanket}. Synthetic generative structure: (left) external input \(s_{\text{ext}}\); (center) mediated blanket variable \(s_{\text{blanket}}\); (right) internal response \(s_{\text{int}}\). This figure illustrates the informational structure of a synthetic agent-environment system composed of three variables: an external state \(s_{\text{ext}}\), a blanket state \(s_{\text{blanket}}\), and an internal state \(s_{\text{int}}\). The left panel shows the mapping from \(s_{\text{ext}}\) to \(s_{\text{blanket}}\), which is generated by a sinusoidal function with added noise. The structured, curved distribution reflects a strong but noisy dependence between external and blanket states. The middle panel displays the relationship between \(s_{\text{blanket}}\) and \(s_{\text{int}}\), also nonlinear and structured, indicating that internal states are tightly coupled to the blanket dynamics. The right panel shows the direct relationship between \(s_{\text{ext}}\) and \(s_{\text{int}}\), which appears more diffuse. Although some dependency remains, the structure is significantly weaker, because the internal state is influenced by the external state only indirectly through the blanket. Together, these plots demonstrate the mediating role of the blanket state in shaping the flow of information from the external to the internal system. This functional mediation is the defining property of a Markov blanket. The figure supports the idea that this mediation can vary in strength across space, and that such variation can be formally quantified as MB density. Using estimators such as the KSG method, this density can be empirically estimated from data, allowing for simulation and validation of the theoretical framework presented in the main text.}
\label{fig:mbd_sim}
\end{figure}

\appendix
\section*{Appendix D: Mathematical Background}
\addcontentsline{toc}{section}{Appendix: Mathematical Background}

\subsection*{Overview}
In this appendix, we collect and summarize the principal mathematical concepts, definitions, and results that underpin the main text. The goal is to provide a concise but self‐contained exposition of the background material required to follow the formal arguments and proofs in this paper. We assume that the reader is familiar with basic real analysis and elementary probability theory; we then introduce, in turn:
\begin{itemize}
    \item The notions of \emph{entropy}, \emph{mutual information}, and \emph{conditional mutual information} in both the discrete and continuous settings.
    \item The \emph{nonparametric} estimation of mutual information via the Kraskov–Stögbauer–Grassberger (KSG) $k$‐nearest‐neighbors (kNN) estimator.
    \item The concept of convergence in the norm $C^1(K)$ for functions defined on compact subsets $K\subset \mathbb{R}^n$.
    \item The theory of \emph{gradient flows} and ordinary differential equations (ODEs) of the form $\dot x(t) = -g(x(t))\,\nabla F(x(t))$, including existence, uniqueness, and basic stability estimates.
    \item Lipschitz continuity, differentiability classes ($C^k$), and regularity properties for functions on Euclidean spaces.
    \item Basic ideas from the theory of \emph{random fields} or \emph{stochastic processes indexed by space}, including covariance functions, stationarity, and concentration inequalities (in particular Hoeffding’s inequality).
    \item Notions from \emph{geometric measure theory} regarding compact domains with smooth boundary, volume (Lebesgue measure), and balls $\mathrm{Ball}(x;r)$ in $\mathbb{R}^n$.
\end{itemize}
Throughout, we adopt the following notational conventions:
\begin{itemize}
    \item $\mathbb{R}^n$ denotes \(n\)--dimensional Euclidean space, with the standard Euclidean norm $\|x\| = \sqrt{x_1^2 + \cdots + x_n^2}$.
    \item $\Omega \subset \mathbb{R}^n$ will denote a compact set with $C^2$ boundary, or more generally a domain (open connected set) whose closure $\overline{\Omega}$ is compact.
    \item Given a probability density $p(x)$ on $\mathbb{R}^n$, $H(p)$ denotes its (differential) entropy, and $I(X;Y)$ denotes mutual information between random variables $X,Y$ (possibly vector‐valued).
    \item For an open set $U\subset \mathbb{R}^n$, $C^k(U)$ is the space of $k$--times continuously differentiable real‐valued functions on $U$, and $\|f\|_{C^1(K)}$ denotes the $C^1$—norm on a compact $K\subset U$.
    \item For random variables indexed by points in space (a “random field”), we often write $\rho(x,\theta)$ where $\theta$ is a point in some probability space $(\Theta,\mathcal{F},\mathbb{P})$.
\end{itemize}

\subsection*{Entropy and Mutual Information}
\label{sec:entropy}
\subsubsection*{Discrete Entropy}
Let $X$ be a discrete random variable taking values in a finite or countable set $\mathcal{X}$, with probability mass function (pmf) $p_X(x) = \mathbb{P}(X=x)$. The \emph{Shannon entropy} of $X$ is
\[
H(X) \;=\; -\sum_{x\in\mathcal{X}} p_X(x)\,\ln p_X(x)\,,
\]
where throughout $\ln$ denotes the natural logarithm. Entropy $H(X)$ measures the expected “surprisal” of $X$ and satisfies $0 \le H(X) \le \ln|\mathcal{X}|$ when $|\mathcal{X}| < \infty$.

\subsubsection*{Differential Entropy}
When $X$ is a continuous random vector in $\mathbb{R}^d$ with probability density function (pdf) $p_X(x)$, its \emph{differential entropy} is defined as
\[
h(X) \;=\; -\int_{\mathbb{R}^d} p_X(x)\,\ln p_X(x)\,dx,
\]
provided the integral exists (i.e., $p_X$ is absolutely continuous and $\int p_X\,|\ln p_X| < \infty$). Unlike discrete entropy, differential entropy can be negative and is not invariant under change of variable.

\subsubsection*{Mutual Information}

Given two random variables (or vectors) \(X\) and \(Y\), with joint distribution \(p_{X,Y}(x,y)\) and marginals \(p_X(x)\), \(p_Y(y)\), the \emph{mutual information} between \(X\) and \(Y\) is defined by
\[
  I(X;Y) \;=\; \int_{\mathbb{R}^d \times \mathbb{R}^{d'}} 
    p_{X,Y}(x,y)\,\ln\!\Bigl(\frac{p_{X,Y}(x,y)}{\,p_X(x)\,p_Y(y)\,}\Bigr)\;dx\,dy
\]
in the continuous case, or the analogous sum in the discrete case. Equivalently, in the discrete setting:
\[
  I(X;Y) \;=\; H(X) \;+\; H(Y) \;-\; H(X,Y),
\]
and in the continuous setting:
\[
  I(X;Y) \;=\; h(X) \;+\; h(Y) \;-\; h(X,Y).
\]
Mutual information is always nonnegative, i.e.\ \(I(X;Y)\ge 0\), and vanishes precisely when \(X\) and \(Y\) are independent. It can also be written as the Kullback–Leibler divergence
\[
  I(X;Y) \;=\; D_{\mathrm{KL}}\bigl(p_{X,Y}\;\big\|\;p_X \otimes p_Y\bigr).
\]

\subsubsection*{Conditional Mutual Information}
For three random variables (vectors) $X$, $Y$, and $Z$, the \emph{conditional mutual information} of $X$ and $Y$ given $Z$ is
\[
I\bigl(X;Y \mid Z\bigr) 
\;=\;
\mathbb{E}_{Z}\Bigl[
D_{\mathrm{KL}}\bigl(p_{X,Y\mid Z}\,\big\|\,p_{X\mid Z}\otimes p_{Y\mid Z}\bigr)
\Bigr].
\]
Equivalently, in terms of (differential) entropies:
\[
I\bigl(X;Y \mid Z\bigr) 
\;=\; H(X \mid Z) + H(Y \mid Z) - H(X,Y \mid Z),
\]
or in the continuous case
\[
I\bigl(X;Y\mid Z\bigr) \;=\; h(X \mid Z) + h(Y \mid Z) - h(X,Y \mid Z).
\]
An equivalent formula in continuous form is
\[
I(X;Y \mid Z) 
= \int\int\int p_{X,Y,Z}(x,y,z)\,\ln \frac{p_{X,Y \mid Z}(x,y\,|\,z)}{\,p_{X\mid Z}(x\,|\,z)\,p_{Y\mid Z}(y\,|\,z)\,}\;dx\,dy\,dz.
\]
Conditional mutual information measures the residual statistical dependence between $X$ and $Y$ once $Z$ is known. In our context, $I$ (internal states), $B$ (blanket) and $E$ (external states) play the roles of $X$, $Z$, and $Y$ respectively.

\subsubsection*{Normalized “Blanket Strength”}
In the paper, the \emph{Markov blanket strength} is defined at a point $x$ by
\[
S(x) \;=\; 1 \;-\; \frac{\,I\bigl(I;E \,\mid\, B\bigr)\,}{\,I\bigl(I;E\bigr)}, 
\]
and the associated MB density by $\rho(x) = S(x)$. Here $I(I;E)$ and $I(I;E\mid B)$ denote the marginal and conditional mutual information restricted to the subsets $I(x)$, $B(x)$, and $E(x)$ around $x$. One must therefore be fluent in all of the foregoing definitions.

\subsection*{Nonparametric Estimation of Mutual Information via KSG--kNN}
\label{sec:ksg}

\subsubsection*{Nearest‐Neighbor Distances and Entropy Estimation}
Given a sample $\{\,z_i\,\}_{i=1}^N \subset \mathbb{R}^d$, consider the distance to the $k$‐th nearest neighbor:
\[
\varepsilon_k(i) \;=\; \min \bigl\{\,r>0:\,|\{\,j\neq i : \|z_j - z_i\| \le r\}| \ge k\,\bigr\}.
\]
The classical \emph{Kozachenko–Leonenko} (KL) estimator for the (differential) entropy $h(Z)$ is
\[
\hat h_{\mathrm{KL}}(Z) 
= \psi(N) - \psi(k) + \ln\bigl(c_d\bigr) + \frac{d}{N} \sum_{i=1}^N \ln \varepsilon_k(i),
\]
where $\psi(\cdot)$ is the digamma function, $c_d = \pi^{d/2}/\Gamma(\frac{d}{2}+1)$ is the volume of the unit ball in $\mathbb{R}^d$, and $\varepsilon_k(i)$ is half the distance to the $k$‐th nearest neighbor when using the maximum norm (or Euclidean norm if appropriate correction is made).

\subsubsection*{Kraskov–Stögbauer–Grassberger (KSG) Estimator}
Kraskov, Stögbauer, and Grassberger (2004) generalized the KL estimator to estimate mutual information between two continuous random vectors $X \in \mathbb{R}^{d_x}$ and $Y \in \mathbb{R}^{d_y}$. Given samples $\{(x_i,y_i)\}_{i=1}^N$, for each $i$ define
\[
\varepsilon_k(i) = \min\{\,\max\{\|x_j - x_i\|_\infty,\,\|y_j - y_i\|_\infty\}: 1 \le j \le N,\; j \neq i,\; \text{rank}(j) = k\,\},
\]
i.e.\ the distance (in the maximum norm) to the $k$‐th nearest neighbor in the joint space $\mathbb{R}^{d_x+d_y}$. Then count
\[
n_{x}(i) = \bigl|\{\,j \neq i : \|x_j - x_i\|_\infty \le \varepsilon_k(i)\}\bigr|,\quad
n_{y}(i) = \bigl|\{\,j \neq i : \|y_j - y_i\|_\infty \le \varepsilon_k(i)\}\bigr|.
\]
The KSG estimator for mutual information is
\[
\hat I_{\mathrm{KSG}}(X;Y)
\;=\; \psi(k) 
- \frac1k 
+ \psi(N)
- \frac{1}{N} \sum_{i=1}^N \bigl[\psi\bigl(n_{x}(i) + 1\bigr) + \psi\bigl(n_{y}(i) + 1\bigr)\bigr],
\]
where $\psi$ is again the digamma function. Under mild regularity conditions on the joint density $p_{X,Y}$, this estimator is (asymptotically) unbiased and consistent for large $N$. An analogous procedure can be applied to estimate conditional mutual information $I(X;Y \mid Z)$ by conditioning on $Z$ in a similar nearest‐neighbor scheme.

\subsubsection*{Convergence Properties and Conditions}
To employ KSG–kNN estimation within a theoretical analysis, one often needs more than mere pointwise consistency $\hat I \overset{p}{\longrightarrow} I_{\mathrm{true}}$. In the paper’s arguments, the key requirement is convergence in the norm
\[
\bigl\|\hat I(\cdot) - I_{\mathrm{true}}(\cdot)\bigr\|_{C^1(K)} \;=\; O_p\bigl(N^{-\alpha}\bigr),
\]
for some $\alpha>0$ and any compact $K$ in the domain. Convergence in $C^1(K)$ means:
\begin{enumerate}
    \item $\sup_{x\in K} |\hat I(x) - I_{\mathrm{true}}(x)| = O_p\bigl(N^{-\alpha}\bigr)$,
    \item $\sup_{x\in K} \bigl\|\nabla \hat I(x) - \nabla I_{\mathrm{true}}(x)\bigr\| = O_p\bigl(N^{-\alpha}\bigr)$,
\end{enumerate}
where $\nabla$ denotes the gradient with respect to the spatial coordinate $x\in\mathbb{R}^n$. Establishing such rates typically requires:
\begin{itemize}
    \item Assumptions that the true densities are bounded away from zero and infinity on $K$, with Lipschitz (or Hölder) continuous derivatives.
    \item Control of the bias and variance of the kNN–KSG estimator and uniformity over $x\in K$.
    \item Strong concentration inequalities for the nearest‐neighbor distances and counts $n_x(i)$, $n_y(i)$.
\end{itemize}

\subsection*{Convergence in the Norm \texorpdfstring{$C^1(K)$}{C1(K)}}
\label{sec:c1}

\subsubsection*{Function Spaces of Class \(C^1\)}
Let \(U \subset \mathbb{R}^n\) be an open set and \(K \subset U\) a compact subset. We say a function \(f:U \to \mathbb{R}\) belongs to \(C^1(U)\) if it is continuously differentiable, i.e., all first partial derivatives \(\partial f / \partial x_i\) exist and are continuous on \(U\). The restriction \(f|_K\) is then in \(C^1(K)\) in the sense that \(f\) and its gradient \(\nabla f\) are continuous on the compact set \(K\).  

\subsubsection*{Definition of the $C^1$--Norm}
For $f \in C^1(U)$ and a compact $K\subset U$, define
\[
\|f\|_{C^1(K)} 
\;=\; 
\sup_{x\in K} \bigl|f(x)\bigr| 
\;+\; 
\sup_{x\in K} \bigl\|\nabla f(x)\bigr\|.
\]
If $\|f - g\|_{C^1(K)} \to 0$ as some parameter (e.g., sample size $N$) grows, we say $f$ converges to $g$ in the $C^1(K)$ norm. This implies uniform convergence of both $f$ and $\nabla f$ on $K$.

\subsubsection*{Implications for Gradient‐Based Dynamics}
Convergence in $C^1(K)$ is crucial when one studies ODEs of the form
\[
\dot x(t) 
= -[\,1 - \rho_N(x(t))\,]\,\nabla F(x(t)),
\]
when $\rho_N(x) \to \rho_{\mathrm{true}}(x)$ in $C^1(K)$. Under such convergence, one can pass to the limit in the vector fields and deduce that solutions to the “estimated” flow approach those of the “true” flow, provided standard conditions of Lipschitz continuity hold. In particular, if $\rho_N\to\rho_{\mathrm{true}}$ and $\nabla\rho_N\to\nabla\rho_{\mathrm{true}}$ uniformly on $K$, then the direction of descent $-[1-\rho_N]\nabla F$ converges uniformly to $-[1-\rho_{\mathrm{true}}]\nabla F$. This is one of the stepping stones in the proof of Theorem~1.

\subsection*{Gradient Flows and Ordinary Differential Equations}
\label{sec:ode}

\subsubsection*{Gradient Descent in $\mathbb{R}^n$}
Given a continuously differentiable function $F:\Omega\subset\mathbb{R}^n \to \mathbb{R}$, the \emph{gradient flow} is the ODE
\[
\dot x(t) = -\nabla F\bigl(x(t)\bigr),
\]
with initial condition $x(0)=x_0\in\Omega$. Under the standard assumption that $\nabla F$ is globally Lipschitz (or at least locally Lipschitz on $\Omega$), the Picard–Lindelöf theorem guarantees the existence and uniqueness of a solution defined on a maximal interval. Moreover, if $\Omega$ is compact and $\nabla F$ is continuous, then the flow exists for all $t\ge 0$ and $F(x(t))$ is nonincreasing (since $\tfrac{d}{dt}F(x(t)) = \nabla F(x)\cdot \dot x = -\|\nabla F(x)\|^2 \le 0$).

\subsubsection*{Modified Gradient Flow with Mobility Function}
In the paper, the dynamics are modified as
\[
\dot x(t) = -M\bigl(x(t)\bigr)\,\nabla F\bigl(x(t)\bigr),
\]
where the \emph{mobility} (or “coupling”) function is
\[
M(x) = 1 - \rho(x), 
\quad 0 \le \rho(x)\le 1.
\]
Hence,
\[
\dot F\bigl(x(t)\bigr) 
\,=\, 
\nabla F(x)\cdot \dot x 
\,=\, 
- \bigl[1 - \rho(x)\bigr]\,\|\nabla F(x)\|^2 
\;\le\; 0.
\]
This shows that $F(x(t))$ is nonincreasing along trajectories. If $\rho(x)=1$ at some $x$, then $M(x)=0$ and $\dot x=0$, so the flow is \emph{frozen} at that point.

\subsubsection*{Existence and Uniqueness under Lipschitz Conditions}
Suppose $F\in C^2(\Omega)$, so $\nabla F$ is Lipschitz continuous on any compact $K\subset\Omega$ with Lipschitz constant $L_F$. If, in addition, $\rho(x)$ is $C^1$ on $K$, then $M(x) = 1 - \rho(x)$ is also Lipschitz on $K$. Consequently, the vector field $v(x) = -M(x)\,\nabla F(x)$ satisfies
\[
\|\,v(x) - v(y)\| 
\;=\; 
\bigl\|M(x)\,\nabla F(x) - M(y)\,\nabla F(y)\bigr\|
\;\le\; 
\bigl\|M(x)\,\bigl(\nabla F(x)-\nabla F(y)\bigr)\bigr\| 
+\|\,\nabla F(y)\|\;\bigl|M(x)-M(y)\bigr|.
\]
Since $M$ and $\nabla F$ are each bounded and Lipschitz on $K$, $v(x)$ is Lipschitz. Hence by the Picard–Lindelöf theorem, for each $x_0\in K$ there is a unique solution $x(t)\in K$ for some maximal interval of existence. If $K=\overline{\Omega}$ is compact and $v$ does not push trajectories outside $\overline{\Omega}$ (e.g., $v$ is tangent at the boundary), the solution exists for all $t\ge 0$ and remains in $\overline{\Omega}$.

\subsection*{Lipschitz Continuity and Differentiability Classes}
\label{sec:lipschitz}

\subsubsection*{$C^k$ Function Spaces}
Let $U\subset \mathbb{R}^n$ be open. We denote by $C^k(U)$ the set of functions $f:U\to\mathbb{R}$ whose partial derivatives up to order $k$ exist and are continuous on $U$. In particular:
\begin{itemize}
    \item $C^0(U)$ is the set of continuous functions.
    \item $C^1(U)$ consists of continuously differentiable functions; i.e., all first partials exist and are continuous.
    \item $C^2(U)$ consists of twice continuously differentiable functions, etc.
\end{itemize}
If $\Omega$ is compact with $C^2$ boundary, and $F\in C^2(\Omega)$, then $\nabla F$ is Lipschitz on $\Omega$ (since a continuously differentiable map on a compact set is automatically Lipschitz). Specifically, there exists $L_F>0$ such that
\[
\|\nabla F(x) - \nabla F(y)\| \;\le\; L_F \,\|x - y\|
\quad
\forall\,x,y\in \Omega.
\]

\subsubsection*{Lipschitz Continuity}
A function $f:K\to\mathbb{R}$ defined on a metric space $(K,d)$ is \emph{Lipschitz continuous} if there exists a constant $L\ge 0$ such that
\[
|f(x) - f(y)| \;\le\; L\,d(x,y)
\quad
\forall\,x,y \in K.
\]
If $f\in C^1(K)$ for a compact $K\subset\mathbb{R}^n$, then the mean value theorem implies $f$ is Lipschitz with Lipschitz constant
\[
L_f = \sup_{x\in K} \|\nabla f(x)\|.
\]
Analogously, a vector field $v:K\to\mathbb{R}^n$ is Lipschitz if $\sup_{x\in K}\|Dv(x)\|<\infty$, where $Dv(x)$ is the Jacobian matrix and $\|\cdot\|$ is the operator norm.

\subsection*{Random Fields and Concentration Inequalities}
\label{sec:randomfields}

\subsubsection*{Random Fields on a Compact Domain}
A \emph{random field} on a compact set $\Omega\subset\mathbb{R}^n$ is a collection of real‐valued random variables $\{\rho(x)\}_{x\in\Omega}$ defined on a common probability space $(\Theta,\mathcal{F},\mathbb{P})$. We denote $\rho(x,\theta)$ when emphasizing the dependence on the random element $\theta\in \Theta$. Conditions often imposed on $\rho(x,\theta)$ in the paper include:
\begin{itemize}
    \item \emph{Boundedness:} $0 \le \rho(x,\theta) \le 1$ for all $x,\theta$.
    \item \emph{Stationarity of mean:} $\mathbb{E}_\theta[\rho(x,\theta)] = \mu$ is constant for all $x\in\Omega$.
    \item \emph{Constant covariance with a deterministic field:} $\mathrm{Cov}(\rho(x),\,\|\nabla F(x)\|^2) = C$ is independent of $x$.
    \item \emph{Decay of spatial correlations:} There exists a length‐scale $\ell>0$ such that
    \[
    \bigl|\mathrm{Cov}\bigl(\rho(x),\,\rho(y)\bigr)\bigr| \;\le\; \sigma^2\exp\bigl(-\|x - y\|/\ell\bigr) 
    \quad \forall\,x,y\in\Omega.
    \]
\end{itemize}
The latter condition is a form of \emph{exponential mixing} or \emph{exponential decay of correlations} and ensures that values of $\rho$ at distant points become nearly independent.

\subsubsection*{Hoeffding’s Inequality for Bounded Random Variables}
Suppose $Z_1,\dots,Z_N$ are independent random variables with $a_i \le Z_i \le b_i$ almost surely. Then for any $\varepsilon>0$,
\[
\mathbb{P}\Bigl(\,\bigl|\tfrac{1}{N}\sum_{i=1}^N Z_i \;-\; \mathbb{E}[Z_i]\bigr|\ge \varepsilon\Bigr)
\;\le\; 
2\exp\!\Bigl(-\frac{2N^2\varepsilon^2}{\sum_{i=1}^N (b_i - a_i)^2}\Bigr).
\]
In the paper, to derive a \emph{uniform} high‐probability bound on 
\[
\varphi(x) = [1 - \rho(x)]\,\|\nabla F(x)\|^2
\]
over all $x\in\Omega$, one discretizes $\Omega$ by a finite grid $\{x^{(1)},\dots,x^{(N)}\}$ of mesh $\delta$. Then Hoeffding’s inequality yields, for each grid point $x^{(j)}$,
\[
\mathbb{P}\Bigl(\bigl|\varphi(x^{(j)}) - \mathbb{E}[\varphi(x^{(j)})]\bigr| \ge \varepsilon\Bigr)
\;\le\;
2\exp\bigl(-C\,\varepsilon^2\,\bigr)
\]
for some constant $C>0$ if $\varphi$ is bounded (since $0 \le \varphi \le \|\nabla F\|_\infty^2$). A union bound over all grid points (and then controlling the remainder of $\Omega$ by Lipschitz continuity) yields
\[
\mathbb{P}\Bigl(\sup_{x\in\Omega} \bigl|\varphi(x) - \mathbb{E}[\varphi(x)]\bigr| \ge \varepsilon\Bigr)
\;\le\;
N\,\cdot 2\exp\bigl(-C\,\varepsilon^2\bigr),
\]
so that with high probability the random field $\varphi(x)$ is uniformly close to its expectation.

\subsection*{Compact Domains and Volume in \texorpdfstring{\(\mathbb{R}^n\)}{R^n}}
\label{sec:geometry}

\subsubsection*{Compact Sets with Smooth Boundary}
Let \(\Omega \subset \mathbb{R}^n\) be a bounded open set whose boundary \(\partial\Omega\) is a \(C^2\) hypersurface. Such an \(\Omega\) is said to be a \emph{compact domain with \(C^2\) boundary} if \(\overline{\Omega}\) is compact and \(\partial\Omega\) is a \(C^2\) manifold. In particular:
\begin{itemize}
    \item There exist coordinate charts \((U_i,\psi_i)\) covering \(\partial\Omega\) such that \(\psi_i(U_i)\) is open in \(\mathbb{R}^{n-1}\) and \(\partial\Omega\) is locally given by \(x_n = \phi_i(x_1,\dots,x_{n-1})\) for some \(C^2\) function \(\phi_i\).  
    \item \(\Omega\) satisfies an \emph{interior sphere condition}: every point on \(\partial\Omega\) has a ball of positive radius contained in \(\Omega\) tangent to \(\partial\Omega\) at that point.
\end{itemize}
These properties guarantee that standard PDE and ODE results (e.g., existence of flows that remain in \(\Omega\) with vector fields tangent at the boundary) apply.

\subsubsection*{Volume of Balls}
The $n$‐dimensional Lebesgue measure (volume) of a ball of radius $r>0$ in $\mathbb{R}^n$ is
\[
\mathrm{Vol}\bigl(\mathrm{Ball}(x;r)\bigr) 
= \mathrm{Vol}\bigl(\mathrm{Ball}(0;r)\bigr) 
= c_n\,r^n,
\]
where 
\[
c_n = \frac{\pi^{n/2}}{\Gamma\bigl(\frac{n}{2}+1\bigr)}.
\]
In many consistency arguments for kNN estimators, one requires that $N\,\mathrm{Vol}(\mathrm{Ball}(x;r_2(N))) \to \infty$ as $N\to\infty$, which ensures that, on average, there are infinitely many sample points in an $r_2(N)$‐neighborhood of any given $x$. Usually $r_2(N)$ is chosen so that $N\,r_2(N)^n\to \infty$ but $r_2(N)\to0$.

\subsection*{Gradient Alignment and Monotonicity Conditions}
\label{sec:alignment}

\subsubsection*{Gradient Conditions for Theorem~1}
In Theorem~1, one assumes that there exist continuous functions
\[
I_{\mathrm{true}}\bigl(I(x);E(x)\bigr),\quad I_{\mathrm{true}}\bigl(I(x);E(x)\mid B(x)\bigr)
\;:\;D\subset\Omega \;\longrightarrow\; \mathbb{R},
\]
which are $C^1$ on an open set $D$, and satisfy
\[
\nabla\bigl[I_{\mathrm{true}}(I(x);E(x)\_]()
\]

\bigskip
\section*{Acknowledgements}

I thank Kobus and Stefan Esterhuysen for their assistance in developing the technical aspects of this paper.

\clearpage
\renewcommand{\refname}{References}

\end{document}